\documentclass[a4paper,11pt]{article}
\usepackage{amssymb,amsmath}
\usepackage{graphicx}\graphicspath{{Figures/}} 
\usepackage{color}
\usepackage{enumitem}
\usepackage{array}
\usepackage{braket}
\usepackage[a4paper,width=15.3cm]{geometry}
\usepackage{wrapfig}
\usepackage{cancel}
\usepackage{float}
\usepackage{slashed}
\usepackage{subcaption}
\usepackage{cases}

\usepackage[table]{xcolor}
\usepackage[normalem]{ulem}
\usepackage{soul}

\newcommand{\abs}[1]{\lvert#1\rvert}

\mathchardef\ordinarycolon\mathcode`\:
\mathcode`\:=\string"8000
\begingroup \catcode`\:=\active
\gdef:{\mathrel{\mathop\ordinarycolon}}
\endgroup

\pdfoutput=1 

\usepackage{jheppub}\makeatletter\gdef\@fpheader{}\makeatother
\usepackage[T1]{fontenc} 
\usepackage{tikz}
\usetikzlibrary{shapes.geometric, arrows}
\tikzstyle{startstop} = [rectangle, rounded corners, minimum width=3cm, minimum height=1cm,text centered, draw=black, fill=red!30]
\tikzstyle{io} = [trapezium, trapezium left angle=70, trapezium right angle=110, minimum width=3cm, minimum height=1cm, text centered, draw=black, fill=blue!30]
\tikzstyle{process} = [rectangle, minimum width=3cm, minimum height=1cm, text centered, draw=black, fill=orange!30]
\tikzstyle{decision} = [diamond, minimum width=3cm, minimum height=1cm, text centered, draw=black, fill=green!30]
\tikzstyle{arrow} = [thick,->,>=stealth]
\tikzstyle{io} = [rectangle, minimum width=3cm, minimum height=1cm, text centered, text width=3cm, draw=black, fill=green!30]
\tikzstyle{start} = [rectangle, rounded corners, minimum width=3cm, minimum height=1cm, text centered, text width=3cm, draw=black, fill=blue!30]

\title{Exploring New Physics in transition $b\to s\,\ell^+\ell^-$  through different $B_c\to D_s^{(\ast)} \,\ell^+\ell^-$ observables}

\author[a,b]{Qazi Maaz Us Salam\footnote{Corresponding author}}
 \author[b]{, Ishtiaq Ahmed}
\author[a]{, Rizwan Khalid}
 \author[a,c]{, Ibad Ur Rehman}

 \affiliation[a]{School of Science and Engineering, Lahore University of Management Sciences (LUMS), Opposite Sector U, D.H.A, Lahore 54792, Pakistan.}
 \affiliation[b]{National Center for Physics, Islamabad 44000, Pakistan.}
 \affiliation[c]{School Education and Literacy Department, Government of Sindh, Pakistan.}

\emailAdd{qazi.salam@lums.edu.pk}
\emailAdd{ishtiaq.ahmed@ncp.edu.pk}
\emailAdd{rizwan\_khalid@lums.edu.pk}
\emailAdd{ibadnustphy@gmail.com}

\abstract{Inspired by the discrepancies observed in the $b\to s\ell^+\ell^-$ neutral current decays, we study the decay channel $B_c\to D_s^{(\ast)} \,\ell^+\ell^-$ ($\ell=\mu,\tau$), which is based on the same flavor changing neutral current (FCNC) transition at the quark level. The current study shows that this decay channel can provide a useful probe for physics beyond the standard model. We use the helicity formalism while employing the effective theory approach where we include the effects of vector and axial vector `new' physics (NP) operators. In this study, we have computed the branching ratio $\mathcal{B}_r$, the $D_s^\ast$ helicity fraction $f_L$, the lepton forward-backward asymmetry $\mathcal{A}_{FB}$, and the lepton flavor universality ratio (LFU) $R^{\tau\mu}_{D_s^*}$. In addition, as a complementary check on the LFU, we also calculate the various other LFU observables, $R_{i}^{\tau\mu}$ where $i=A_{FB}$, $f_L$. We assume that the NP universal coupling is present for both muons and tauons, while the non-universal coupling is only present for muons. Regarding these couplings, we employ the latest global fit to the $b\to s\ell^+\ell^-$ data, which is recently computed in \cite{Alguero:2023jeh}. We give predictions of some of the mentioned observables within the SM and the various NP scenarios. We have found that not only are the considered observables sensitive to NP but are also helpful in distinguishing among the different NP scenarios. These results can be tested at the LHCb, HL-LHC, and FCC-ee, and therefore, a precise measurements of these observables not only deepens our understanding of the $b\to s\ell^+\ell^-$ process but also provides a window of opportunity to possibly study various NP scenarios.
}
\begin{document} 
\maketitle

\section{Introduction}
The flavor changing neutral current (FCNC) processes based on the 
$b\to s\ell^+\ell^-$ transitions are forbidden in the standard model of particle physics (SM) at the tree level and occur at the loop level. 
Due to this fact, the leading 
order contributions to these processes can, in principle, receive corrections from so-called new physics (NP) scenarios beyond the SM. Therefore, FCNC 
processes provide an attractive theoretical and experimental tool to test the SM and explore any NP~\cite{Albrecht:2021tul, London:2021lfn}.

In this context, the FCNC decay process $B\to K^{(\ast)}\ell^+\ell^-$ has been widely
investigated experimentally~\cite{CDF:1999uew, BaBar:2000jlq, Belle:2001oey, BaBar:2003szi, BaBar:2008jdv, Belle:2003ivt, Belle:2016fev, BELLE:2019xld, Belle:2021ecr, Belle:2009zue, Belle:2019oag, BaBar:2012mrf, CDF:2011buy, CMS:2015bcy, LHCb:2014vgu, LHCb:2016ykl, LHCb:2017avl, LHCb:2013ghj}. The LHCb and Belle collaborations have also reported important data on the $b\to d\ell^+\ell^-$ decay channels such as $B\to (\rho,\omega,\pi,\eta)\ell^+\ell^-$, 
$B_s^0\to\bar K^{*0}\mu^+\mu^-$ as well as various 
ratios related to $b\to d$ and $b\to s$ transitions~\cite{Belle:2024cis, LHCb:2017lpt, LHCb:2015hsa, LHCb:2012de, LHCb:2018rym}. In this regard, the lepton flavor universality LFU ratios $R_K$ and $R_{K^*}$ measured at LHCb indicated deviations from their SM values even after including quantum electrodynamic (QED) corrections~\cite{LHCb:2014vgu, LHCb:2017avl, LHCb:2022vje, Isidori:2022bzw}.
 Whereas the `anomalies' in $R_K$ and $R_{K^*}$, have gone away with more statistics~\cite{LHCb:2022vje, LHCb:2022qnv}, 
 other measurements, for instance, the branching ratios of decays like $B \to K^{(*)}\mu^+\mu^-$, $B_s \to \phi\mu^+\mu^-$ \cite{LHCb:2014cxe, LHCb:2016ykl, LHCb:2021zwz}, and $B \to K^*\mu^+\mu^-$ angular 
 observables, particularly, in $P_5^\prime$ \cite{LHCb:2020lmf, Descotes-Genon:2012isb} still show considerable (up to 2.5 $\sigma$) deviation from the SM values predictions~\cite{Alguero:2023jeh}.\footnote{The details 
 about these anomalies can be perused in the book by Artuso \emph{et al}~\cite{Artuso:2022ijh}
 and the excellent review article by London \emph{et al}~\cite{London:2021lfn} as well as references therein.} Moreover, the measurements 
of $R_{K^0_s}$ and $R_{K^{*+}}$ show some deviations from the SM~\cite{LHCb:2021lvy, LHCb:2017avl, CMS:2024syx}. 
In addition, regarding the semi-leptonic $B_s\to\phi\ell^+\ell^-$ decays, the D$\varnothing$ collaboration set some limits on the decay, CDF made a first observation on it and then LHCb collaborations also measured and reported the detailed
studies on it~\cite{CDF:2001yrm, CDF:2008zhr, D0:2006pmq, CDF:2011grz, LHCb:2013tgx, LHCb:2015wdu}. The LHCb has also reported the $B_s\to f_2^\prime\ell^+\ell^-$ decay 
process~\cite{LHCb:2021zwz}.

Similarly, in the last decades, the FCNC $b\to s\ell^+\ell^-$ process has also 
been studied theoretically in detail within the 
SM~\cite{Horgan:2015vla, Bobeth:2008ij, Bharucha:2015bzk, Gao:2019lta, Li:2009tx, Deandrea:2001qs, Dubnicka:2016nyy, Issadykov:2022imz, Bailey:2015dka, Ball:2004rg, Wu:2006rd, Cheng:2017bzz, Wang:2017jow, Lu:2018cfc, Gao:2021sav, Cui:2022zwm, Wang:2007an, Xiao:2013lia, Jin:2020jtu, Jin:2020qfp, Soni:2020bvu, Lu:2011jm, Ahmady:2019hag}, 
and to address the above mentioned deviations from the SM that occurred in these FCNC process, they are also studied in the various SM 
extensions~\cite{Li:2018rax, Barman:2018jhz, DelleRose:2019ukt, Ordell:2019zws, Marzo:2019ldg, Iguro:2018qzf, Iguro:2023jju, Aslam:2009cv, Trifinopoulos:2019lyo, Shaw:2019fin, Altmannshofer:2014cfa, Bhattacharya:2014wla, Crivellin:2015lwa, Falkowski:2015zwa, Bhattacharya:2016mcc, Falkowski:2018dsl, Dwivedi:2019uqd, Capdevila:2020rrl, Hiller:2014yaa, Gripaios:2014tna, Becirevic:2017jtw, Cornella:2019hct, DaRold:2019fiw, Popov:2019tyc, Datta:2019bzu, Crivellin:2019dwb, Iguro:2021kdw}.
Even though significant progress in the domain of $B_{d,s}$ rare semileptonic 
decays has been made both experimentally and theoretically, 
however, a consistent framework of NP is yet to be that can address all discrepancies in the current data. This situation has lead to the study of various additional decay channels corresponding to these FCNC transitions that can provide a complementary
check and further explore the same. In this regard, the decay channels $B\to (K_1,K_2)\mu^+\mu^-$ 
and $B_s\to f_2^\prime\mu^+\mu^-$ have been carefully analyzed from the point of 
view of in different NP scenarios~\cite{Huang:2018rys, Das:2018orb, Mohapatra:2021izl, Rajeev:2020aut, Bhutta:2024zwj, Ishaq:2013toa, MunirBhutta:2020ber}.

In a similar fashion, the semi leptonic decays of the $B_c$ meson, and, in particular, $B_c\to D_s^{(*)}\ell^+\ell^-$, can also be a good candidate to check the SM predictions regarding $b\to s\ell^+\ell^-$ transitions, in addition to serving as an ideal tool to probe any possible NP in this sector. The decay channel $ B_c \to D_s^{(*)} \ell^+ \ell^- $, although it has significantly smaller sample sizes compared to $ B \to K^{(*)} \ell^+ \ell^- $ and $ B_s \to \phi \ell^+ \ell^- $, offers unique theoretical and phenomenological advantages. Unlike the decays of lighter $ B $ mesons, which involve heavy-to-light transitions (e.g. $ b \to s $ with light hadronic final states such as $ K^* $ or $ \phi $, the $ B_c \to D_s^{(*)} \ell^+ \ell^- $ decay is a heavy-to-heavy transition. As in the heavy-to-light regime, the weak-decay form factors lose sensitivity to both the flavor and spin orientations of the heavy quark. Instead, they can all be encapsulated within a single universal function, known as the Isgur-Wise function \cite{Isgur:1990yhj}. However, when dealing with the $ B_c $ meson, the usual heavy flavor and spin symmetries require a more nuanced approach, as both the $ b $ and $ c $ quarks are heavy. This requires an analysis that adequately accounts for finite-quark-mass effects, providing a more precise and physically meaningful description \cite{Geng:2001vy}. This leads to better control over form factor calculations where theoretical uncertainties in form factors are typically reduced in heavy-to-heavy transitions.
Furthermore, the hadronic resonance effects in $ B \to K^* \ell^+ \ell^- $ and $ B_s \to \phi \ell^+ \ell^- $ are complicated by the broad widths of $ K^* $ ($\sim$ 50 MeV) and $ \phi $ ($\sim$ 4 MeV), leading to significant nonperturbative effects such as long-distance contributions from charm loops. In contrast,  The $ D_s^* $ width is much narrower (experimental upper limit $<$ 1.9 MeV) and the theoretical prediction is orders of magnitude smaller, reducing hadronic uncertainties. This allows for a cleaner extraction of short-distance effects and possible NP contributions in the Wilson coefficients $ C_9 $ and $ C_{10}$.

However, the charmed $B$ meson decays have not been as extensively studied as the strange $B$ decays for a variety of reasons. On the experimental front, the LHCb collaboration found that the fragmentation of the $B_c$ meson, $f_c$, is approximately a thousand 
times lesser than the fragmentation of the $B_u$ meson, $f_u$~\cite{LHCb:2019tea}. In particular, LHCb 
has set an upper limit for the decay $B_c^+\to D_s^+\ell^+\ell^-$ as $(f_c/f_u)\times\mathcal{B}_r(B_c^+\to D_s^+\ell^+\ell^-)<9.6\times10^{-8}$~\cite{LHCb:2023lyb}. 
Therefore, even though the branching ratios of $B_c\to D_s^{(*)}\ell^+\ell^-$ are of the same order of 
magnitude as the corresponding $B_s$ decays,~\cite{Wang:2014yia} the reconstruction of $D_s^*\to D_s\gamma$ makes this decay channel more 
difficult to measure at LHCb with the current luminosity. However, it has been demonstrated that partial reconstruction can be used to
cleanly reconstruct $B_c \to J/\psi D_s^*$ at LHCb \cite{LHCb:2013kwl}, and the same method
could also be applied by ATLAS and CMS. This can equally well be used for nonresonant $B_c \to
D_s^*\ell^+\ell^-$.  

The decay \( B_c \to D_s^* \ell^+ \ell^- \) faces experimental challenges due to the \( B_c \) meson’s short lifetime (~0.51 ps), limiting reconstructed events and precise measurements. Additionally, detecting neutrals and soft photons from \( D_s^* \) is inefficient, reducing signal yield. High-luminosity upgrades and future \( e^+e^- \) colliders could enhance sensitivity through improved reconstruction techniques and theoretical refinements. Future colliders like HL-LHC and FCC-ee will provide better platforms for studying semi-leptonic \( B_c \) decays.

From a theoretical perspective, the $B_c$ rare semileptonic decays, 
$B_c^+\to D_s^+\ell^+\ell^-$, have been studied using several approaches including the relativistic quark model, the light-front quark model, and QCD sum rules, \emph{etc}~\cite{Geng:2001vy,Ebert:2010dv,Azizi:2008vv, Wang:2014yia, Ivanov:2024iat}. Additionally, a host of NP implications for this decay have been studied in various extensions of SM such as a single Universal extra dimension, non-Universal $Z^\prime$, two Higgs Doublet Models, in addition to employing a model-independent approach~\cite{Yilmaz:2012ah, Maji:2020zlq, Maji:2020wer, Mohapatra:2021izl, Mohapatra:2021ynn, Dutta:2019wxo, Li:2023mrj, Mohapatra:2024lmp, Zaki:2023mcw}. Ref.~\cite{Maji:2020zlq} and Ref.~\cite{Maji:2020wer} explore non-universal $Z'$ and charged Higgs boson effects, respectively, while this work considers also the model independent vector and axial-vector operators, allowing for a more comprehensive analysis of $B_c \to D_s^{(*)} \mu^+ \mu^-$. Ref.~\cite{Mohapatra:2021ynn} considers both $B_c \to D_s^{(*)} \mu^+ \mu^-$ and $B_c \to D_s^{(*)} \nu \bar{\nu}$ decays within $Z'$ and leptoquark models. Our work complements this study by providing additional constraints on NP scenarios. Similarly, Ref.~\cite{Dutta:2019wxo} performs a model-independent analysis of NP contributions in $B_c \to D_s^{(*)} \mu^+ \mu^-$ decays, however, our work builds on this by including updated theoretical and experimental inputs, including more observables. Ref.~\cite{Li:2023mrj} analyzes angular distributions in $B_c \to D_s^*(\to D_s \pi) \ell^+ \ell^-$ decays within the SM, whereas the current study focus on different observables and their sensitivity to NP effects. In short, our study not only builds upon but also extends the existing literature by incorporating a more comprehansive analysis of NP scenarios, including a wider class of operators and a more detailed study of the observables of $B_c \to D_s^{(*)} \mu^+ \mu^-$ decays.

The latest LHCb results (Dec 2022) \cite{LHCb:2022qnv,LHCb:2022vje} and (Oct 2024) \cite{LHCb:2024rto} show \( R_K^{(*)} \) and \( R_\phi \) aligns with SM predictions, limiting LFU violation in the \(\mu/e\) sector \cite{SinghChundawat:2022ldm,Ciuchini:2022wbq,Wen:2023pfq}. This motivates exploring whether \( b \to s e^+ e^- \) and \( b \to s \mu^+ \mu^- \) data still allow LFU violation in the \(\tau - \mu\) sector, where new physics effects may arise. Therefore, in this manuscript, we focus on exploration of possible NP signatures in the decay $B_c\to D_s^{(\ast)} \,\ell^+\ell^-$ ($\ell=\mu,\tau$). It is significant to mention here that the tau pair mode is less studied in the literature; for instance, a recent study of the decay $B^0\to K^{\ast 0}\tau^+\tau^-$ \cite{SinghChundawat:2022ldm} shows that its branching fraction is still several orders of magnitude smaller than the upper limit set by the Belle experiment ($< 3.1\times 10^{-3}$ at a 90\% confidence level)  \cite{Belle:2021ecr}. However, this situation is expected to improve at FCC-ee colliders where these 
results hopefully be tested.

In this study, we use the helicity formalism for this decay by employing the effective theory approach where both the vector and axial vector NP operators are taken into account. 
In particular, we have calculated several observables, such as the branching ratio 
$\mathcal{B}_r$, the helicity fraction $f_L$, the lepton forward-backward asymmetry $A_{FB}$, and the lepton flavor universality ratio (LFU) $R^{\tau\mu}_{D_s^*}$.

In addition, as a complementary check on the LFU, we also calculate the ratio 
of different observables $R_{i}^{\tau\mu}$ where $i=A_{FB}$, $f_L$. We note that $B_c\to D_s^{(\ast)} \,\ell^+\ell^-$ has been studied within the 
context of universal couplings to leptons in the NP sector~\cite{Mohapatra:2024lmp}.
In the current study, we address the possibility of a non-universal NP 
coupling for the muon. We employ the latest global fit to the $b\to s\ell^+\ell^-$ data, which has most recently been computed in~\cite{Alguero:2023jeh}. 
Furthermore, to check the sensitivity of the NP couplings to the observable as a function of $q^2$, we set them by optimizing within their 1$\sigma$ ranges, which give the maximum and minimum deviation from their SM values. Moreover, we have also calculated the maximum and minimum variation after the integration over the low $q^2$ bin for $\mu$ and high $q^2$ bin for both $\mu$ and $\tau$.
In addition, to see the explicit dependence on the couplings, 
we have calculated the analytical expressions of these observables in terms of NP 
Wilson coefficients (WCs) and plotted them against the NP couplings in their 
$1\sigma$ range. These expressions should prove very useful for determining the precise values of the universal and non-universal couplings whenever needed.   

We use FeynCalc, a Mathematica package to solve the hadronic and leptonic parts, traces appearing in the 
analytical expressions, and to get the numerical values of the observables. We 
give our predictions of the mentioned observables for both $\mu$ and $\tau$ lepton final states within the SM and various NP scenarios. We identify observables that are not only sensitive to NP but also helpful in distinguishing among different NP scenarios. 

We now give the structure of the rest of this paper. In \S\ref{theoretical description}, we outline the theoretical framework that we have used to study the decay $B_c\to D_s^{\ast} \,\ell^+\ell^-$. We have given the parameterization of the $B_c\to D_s^{\ast}$ hadronic matrix elements in terms of the form factors, the outlined the helicity formalism, and used this to give the branching ratio and helicity fraction. We have also defined the forward-backward asymmetry, and a lepton flavor universality ratio.
We begin \S\ref{pheno analysics} by giving the values of various parameters and the form factors (which are calculated in the relativistic quark model) as well as 
describing the various NP scenarios that we consider. We then go on to also discuss the phenomenology of the observables considered in the presence of different NP scenarios. Finally, in \S\ref{conclusion}, we provide a summary and present the 
conclusion.

\section{Theoretical Description\label{theoretical description}}
In this section, we outline the theoretical framework that we use to compute the helicity amplitudes of $B_c \to D^{\ast}_{s}\ell^{+}\ell^{-}$ meson decays. We then write the differential decay rate in terms of the helicity amplitudes, and define the observables of interest, \emph{i.e.} the forward-backward asymmetry ($A_{FB}$), longitudinal 
helicity fraction ($f_L$), and lepton flavor universality ratio ($R_{D^{*}_s}$). 

\subsection{The Decay Amplitude for $ B_c \to D^{\ast}_{s}\ell^{+}\ell^{-}$}
First, we would like to mention here that $B_c \to D_s^* \ell^+ \ell^-$ can also occur through weak annihilation where its contribution corresponds to four-quark operators like $O_1$ and $O_2$ with additional photon/gluon emissions. These contributions appear only at higher orders in QCD corrections and are typically around a few percent compared to the dominant $C_9$, $C_{10}$, and $C_7$ operators from the loop-induced FCNC process. In addition, weak annihilation contributions typically suffer from color suppression in mesonic decays as in $B_c \to D_s^* \ell^+ \ell^-$, the initial $b \bar{c}$ system must annihilate into a color-singlet configuration, which may reduces the amplitude significantly. Therefore, the available phase space for WA diagrams is constrained compared to the dominant FCNC processes, which involve a direct loop-induced transition from $b \to s \ell^+ \ell^-$. Due to these facts, it is reasonable to assume that WA effects are negligible and do not significantly alter the result of the physical observables and have not been taken in the studies \cite{Geng:2001vy,Li:2023mrj,Zaki:2023mcw,Ivanov:2024iat}. 

\begin{figure}[H]
	\centering	
	\includegraphics[width=0.5\linewidth]{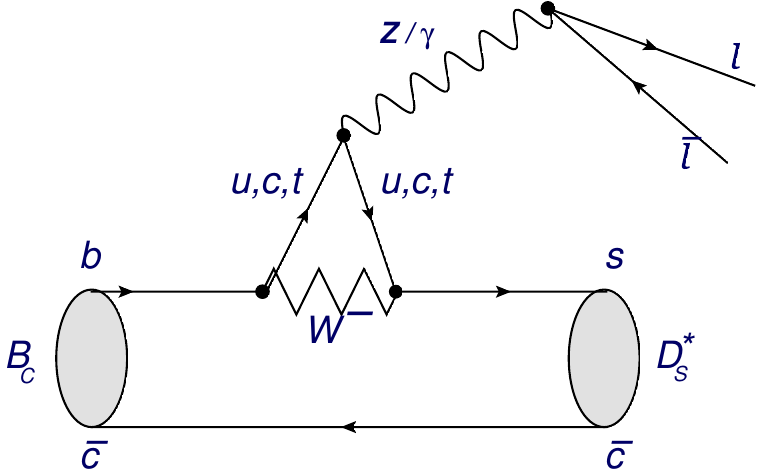}
	\caption{Penguin diagram for $B_{c} \to D_{s}^{*}\ell^{+}\ell^{-}$  decay.
 }
	\label{fig:penguindiagram}
\end{figure}

It is also important to mention here that $b\to s \ell^+\ell^-$ transition can also occur via box diagram but it contains two $W$-boson propagators and are typically suppressed by additional power of $\frac{1}{M_W^2}$ \cite{Buchalla:1995vs,Buras:1998raa}. Therefore, for the FCNC processes ( $b \to s \ell^+ \ell^-$ ), the penguin diagrams provide the dominant contribution. Consequently, we focus on calculating the exclusive semileptonic $B_c$ meson decay $B_{c} \to D_{s}^{*}\ell^{+}\ell^{-}$ for which  
quark level Feynman diagram in the SM for this process is shown in Fig.~(\ref{fig:penguindiagram}). 
The decay amplitude for this process can be written as;  
\begin{align}
	\mathcal{M} =&
       -\frac{G_F\alpha}{\sqrt{2}\pi}V_{tb}V^*_{ts} \{\mathcal{C}^{\text{eff}}_9\bra{D_s^*(k,\epsilon)}{\bar s \gamma_{\mu}P_L b\ket{B_c(p)}}(\bar l\gamma^{\mu}l) +\mathcal{C}_{10}^{\text{eff}}\bra{D_s^*(k,\epsilon)}(\bar s \gamma_{\mu}P_L b)\ket{B_c(p)}(\bar l\gamma^{\mu }\gamma^5l) \nonumber \\ 
	&- 2\thinspace\mathcal{C}^{\text{eff}}_7\dfrac{m_b}{q^2}\bra{D_s^*(k,\epsilon)}(\bar si\sigma_{\mu \nu}q^\nu P_R b)\ket{B_c(p)}\bar l\gamma^{\mu}l\nonumber\\
 &+ \mathcal{C}_{9\ell}\bra{D_s^*(k,\epsilon)}{\bar s \gamma_{\mu}P_L b\ket{B_c(p)}}(\bar l\gamma^{\mu}l) +\mathcal{C}_{10\ell}\bra{D_s^*(k,\epsilon)}(\bar s \gamma_{\mu}P_L b)\ket{B_c(p)}(\bar l\gamma^{\mu }\gamma^5l)\nonumber\\
 &+ \mathcal{C}^{\prime}_{9\ell}\bra{D_s^*(k,\epsilon)}{\bar s \gamma_{\mu}P_R b\ket{B_c(p)}}(\bar l\gamma^{\mu}l) +\mathcal{C}_{10\ell}^{\prime}\bra{D_s^*(k,\epsilon)}(\bar s \gamma_{\mu}P_R b)\ket{B_c(p)}(\bar l\gamma^{\mu }\gamma^5l)\}.\label{eq:TotalAmplitude}
	\end{align}   
 
Here $P_{L,R}=\frac{1}{2}(1\mp\gamma_5)$ and we have incorporated the effect of `new physics' vector and axial vector couplings. The $\mathcal{C}^{\rm eff}_i$ are the SM effective Wilson coefficients while the $\mathcal{C}_{9\ell}^{(\prime)}$ and $\mathcal{C}_{10\ell}^{(\prime)}$ correspond to the Wilson coefficients of the new vector and axial vector operators with the $\mathcal{C}_{j\ell}^\prime$~($j=9,10$) corresponding to the respective helicity-flipped operators. 

\subsection{Matrix Elements and Form Factors}
The matrix elements appearing in Eq.~(\ref{eq:TotalAmplitude}) can be parameterized in terms of the 
form factors which are scalar functions of the square of the momentum transfer
$q^2=(p-k)^2$,
\begin{align}
 \bra{D_s^*(k,\epsilon)}\bar s\gamma _{\mu} b\ket{B_c(p)}  &= \frac{2i \varepsilon_{\mu \nu \alpha \beta}}{M_{B_c} + M_{D^*_s}} \ \epsilon^{*\nu} p^\alpha k^\beta A_{V}(q^2), \nonumber \\
 \bra{D_s^*(k,\epsilon)}\bar s\gamma _{\mu} \gamma_5 b\ket{B_c(p)}  &=  (M_{B_c}+M_{D^*_s})\varepsilon^{*\mu}A_0(q^2) -\frac{(\varepsilon^*.q)A_+(q^2)}{M_{B_c}+M_{D^*_s}}(p+k)^\mu  \nonumber\\ 
 & -\frac{A_{-}(q^2)}{M_{B_c}+M_{D^*_s}}(\varepsilon^*.q)q^\mu,  \nonumber\\  
 \bra{ D^*_s(k,\epsilon)}\bar si\sigma_{\mu \nu}q^\nu b\ket{B_c(p)}  &= 2i\epsilon_{\mu \nu \alpha \beta} \varepsilon^{*\nu}p^\alpha k^\beta T_1(q^2),  \nonumber\\
 \bra{D^*_s(k,\epsilon)}\bar si\sigma_{\mu \nu}q^\nu \gamma^5 b\ket{B_c(p)} &= \left[(M^2_{B_c} + M^2_{D^*_s})\varepsilon^*_\mu - (\varepsilon^*\cdot q)(p+k)_\mu\right]T_2(q^2), \nonumber\\  
 & + (\varepsilon^*\cdot q)\left[q_\mu - \frac{q^2}{(M^2_{B_c} + M^2_{D^*_s})}(p+k)_\mu\right]T_3(q^2).
\end{align}

Here $p$ denotes the momentum of the $ B_c $ meson, while $k$ and $\varepsilon (k) $ are the momentum and polarization vectors 
of the $ D^*_s $ meson. The $ A_V(q^2), A_0(q^2),A_+(q^2),A_-(q^2), T_1(q^2), T_2(q^2)$ and $T_3(q^2) $ are the seven 
independent form factors. 

\subsection{Helicity Amplitude of $ B_c \to D^{\ast}_{s}\,\ell^{+}\ell^{-}$}
The decay rate of $B_c\to D_s^\ast \ell^+\ell^-$ is given by, 
\begin{align}
	\frac{d^2\Gamma(B_c \to D^*_s \ell^+ \ell^-)}{dq^2 d(\cos\theta)}  &=  \dfrac{1}{(2\pi)^3} \frac{1}{32 M_{B_c}^3}|\mathcal{M}|^2,
\end{align}

Since we intend to discuss the longitudinal helicity fraction ($f_L$), it is convenient to express 
this decay rate in the helicity basis. We begin by rewriting the amplitude (Eq.~(\ref{eq:TotalAmplitude}))
in the form:
\begin{eqnarray}
\mathcal{M}  =  -\frac{G_F\alpha}{2\sqrt{2}\pi}V_{tb}V^*_{ts}[Q^1_{\mu}(\bar l\gamma^{\mu}l)+Q^2_{\mu\nu}(\bar l\gamma^{\nu}\gamma^5l)],
\end{eqnarray}  
where,
\begin{eqnarray} Q^1_{\mu\nu} =& -i\epsilon_{\mu \nu \alpha \beta}\varepsilon^{*\nu}p^{\alpha}k^{\beta}\mathcal{F}_1(q^2) - g_{\mu \nu}\mathcal{F}_2(q^2) + q_{\mu}q_{\nu}\mathcal{F}_3(q^2) + P_{\mu}q_{\nu}\mathcal{F}_4(q^2), \nonumber \\ 
Q^2_{\mu\nu} =& -i\epsilon_{\mu \nu \alpha \beta}\varepsilon^{*\nu}p^{\alpha}k^{\beta}\mathcal{F}_5(q^2) - g_{\mu \nu}\mathcal{F}_6(q^2) + q_{\mu}q_{\nu}\mathcal{F}_7(q^2) + P_{\mu}q_{\nu}\mathcal{F}_8(q^2). \label{HP}
\end{eqnarray} 
The functions $ \mathcal{F}_1 $ to  $ \mathcal{F}_8 $ are the so-called auxiliary functions that contain the hadronic form factors as well as the Wilson coefficients and are
defined as: 
\begin{align}    
\mathcal{F}_1  =& \frac{2C_9^{\text{eff}}A_V(q^2)}{M_{D_s^{*}}+M_{B_c}}+ \frac{4m_b}{q^2}C_7^{\text{eff}}T_1(q^2),\nonumber\\
\mathcal{F}_2  =& C_9^{\text{eff}}A_0(q^2)({M_{D_s^{*}}+M_{B_c}})+\frac{2m_b}{q^2}C_7^{\text{eff}}T_2(q^2)({M_{D_s^{*}}+M_{B_c}}), \nonumber\\ 
\mathcal{F}_3  =& \frac{A_-(q^2)C_9^{\text{eff}}}{M_{D_s^{*}}+M_{B_c}}+\frac{2m_b}{q^2}C_7^{\text{eff}}T_3(q^2), \quad
\mathcal{F}_4  = \frac{A_+(q^2)C_9^{\text{eff}}}{M_{D_s^{*}}+M_{B_c}}+\frac{2m_b}{q^2}(T_2(q^2)+\frac{q^2T_3(q^2)}{M_{D_s^{*}}+M_{B_c}}), \nonumber\\
\mathcal{F}_5  =& \frac{2C_{10}^{\text{eff}}A_V(q^2)}{M_{D_s^{*}}+M_{B_c}}, \quad
\mathcal{F}_6  = C_{10}^{\text{eff}}({M_{D_s^{*}}+M_{B_c}})A_0(q^2), \quad 
\mathcal{F}_7  = \frac{C_{10}^{\text{eff}}A_{-}(q^2)}{M_{D_s^{*}}+M_{B_c}}, \nonumber\\ 
\mathcal{F}_8  =& \frac{C_{10}^{\text{eff}}A_{+}(q^2)}{M_{D_s^{*}}+M_{B_c}}.  \label{eq:auxilary}
\end{align}

We define the hadronic vectors and tensors in the helicity basis via~\cite{Faessler:2002ut}: 
\begin{align}
&H^{i}_m = \varepsilon^{\dagger \mu}(m)Q^{(i)}_\mu, \nonumber \\ 
&H^{i}_{m,n}= \varepsilon^{*\mu}(m)\varepsilon^{* \nu}(n)Q^{(i)}_{\mu\nu}, 
\end{align}  
where $ Q^{(i)}_\mu = \varepsilon^{* \nu}(n)Q^{(i)}_{\mu\nu}$, \textit{i} = 1, 2,
and  $ \varepsilon^{\mu} (m)$ are the polarization vectors of the final state $ D^*_s $ meson in the helicity basis. The $m$ and $n$ labels take on the values
$0, \pm, \textrm{ and }t$ corresponding to the longitudinal, transverse and time-like polarizations, respectively. Explicitly, the helicity 
eigenbasis in the  rest frame of the \textit{B}-meson is: 
\begin{align}
\varepsilon^{\mu}(t) = \frac{1}{\sqrt{q^2}}(q_0,0,0,|k|),\quad
\varepsilon^{\mu}(\pm) = \frac{1}{\sqrt{2}}(0,\mp 1,i,|k|),\quad
\varepsilon^{\mu}(0) = \frac{1}{\sqrt{q^2}}(|k|,0,0,q_0),\nonumber 
\end{align}
whereas, the momentum 4-vectors of the $B$ and $D^*_s$ measons are 
$p^\mu = (m_B,0,0,0)$, and $k^\mu = (E_{k},0,0,|k|)$, and the momentum 
transfer is $q^\mu = (q_0,0,0,|k|)$. Here $|k|={\sqrt{\lambda}}/{2m_{B}}$ while 
$\lambda = m^4_{B}+ m^4_{D^*_s}+q^4-2(m^2_{B}m^2_{D^*_s}+m^2_{D^*_s}q^2+m^2_{B}q^2)$. We are now 
in a position to give the vectors and tensors corresponding to the hadronic part explicitly in terms of kinematic variables and the auxiliary functions:
\begin{align}
H^{(1)}_0 =& \frac{1}{m_{D^*_s}\sqrt{q^2}}[2q_0|k|^2(q_0-E_{D^*_s})\mathcal{F}_2+(|k|^2+q_0E_{D^*_s})\mathcal{F}_3\nonumber \\ 
&+|k|^2(q_0(m_B+2E_{D^*_s})
-q^2_0-E_{D^*_s}(m_B+E_{D^*_s}))\mathcal{F}_4], \nonumber \\ 
H^{(2)}_0 =& \frac{1}{m_{D^*_s}\sqrt{q^2}}[2q_0|k|^2(q_0-E_{D^*_s})\mathcal{F}_6+(|k|^2+q_0E_{D^*_s})\mathcal{F}_7 \nonumber \\ 
&+|k|^2(q_0(m_B+2E_{D^*_s}) 
-q^2_0-E_{D^*_s}(m_B+E_{D^*_s}))\mathcal{F}_8] ,\nonumber \\ 
H^{(1)}_+ =&-i|k|m_B\mathcal{F}_1+\mathcal{F}_3,\quad
H^{(2)}_+ =-i|k|m_B\mathcal{F}_5+\mathcal{F}_7, \quad
H^{(1)}_- =i|k|m_B\mathcal{F}_1+\mathcal{F}_3,\nonumber\\
H^{(2)}_- =&i|k|m_B\mathcal{F}_5+\mathcal{F}_7,\label{had}
\end{align}
where $ E_{D^*_s}={(m_{B}^2+m_{D^*_s}^2-q^2)}/{2m_{B}}$. The subscripts $ \pm,0$ denote, respectively, the transverse and longitudinal helicity components as before.

Similarly, we can define the leptonic tensors $ L_{\mu \nu}^{(k)} $. In the $\bar ll$ center of mass frame (COM) the 4-momenta of the lepton pair are:
\begin{align}
	p^\mu_1 =& (E_l, |p_1|\sin\theta ,0,|p_1|\cos\theta) \nonumber \\
	p^\mu_2 =& (E_l, -|p_1|\sin\theta ,0,-|p_1|\cos\theta)\nonumber,
\end{align} 
 where $E_l=\sqrt{q^2}/2$, $q^\mu = (\sqrt{q^2},\vec{0})$, and $ |p_1| = \sqrt{q^2-4m_l^2}/2$ and the polarization vectors are: $\epsilon^\mu(\pm) = \frac{1}{\sqrt{2}}(0,\pm 1,i,0)$, $\epsilon^\mu(0) = (0,0,0,1)$, and $\epsilon^\mu(t) = (1,0,0,0)$. These kinematic definitions allow us to write the leptonic part as:
\begin{align}
L^1_{00} =& -2|p_1|^2 \cos^2\theta,\quad  L^2_{00} = -1,\quad L^3_{00} = 0 \nonumber \\
L^1_{++} =& E_l-|p_1|^2 \sin^2\theta,\quad L^2_{++} = -1 ,\quad L^3_{++} = -2E_l|p_1|\cos\theta \nonumber \\
L^1_{--} =& E_l^2,\quad L^2_{--} = -1,\quad L^3_{--} = 2E_l|p_1|\cos\theta,\label{lep}
\end{align}
 
We are now in a position to express the amplitude squared $\abs{\mathcal{M}}^2$ for the $B_c\to D_s^\ast \ell^+\ell^-$ decay in the helicity basis as:
\begin{align}
    \abs{\mathcal{M}}^2 &= \frac{G_F^2}{(2\pi)^3}\left(\frac{\alpha |\lambda_t|}{2\pi}\right)^2\frac{|k|\sqrt{1-4m_l^2/q^2}}{8m_l^2}\frac{1}{2}[\ L_{\mu \nu}^{(1)}(H^{\mu\nu}_{11}+H^{\mu\nu}_{22})\nonumber\\
&-\frac{1}{2}L_{\mu \nu}^{(2)}(q^2H^{\mu\nu}_{11}+(q^2-m_l^2)H^{\mu\nu}_{22}) + L_{\mu \nu}^{(3)}(H^{\mu\nu}_{12}+H^{\mu\nu}_{21})\ ],
\label{eq:dobulediffdecay}
\end{align}
where $\lambda_t$ is related to the CKM matrix elements $V_{ts}$ and $V_{tb}$ via $ \lambda_t = |V_{ts}^\dagger V_{tb}| $. 

We can now write the differential decay rate after integration over the $\cos\theta$ as: 
\begin{align}
\frac{d\Gamma (B \to D^*_s \ell^+ \ell^- )}{dq^2} = \frac{G_F^2}{(2\pi)^3}\left(\frac{\alpha |\lambda_t|}{2\pi}\right)^2\frac{\lambda^{1/2}q^2}{48M_B^3}&\sqrt{1-4m_l^2/q^2}\Big[ H^1H^{1\dagger}(1+4m_l^2/q^2)+ \nonumber\\
&+H^2H^{2\dagger}(1-4m_l^2/q^2)\Big ],
\end{align}
where $ m_{l} $ is the lepton mass. 

\subsection{Forward Backward Asymmetry $ \mathcal{A}_{FB} $}
The $ \mathcal{A}_{FB} $ of leptons is defined as;
\begin{align}
\mathcal{A}_{FB}=\frac{\mathcal{N}^F-\mathcal{N}^B}{\mathcal{N}^F+\mathcal{N}^B},
\end{align}
where $ \mathcal{N}^F(\mathcal{N}^B) $ is the probability of the lepton moving in
the forward (backwards) direction. In terms of the differential decay rate, these probabilities can be written as
\begin{align}
\mathcal{N}^F=\int_{0}^{1}d\cos\theta \frac{d^2\Gamma(q^2, \cos\theta)}{dq^2d\cos\theta},\quad
\mathcal{N}^B=\int_{-1}^{0}d\cos\theta \frac{d^2\Gamma(q^2, \cos\theta)}{dq^2d\cos\theta}.
\end{align}

By using the helicity amplitudes, one can write the analytical expression for $ \mathcal{A}_{FB} $ as; 
\begin{align}
\mathcal{A}_{FB}=&\frac{3}{4}\sqrt{1-\frac{4m_l^2}{q^2}}\frac{Re(H^{(1)}_+H^{\dagger(2)}_+)-Re(H^{(1)}_{-}H^{\dagger(2)}_{-})}{H^{(1)}H^{\dagger(1)}(1+{4m_l^2}/{q^2})+H^{(2)}H^{\dagger(2)}(1-{4m_l^2}/{q^2})}.
\end{align}

\subsection{Helicity Fraction}
The longitudinal helicity fraction as a function of the momentum transfer is defined as;
\begin{align}
f_L =&\frac{d\Gamma_L(q^2)/dq^2}{d\Gamma(q^2)/dq^2},
\label{eq:longitudinalHF}
\end{align}
where $ d\Gamma_L(q^2)/dq^2 $ is the longitudinal component of the decay rate. We 
can use our expressions above to write the $f_L$ for the decay $B_c \to D^*_s l^+ l^-$ as:
\begin{align}
f_L(q^2)=\frac{H^{(1)}_0H^{(1)\dagger}_0(1+{4m_l^2}/{q^2})+H^{(2)}_0H^{(2)\dagger}_0(1-{4m_l^2}/{q^2})}{H^{(1)}H^{(1)\dagger}(1+\frac{4m_l^2}{q^2})+H^{(2)}H^{(2)\dagger}(1-\frac{4m_l^2}{q^2})}.
\end{align}
The longitudinal helicity fraction of $ D^*_s $ is potentially sensitive to NP contributions in $B_c\to D_s^\ast \ell^+\ell^-$ decays.

\subsection{Lepton Flavor Universality  Ratios}\label{LFU r}
Lepton Flavor Universality~(LFU) ratios are designed to the test the gauge universal 
nature of the effective electroweak interaction. Specifically, we construct 
ratios of branching fractions to various lepton generations. In the present context for the $B_c\to D^{(\ast)}_s\ell^+\ell^-$ decay, we define the ratio $R_{D^{(\ast)}_s}$ as,
\begin{align}
	R_{D^{(\ast)}_s}=\frac{\int_{q^2_{min}}^{q^2_{max}}{\frac{d\mathcal{B}(B_c\to D^{(\ast)}_s\tau^+\tau^-)}{dq^2}dq^2}}{\int_{q^2_{min}}^{q^2_{max}}{\frac{d\mathcal{B}(B_c\to D^{(\ast)}_s\mu^+\mu^-)}{dq^2}dq^2}},
\end{align}
where the integration is over the appropriate $q^2$ bin for comparison with 
experimental findings.

\subsection{Differential decay rate of $ B_c \to D_{s}\ell^{+}\ell^{-}$}

While our primary focus is on the decay of $B_c$ to the vector meson $D_s^*$, we also discuss the decay to the scalar: $B_c \to D_s\ell^+\ell^-$. The branching ratio of this channel can be written as \cite{Bouchard:2013eph}: 
\begin{align}
    \frac{d \mathcal{B}_r}{dq^2}=2\,a_l+\frac{2\,c_l}{3},
\end{align}
where $a_l$ and $c_l$ are defined via:
\begin{align}
    a_l&=\mathcal{C_o}\left[q^2|\mathcal{F}_P|^2 +\frac{\lambda}{4}(\left |\mathcal{F}_A|^2+|\mathcal{F}_V|^2\right)+4m_l^2M_{B_c}^2|\mathcal{F}_A|^2+2m_l(M_{B_c}^2-M_{D_s}+q^2)\text{Re}(\mathcal{F}_P\mathcal{F}_A^*) \right],\nonumber\\
    c_l&=-\mathcal{C_o}\frac{\lambda}{4}\left(1-\frac{4m_l^2}{q^2}\right)(|\mathcal{F}_A|^2+|\mathcal{F}_V|^2).
\end{align}

Here, the $\mathcal{C_o}$ is:
\begin{align*}
    \mathcal{C_o}=\frac{(\alpha|\lambda_t|)^2}{2^9\,\pi^5\,M_{B_c}^3}\sqrt{1-\frac{4m_l^2}{q^2}}\,\sqrt{\lambda},
\end{align*}
with $\lambda=  M_{B_c}^4 + M_{D_s}^4 + q^4 - 2(M_{B_c}^2 M_{D_s}^2+M_{D_s}^2 q^2 + M_{B_c}^2q^2)$. 
The expressions for the form factors $\mathcal{F}_P$, $\mathcal{F}_V$, and $\mathcal{F}_A$ are given by:
\begin{align*}
    \mathcal{F}_P &=-m_l \left(C_{10}^{\text{eff}} +C_{10\ell}+C_{10\ell}^\prime\right) \left[f_+ -\frac{M_{B_c}^2-M_{D_s}^2}{q^2}(f_0-f_T)\right],\\
    \mathcal{F}_V &= \left(C_{9}^{\text{eff}} +C_{9\ell}+C_{9\ell}^\prime\right)f_+ +\frac{2m_b}{M_{B_c}+M_{D_s}}C_7^{\text{eff}}f_T,\\
    \mathcal{F}_A &= \left(C_{10}^{\text{eff}} +C_{10\ell}+C_{10\ell}^\prime\right)f_+.
\end{align*}
It is worth mentioning here that $D_s$ is a scalar meson, and the $\mathcal{A}_{FB}$ for this channel is zero unless we include scalar-type couplings. We have, however, considered only vector and axial vector type NP couplings. Therefore, we have commented on the branching ratio and the LFU ratio 
for this decay.

\section{Phenomenological Analaysis\label{pheno analysics}}

\subsection{Input Parameters}
We begin by listing all the input parameters relevant to our analysis where we have 
used $\mu\simeq m_b$ as the renormalization scale. In Table~\ref{Input} we show various masses and SM couplings used as well as the the decay time of $B_c$.   
\begin{table}[H]
\centering
\resizebox{\textwidth}{!}{
\begin{tabular}{|ccc|}
\hline
$M_{B_{c}}=6.275$ GeV& $M_{D^{\ast}_{s}} = 2.1123$ GeV& $M_{D_{s}} = 1.968$ GeV\\ $m_{b}(\overline{\text{MS}})=4.2$ GeV& $m_{c}(\overline{\text{MS}})=1.28$ GeV& $m_b(\text{pole}) = 4.8$ GeV \\
$m_e=0.51099895000$ MeV& $m_{\mu}=105.6583755$ MeV& $m_{\tau}=1776.93$ MeV \\
$|V_{tb}V_{ts}^{\ast}|=0.0401 \pm 0.0010$ & $\alpha^{-1}=137.035 999$ & $G_{F}=1.166 378\times 10^{-5}$ GeV$^{-2}$ \\
& $\tau_{B_{c}}=0.51\times 10^{-12}$ sec &  \\ \hline
\end{tabular}}\caption{Numerical values of various input parameters used in numerical analysis~\cite{ParticleDataGroup:2022pth,Altmannshofer:2008dz}.}\label{Input}
\end{table}

The hadronization of quarks and gluons is described by the form factors that are computed employing non-perturbative 
methods QCD and are a source of theoretical uncertainties. The form factors for $B_c \to D_s^{(\ast)}$ are 
taken from Ebert~\emph{et al}~\cite{Ebert:2010dv};
\begin{align}
\mathcal{F}(q^2) =\begin{cases}
	\frac{\mathcal{F}(0)}{\left(1-{q^2}/{M^2}\right)\left(1-\alpha{q^2}/{M^2_{B^{(\ast)}_s}} + \beta{q^4}/{M_{B^{(\ast)}_s}^4}\right)} 
	&	\mathcal{F}(q^2)\in\{A_V(q^2), A_0(q^2),T_1(q^2),f_+(q^2), f_T(q^2)\}\\
	 \frac{\mathcal{F}(0)}{\left(1-\alpha{q^2}/{M_{B^{(\ast)}_s}^2} + \beta{q^4}/{M_{B^{(\ast)}_s}^4}\right)} 
	& \mathcal{F}(q^2)\in\{A_+(q^2), A_-(q^2), T_2(q^2), T_3(q^2), f_0(q^2)\}, \nonumber
\end{cases}
\end{align}
where $q^2$ is the momentum transfer and the values of $\mathcal{F}(0), \alpha $ and $ \beta $ corresponding to the varrious form factors are given in Table~\ref{table:formfactors}. 
The form factor for $A_0(q^2)$ has $M=M_{B_{s}}=5.36692$ GeV~\cite{ParticleDataGroup:2022pth}, while all other form factors have $M=M_{B^{(\ast)}_{s}}=5.4154$ GeV~\cite{ParticleDataGroup:2022pth}. In order to ameliorate the effect of the form factor uncertainties on different observables, we have used $\pm5\%$ uncertainty in $\mathcal{F}(0)$, $\alpha$, and $\beta$ in our calculations~\cite{Dutta:2019wxo}.   
\begin{table}[H]
\centering
\resizebox{1\textwidth}{!}{
\begin{tabular}{|c|c|c|c|c|c|c|c|c|c|c|}
	\hline
$  $	& $ A_V(q^2) $ &$ A_0(q^2) $ & $ A_+(q^2) $ &$ A_-(q^2) $&$ T_1(q^2) $&$ T_2(q^2) $&$ T_3(q^2)$& $f_+(q^2)$&$f_0(q^2)$ &$f_T(q^2)$	\\
\hline
 $ \mathcal{F}(0) $
	&0.182 &0.070&0.089 &0.110&0.085&0.085 &0.051&0.129  &0.129 & 0.098 \\
	 
$\alpha$	&2.133 &1.561&2.479&2.833 &1.540&2.577& 2.783 &2.096 &2.331 &1.412 \\
	
$\beta$&1.183&0.192&1.686&2.167&0.248&1.859& 2.170&1.147 & 1.666& 0.048 \\
	\hline
\end{tabular} }
\caption{Form factors of $B_{c}\to D_{s}^{(*)}$ decays which are calculated by relativistic quark model \cite{Ebert:2010dv}.}
\label{table:formfactors}
\end{table} 
In Table~\ref{wilson}, we list the SM Wilson coefficients computed at the scale $\mu\simeq m_b$ that we use~\cite{Blake:2016olu}. 

\begin{table}[H]
\centering
\resizebox{1\textwidth}{!}{
\begin{tabular}{|c|c|c|c|c|c|c|c|c|c|}
	\hline
$C_1$& $ C_2 $ &$ C_3 $ & $ C_4 $ &$ C_5 $&$ C_6 $&$ C_7 $&$ C_8 $&$C_9$&$C_{10}$	\\
\hline
 $-0.294$&1.017& $-0.0059$&$-0.087$&0.0004&0.0011&$-0.295$&$-0.163$&4.114&$-4.193$  \\
	\hline
\end{tabular} }
\caption{The NNLL Wilson coefficients evaluated at the renormalization scale $\mu\simeq m_b$~\cite{Blake:2016olu}.}
\label{wilson}
\end{table} 

\subsection{NP Scenarios}
Our goal is to specify the effect of NP on $B_c\to D_s^{(\ast)} l^+l^-$ decay observables. Our study uses an effective theory formalism in the presence of new vector ($V$) and axial vector ($A$) couplings which go on to introduce the WCs $C_{(9,10)_l}$ and $C'_{(9,10)_l}$ with $l=e,\mu,\tau$ in 
Eq.~(\ref{eq:TotalAmplitude}). We consider the best-fit data of these NP couplings in different scenarios from the current global fit analysis \cite{Alok:2023yzg} and ask what these
predict for various observables in $B_c\to D_s^{(\ast)} l^+l^-$ decays. 
In discussing possible NP scenarios, there are two fundamentally different 
possibilities; (a) we can have flavor universal couplings, or (b)there may 
be some non-universal flavor structure in the couplings which manifests in the 
WCs. We opt for a minimalist description in terms of WCs which follows:
\begin{eqnarray}
\mathcal{C}_{(9,10)e}&=&\mathcal{C}_{(9,10)\tau}=\mathcal{C}_{(9,10)}^U\,,\quad\quad\mathcal{C}_{(9,10)e}^{\prime}=\mathcal{C}_{(9,10)\tau}^{\prime}=\mathcal{C}_{(9,10)}^{\prime U}\,,\nonumber\\
\mathcal{C}_{(9,10)\mu}&=& \mathcal{C}_{(9,10)}^U +\mathcal{C}_{(9,10)\mu}^V\,,\quad\quad\mathcal{C}_{(9,10)\mu}^{\prime}= \mathcal{C}_{(9,10)}^{\prime U} +\mathcal{C}_{(9,10)\mu}^{\prime V}\, .
\end{eqnarray}

Clearly, $\mathcal{C}_{(9,10)}^{U}$ and $\mathcal{C}_{(9,10)}^{\prime U}$ are the universal contributions to WCs that equally contribute to $b\to s \ell^+\ell^-$ transitions, and, $\mathcal{C}_{(9,10)}^{V}$ and $\mathcal{C}_{(9,10)}^{\prime V}$ are non-universal contributions, which only affect the $b\to s \mu^+\mu^-$ decay. In this regard, we define two frameworks named F-I and 
F-II. F-I has condition $\mathcal{C}^V_{(9,10)\mu}=\mathcal{C}^{\prime 
V}_{(9,10)\mu}=0$, and so only allows universal couplings F-II incorporates non-universal couplings as well. Within F-I, the scenarios S1~\cite{Alguero:2023jeh}, S2~\cite{Alok:2023yzg}, and S3~\cite{Alok:2023yzg} preferred by the recent data and the 1$\sigma$ range of the Wilson coefficients are grouped in Table~\ref{framework-I}. A complete set of 
preferred scenarios S5, S6, S7, S8, S9, S10, S11 and S13 has been reported for F-II 
based on a global fit of 254 precision observables~\cite{Alguero:2021anc}. These NP scenarios, along with the updated 1$\sigma$ range 
of the Wilson coefficients, as mentioned in~\cite{Alguero:2023jeh}, are grouped in 
Table~\ref{framework-II} and Table~\ref{framework-II1}. 
\begin{table}[H]
\centering
\begin{tabular}{|c|c|c|}
	\hline
F-I Solutions 	 & Wilson Coefficients & 1$\sigma$ range   \\
  \hline\hline
S1  & $\mathcal{C}^{U}_9$ & $(-1.00,\, -1.33)$ \\ 

S2 & $\mathcal{C}^{U}_9 = - \mathcal{C}^{U}_{10}$ & $(-0.38,\, -0.62)$  \\ 

S3  & $\mathcal{C}^{U}_9 = - \mathcal{C}^{\prime U}_{9}$ &$(-0.72,\, -1.04)$ \\
 \hline\hline
\end{tabular} 
\caption{Allowed NP $1\sigma$ parametric range of $D=1$ NP Universal couplings. The S1 scenario is introduced in \cite{Alguero:2023jeh}, and the S2 and S3 scenarios are introduced in \cite{Alok:2023yzg}.}
\label{framework-I}
\end{table}

\begin{table}[H]
\centering
\begin{tabular}{|c|c|c|}
	\hline
F-II Solutions	 & Wilson Coefficients& 1$\sigma$ range    \\
  \hline\hline
 S5  & $\mathcal{C}^{V}_{9\mu}$ & (-1.43, -0.61 )   \\ 
     & $\mathcal{C}^{V}_{10\mu}$ & (-0.75, 0.00)  \\
   & $\mathcal{C}^{U}_9 = \mathcal{C}^{U}_{10}$ &  (-0.16, 0.58) \\
\hline S6 & $\mathcal{C}^{V}_{9\mu} = - \mathcal{C}^{V}_{10\mu}$ & (-0.34, -0.20)   \\ 
     & $\mathcal{C}^{U}_9 = \mathcal{C}^{U}_{10}$ & (-0.53, -0.29)  \\
\hline S7  & $\mathcal{C}^{V}_{9\mu}$ & (-0.39, -0.02) \\ 
  & $\mathcal{C}^{U}_9$ & (-1.21, -0.72) \\
\hline S8  & $\mathcal{C}^{V}_{9\mu} = - \mathcal{C}^{V}_{10\mu}$  & (-0.14, -0.02)  \\ 
     & $\mathcal{C}^{U}_9$ & (-1.27, -0.91) \\
 \hline\hline
\end{tabular} 
\caption{
Allowed $1\sigma$ parametric space of $D>1$ universal and non-universal NP scenarios. These scenarios from S5 to S8 are introduced in \cite{Alguero:2023jeh}.}
\label{framework-II}
\end{table}

\begin{table}[H]
\centering
\scalebox{1}{
\begin{tabular}{|c|c|c|}
	\hline
 F-II Solutions	 & Wilson Coefficients & 1$\sigma$ range   \\
  \hline\hline
  S9  & $\mathcal{C}^{V}_{9\mu} = - \mathcal{C}^{V}_{10\mu}$ &(-0.29, -0.13)   \\ 
      & $\mathcal{C}^{U}_{10}$  & (-0.23, 0.11)   \\
 \hline S10  & $\mathcal{C}^{V}_{9\mu}$ &(-0.81, -0.50)   \\ 
  & $\mathcal{C}^{U}_{10}$   &(-0.08, 0.18)  \\
\hline S11 & $\mathcal{C}^{V}_{9\mu}$ & (-0.84, -0.52)  \\ 
     & $\mathcal{C}^{\prime U}_{10}$  & (-0.15, 0.09)  \\
\hline S13  & $\mathcal{C}^{V}_{9\mu}$ & (-0.97, -0.60)   \\ 
     & $\mathcal{C}^{\prime V}_{9\mu}$ &(0.10, 0.57)   \\
& $\mathcal{C}^{ U}_{10}$ &(-0.04, 0.26) \\
& $\mathcal{C}^{\prime U}_{10}$ &(-0.03, 0.30)  \\
 \hline\hline
\end{tabular} }
\caption{Allowed $1\sigma$ parametric space of $D>1$ universal and non-universal NP scenarios. The S9 scenario is inspired by 2HDMs \cite{Crivellin:2019dun}, scenarios S10 to S13 are inspired by the Z$^\prime$ model, and vector-like quarks \cite{Bobeth:2016llm} that are introduced in \cite{Alguero:2023jeh}.}
\label{framework-II1}
\end{table} 

By using the formulae of observables defined in section \S\ref{theoretical description}, we have found the expressions of the branching ratios in terms of NP WCs by integrating over the low $q^2$ bin for muon and high $q^2$ bins both for the muon and tauon as the final state leptons, which read as follows:
\begin{align}
    10^{7}&\times\mathcal{B}_r^{[14,s_{max}]}(B_c\to D_s^{\ast} \,\tau^+\tau^-)=\nonumber\\
    & 0.69^{+0.08}_{-0.07}+0.02^{+0.00}_{-0.00}\{\,(C_{9}^{\prime U })^2+\,(C_{9}^{U})^2\}+0.01^{+0.00}_{-0.00}\{\,C_{10}^{\prime U }\,C_{9}^{U}-\,C_{10}^{ U }\,C_{9}^{U}-\,C_{10}^{ U }\,C_{9}^{\prime U }\}\nonumber\\
    &-0.04^{+0.00}_{-0.00}\,C_{9}^{\prime U }\,C_{9}^{U}+0.09^{+0.01}_{-0.01}\{\,C_{10}^{\prime U}+\,C_{10}^{U}\}-0.23^{+0.02}_{-0.02}\,C_9^{\prime U}+0.25^{+0.03}_{-0.03}\,C_9^{ U}.\label{tua br expr}
\end{align}
\begin{align}
10^{8}&\times\mathcal{B}_r^{[s_{min},6]}(B_c\to D_s^{\ast} \,\mu^+\mu^-)=\nonumber\\
&0.97^{+0.01}_{-0.01}+0.02^{+0.00}_{-0.00}\{\,(C_{10}^{\prime U})^2 + \,(C_{10}^{\prime V})^2  + \,(C_{9}^{\prime V})^2+ \,(C_{9}^{\prime U})^2 + \,(C_{9}^{ U})^2+ \,(C_{10}^{\prime U})^2+ \,(C_{10}^{V})^2 \nonumber\\
&+\,(C_{9}^{V})^2\} + 0.03^{+0.00}_{-0.00}\{\,C_{10}^{\prime V}\,C_{9}^{U} + \,C_{10}^{\prime V}\,C_{9}^{V}- \,C_{10}^{\prime U}\,C_{10}^{U} - \,C_{10}^{\prime U}\,C_{10}^{V} - \,C_{10}^{\prime U}\,C_{9}^{\prime U}- \,C_{10}^{\prime U}\,C_{9}^{\prime V}\nonumber\\
&- \,C_{10}^{\prime V}\,C_{10}^{U}- \,C_{10}^{\prime V} \,C_{10}^{V}- \,C_{10}^{\prime V} \,C_{9}^{\prime U} - \,C_{10}^{\prime V}\,C_{9}^{\prime V}+ \,C_{10}^{\prime U}\,C_{9}^{U}+ \,C_{10}^{\prime U}\,C_{9}^{V} + \,C_{10}^{\prime U}\,C_{9}^{\prime U} + \,C_{10}^{U}\,C_{9}^{\prime V} \nonumber\\
&- \,C_{10}^{U}\,C_{9}^{U}- \,C_{10}^{U}\,C_{9}^{V}+ \,C_{10}^{V}\,C_{9}^{\prime U} + \,C_{10}^{V}\,C_{9}^{\prime V}  - \,C_{10}^{V}\,C_{9}^{U} - \,C_{10}^{V}\,C_{9}^{V}- \,C_{9}^{\prime U} \,C_{9}^{ U} - \,C_{9}^{\prime U}\,C_{9}^{V} \nonumber\\
&- \,C_{9}^{\prime V}\,C_{9}^{U} - \,C_{9}^{\prime V}\,C_{9}^{V}\}+0.04^{+0.00}_{-0.00}\{\,C_{10}^{\prime U}\,C_{10}^{\prime V} + \,C_{10}^{U}\,C_{10}^{V} + \,C_{9}^{\prime U}\,C_{9}^{\prime V}+ \,C_{9}^{U} \,C_{9}^{V}\}\nonumber\\
&+ 0.18^{+0.00}_{-0.00}\{\,C_{10}^{\prime U}  +\,C_{10}^{\prime V}\}
+ 0.25^{+0.00}_{-0.00}\{\,C_{9}^{V}-\,C_{10}^{U} - \,C_{10}^{V}+\,C_{9}^{U}\}+0.22^{+0.00}_{-0.00}\{-\,C_{9}^{\prime U}-\,C_{9}^{\prime V}\}.
\end{align}

\begin{align}
10^{7}&\times\mathcal{B}_r^{[14,s_{max}]}(B_c\to D_s^{\ast} \,\mu^+\mu^-)=\nonumber\\
    & 2.86^{+0.34}_{-0.32} +0.04_{-0.01}^{+0.01}\{\,(C_{10}^{\prime U })^2+\,(C_{10}^{\prime V })^2+\,(C_{10}^{U})^2+\,(C_{10}^{V})^2+ \,(C_{9}^{\prime U})^2+ \,(C_{9}^{\prime V})^2+ \,(C_{9}^{ U})^2\nonumber\\
    &+ \,(C_{9}^{V})^2\}+ 0.08^{+0.01}_{-0.01}\{\,C_{10}^{U}\,C_{9}^{\prime U} - \,C_{10}^{\prime U}\,C_{9}^{\prime U} - \,C_{10}^{\prime V}\,C_{9}^{\prime U} +\,C_{10}^{V}\,C_{9}^{\prime U} - \,C_{10}^{\prime U}\,C_{9}^{\prime V} - \,C_{10}^{\prime V} \,C_{9}^{\prime V} \nonumber\\
    &+ \,C_{10}^{U}\,C_{9}^{\prime V} + \,C_{10}^{V}\,C_{9}^{\prime V} + \,C_{10}^{\prime U}\,C_{9}^{ U}+ \,C_{10}^{\prime V}\,C_{9}^{U} - \,C_{10}^{U}\,C_{9}^{U} - \,C_{10}^{V}\,C_{9}^{ U}- \,C_{10}^{\prime U}\,C_{9}^{\prime U}+ \,C_{10}^{\prime U}\,C_{9}^{V}\nonumber\\
    &+ \,C_{10}^{\prime V}\,C_{9}^{V}- \,C_{10}^{U}\,C_{9}^{V}- \,C_{10}^{V}\,C_{9}^{V}\}+0.09^{+0.01}_{-0.01}\{\,C_{10}^{\prime U}\,C_{10}^{\prime V}-\,C_{10}^{\prime U}\,C_{10}^{V}-\,C_{10}^{\prime V}\,C_{10}^{V}+\,C_{10}^{U}\,C_{10}^{V}\nonumber\\
    &- \,C_{9}^{\prime U}\,C_{9}^{U} - \,C_{9}^{\prime V} \,C_{9}^{U}- \,C_{9}^{\prime U} \,C_{9}^{V}-\,C_{9}^{\prime V}\,C_{9}^{V}\}-\,C_{10}^{\prime U}\,C_{10}^{U}-\,C_{10}^{\prime V}\,C_{10}^{ U}\} + 0.10^{+0.01}_{-0.01}\{\,C_{9}^{\prime U}\,C_{9}^{\prime V}\nonumber\\
    &+\,C_{9}^{U} \,C_{9}^{V}\} 
    +0.69^{+0.08}_{-0.07}\{\,C_{10}^{\prime U }
   +\,C_{10}^{\prime V}-\,C_{9}^{\prime U}-\,C_{10}^{\prime U}  -\,C_{9}^{\prime V}\}+
0.72^{+0.09}_{-0.08}\{-\,C_{10}^{U}-\,C_{10}^{V} \nonumber\\
&+\,C_{9}^{ U} + \,C_{9}^{V}\}.
\end{align}

\begin{align}
10&\times\mathcal{A}_{FB}^{[14,s_{max}]}(B_c\to D_s^{\ast} \,\mu^+\mu^-)=\nonumber\\
   &0.85+
 0.01\{\,C_{10}^{U}-\,(C_{10}^{\prime U})^2- \,(C_{10}^{\prime V})^2 - 
 \,(C_{9}^{ U})^2- \,(C_{9}^{V})^2- \,(C_{10}^{ U})^2- \,(C_{10}^{\prime V})^2+ \,C_{10}^{V} \nonumber\\
 &- \,(C_{9}^{\prime U})^2- \,(C_{9}^{\prime V})^2\} + 0.02\{\,C_{10}^{\prime U}\,C_{10}^{U} + 
 \,C_{10}^{\prime U}\,C_{10}^{V} - \,C_{10}^{\prime U}\,C_{9}^{U} - \,C_{10}^{\prime U}\,C_{9}^{V}+ \,C_{10}^{\prime V}\,C_{10}^{U}\nonumber\\
 &+ \,C_{10}^{\prime V} \,C_{10}^{V} - \,C_{10}^{V}\,C_{9}^{\prime U}- \,C_{10}^{V}\,C_{9}^{V}+ 
 \,C_{9}^{\prime V} \,C_{9}^{U} + \,C_{9}^{\prime V} \,C_{9}^{V}- \,C_{10}^{\prime V}\,C_{9}^{U}- \,C_{10}^{\prime V}\,C_{9}^{\prime V}+\,C_{9}^{U}  \nonumber\\
 &+ \,C_{9}^{V} - \,C_{10}^{U} \,C_{9}^{\prime U}- \,C_{10}^{U} \,C_{9}^{\prime V} + \,C_{9}^{\prime U} \,C_{9}^{U} + \,C_{9}^{\prime U}\,C_{9}^{V}\}+ 0.03\{-\,C_{10}^{\prime U}\,C_{10}^{\prime V} -\,C_{10}^{U}\,C_{10}^{V}\nonumber\\
 &- \,C_{10}^{U} \,C_{9}^{U} - \,C_{10}^{U} \,C_{9}^{V}- 
 \,C_{10}^{V}\,C_{9}^{U} - \,C_{10}^{V}\,C_{9}^{V} - \,C_{9}^{\prime U} \,C_{9}^{\prime V}  - \,C_{9}^{U} \,C_{9}^{V}\}+ 0.08\{\,C_{10}^{\prime U}\,C_{9}^{\prime U} \nonumber\\
 &+\,C_{10}^{\prime U}\,C_{9}^{\prime V}+ \,C_{10}^{\prime V}\,C_{9}^{\prime U} + \,C_{10}^{\prime V}\,C_{9}^{\prime V}\}+ 0.20\{-\,C_{10}^{\prime U} - \,C_{10}^{\prime V}+ \,C_{9}^{\prime U}+\,C_{9}^{\prime V}\}.
\end{align}

\begin{align}
10&\times\mathcal{A}_{FB}^{[s_{min},6]}(B_c\to D_s^{\ast} \,\mu^+\mu^-)=\nonumber\\
  &  - 0.66+ 0.02\{\,(C_{10}^{\prime U})^2 + 
 \,(C_{10}^{\prime V})^2 +  \,(C_{10}^{V})^2+  \,(C_{9}^{\prime U})^2 + 
  \,(C_{9}^{ U})^2+  \,(C_{9}^{V})^2 + \,(C_{10}^{U})^2 \nonumber\\
  &+  \,(C_{9}^{\prime V})^2\} + 0.03\{\,C_{10}^{\prime U}\,C_{9}^{U} + \,C_{10}^{\prime U}\,C_{9}^{V} + \,C_{10}^{\prime V} \,C_{9}^{V} + 0.11\,C_{10}^{\prime V}+  \,C_{10}^{V} \,C_{9}^{\prime U} +  \,C_{10}^{V} \,C_{9}^{\prime V} \nonumber\\
  &+\, C_{10}^{U}\, C_{9}^{\prime U} +  \,C_{10}^{U} \,C_{9}^{\prime V}+ \,C_{10}^{\prime V}\,C_{9}^{U}\}+ 0.04\{-\,C_{10}^{\prime U}\,C_{10}^{U} - \,C_{10}^{\prime U}\,C_{10}^{V} -  \,C_{9}^{\prime U} \,C_{9}^{V}- \,C_{9}^{\prime V} \,C_{9}^{ U}\nonumber\\
  &-  \,C_{9}^{\prime V} \,C_{9}^{V}-  \,C_{9}^{\prime U} \,C_{9}^{ U} - \,C_{10}^{\prime V}\,C_{10}^{U} - \,C_{10}^{\prime V} \, C_{10}^{V} \}+ 0.05\{\,C_{10}^{\prime U}\,C_{10}^{\prime V} + 
 \,C_{10}^{U} \,C_{10}^{V} +  \,C_{9}^{\prime U} \,C_{9}^{\prime V} \nonumber\\
 &+  \,C_{9}^{ U} \,C_{9}^{V}\} + 0.19 \{\,C_{10}^{U} + \,C_{10}^{V}-  \,C_{9}^{\prime U}  -  \,C_{9}^{\prime V} \}+ 0.20\{\,C_{10}^{\prime U}\,C_{9}^{\prime U} + \,C_{10}^{\prime U}\,C_{9}^{\prime V}+\,C_{10}^{\prime V}\,C_{9}^{\prime V}\nonumber\\
 &+\,C_{10}^{\prime V}\,C_{9}^{\prime U}\} - 
 0.28 \{\,C_{10}^{U} \,C_{9}^{ U} - \, C_{10}^{U} \,C_{9}^{V}   -  \,C_{10}^{V} \,C_{9}^{ U} -  \,C_{10}^{V} \,C_{9}^{V}  \}  + 0.11\,C_{10}^{\prime U}  \nonumber\\
 &+ 1.06 \{\,C_{9}^{ U} + \,C_{9}^{V} \}.
\end{align}

\begin{align}
10&\times\mathcal{A}_{FB}^{[14,s_{max}]}(B_c\to D_s^{\ast} \,\tau^+\tau^-)=\nonumber\\
  &0.27  + 0.01\{ -  \,C_{10}^{U} \,C_{9}^{ U} -  \,(C_{9}^{ U})^2 +  \,C_{9}^{\prime U} \,C_{9}^{ U}  -  \,(C_{9}^{\prime U})^2\} + 0.02 \{\,C_{10}^{\prime U} \,C_{9}^{\prime U}-  \,C_{10}^{U} \} \nonumber\\
  &- 0.03 \,C_{10}^{\prime U}+ 0.09\,C_{9}^{\prime U}.
\end{align}

\begin{align}
    10&\times{f}_{L}^{[14,s_{max}]}(B_c\to D_s^{\ast} \,\mu^+\mu^-)=\nonumber\\
   &5.04 +0.06\{\,(C_{10}^{\prime U})^2 + \,(C_{10}^{\prime V})^2 + 
\,(C_{10}^{V})^2+  \,(C_{9}^{\prime U})^2+  \,(C_{10}^{U})^2 \} + 0.13\{ \,C_{10}^{\prime U} \,C_{10}^{\prime V}\nonumber\\
&+  \,C_{10}^{U} \,C_{10}^{V}\}+ 0.04\{ \,C_{9}^{\prime U}-  \,C_{10}^{V} +  \,C_{9}^{\prime V}- \,C_{10}^{U}\}+ 0.08 \{\,(C_{9}^{\prime V})^2+ 
\,(C_{9}^{U})^2\nonumber\\
&- \,C_{9}^{U} + 
 \,(C_{9}^{V})^2 - \,C_{9}^{V}\} + 0.14 \{\,C_{9}^{\prime U} \,C_{9}^{\prime V}  +\,C_{9}^{U} \,C_{9}^{V} \}+ 0.10 \{\,C_{10}^{\prime V}+  \,C_{10}^{\prime U}\} \nonumber\\
 &+ 0.15\{- \,C_{10}^{\prime U} \,C_{10}^{U} - 
  \,C_{10}^{\prime U} \,C_{10}^{V} +  \,C_{10}^{\prime U} \,C_{9}^{\prime U} +  \,C_{10}^{\prime U} \,C_{9}^{\prime V} - 
  \,C_{10}^{\prime U} \,C_{9}^{U} -  \,C_{10}^{\prime U} \,C_{9}^{V}  \nonumber\\
  &-  \,C_{10}^{\prime V} \,C_{10}^{U} -  \,C_{10}^{\prime V} \,C_{10}^{V} + 
  \,C_{10}^{\prime V} \,C_{9}^{\prime U} +  \,C_{10}^{\prime V} \,C_{9}^{\prime V} -  \,C_{10}^{\prime V} \,C_{9}^{U} - 
  \,C_{10}^{\prime V} \,C_{9}^{V}    -  \,C_{10}^{U} \,C_{9}^{\prime U}\nonumber\\
  &-  \,C_{10}^{U} \,C_{9}^{\prime V} + 
  \,C_{10}^{U} \,C_{9}^{U} +  \,C_{10}^{U} \,C_{9}^{V}   -  \,C_{10}^{V} \,C_{9}^{\prime U} -  \,C_{10}^{V} \,C_{9}^{\prime V} + 
  \,C_{10}^{V} \,C_{9}^{U} +  \,C_{10}^{V} \,C_{9}^{V} \nonumber\\
  &-  \,C_{9}^{\prime U} \,C_{9}^{U} - \,C_{9}^{\prime U} \,C_{9}^{V}- \,C_{9}^{\prime V} \,C_{9}^{U} -  \,C_{9}^{\prime V} \,C_{9}^{V} \}.
\end{align}

\begin{align}
    10&\times f_{L}^{[s_{min},6]}(B_c\to D_s^{\ast} \,\mu^+\mu^-)=\nonumber\\
    &24 +0.1\{ \,(C_{10}^{\prime U})^2 +  \,(C_{9}^{U})^2 +   \,(C_{10}^{V})^2+  \,(C_{9}^{\prime U})^2+  \,(C_{10}^{\prime V})^2 +  \,(C_{9}^{V})^2+  \,(C_{10}^{U})^2+ \,(C_{9}^{\prime V})^2\}\nonumber\\
    &+ 0.2 \{\,C_{10}^{\prime U} \,C_{10}^{\prime V} + \,C_{10}^{U} \,C_{10}^{V}+ \,C_{9}^{\prime U} \,C_{9}^{\prime V}+ \,C_{9}^{U} \,C_{9}^{V}\}+ 0.7\{- \,C_{10}^{\prime U} \,C_{10}^{U} -  \,C_{10}^{\prime U} \,C_{10}^{V}-  \,C_{10}^{\prime V} \,C_{10}^{U}\nonumber\\
    &-  \,C_{10}^{\prime V} \,C_{10}^{V}-  \,C_{9}^{\prime U} \,C_{9}^{U}-  \,C_{9}^{\prime U} \,C_{9}^{V}  -  \,C_{9}^{\prime V} \,C_{9}^{U} - \,C_{9}^{\prime V} \,C_{9}^{V} \}+ 0.4 \{-\,C_{9}^{U} - \,C_{9}^{V}\}\nonumber\\
    &+ 0.8\{ \,C_{9}^{\prime V}  -  \,C_{10}^{U}-  \,C_{10}^{V}   +  \,C_{9}^{\prime U}\}+ 1.1 \{\,C_{10}^{\prime U} \,C_{9}^{\prime U} +  \,C_{10}^{\prime U} \,C_{9}^{\prime V} - 
  \,C_{10}^{\prime U} \,C_{9}^{U} -  \,C_{10}^{\prime U} \,C_{9}^{V}   \nonumber\\
  &+ \,C_{10}^{\prime V} \,C_{9}^{\prime U} +  \,C_{10}^{\prime V} \,C_{9}^{\prime V} -  \,C_{10}^{\prime V} \,C_{9}^{U} - 
  \,C_{10}^{\prime V} \,C_{9}^{V}  -  \,C_{10}^{U} \,C_{9}^{\prime U} -  \,C_{10}^{U} \,C_{9}^{\prime V} + 
  \,C_{10}^{U} \,C_{9}^{U} +  \,C_{10}^{U} \,C_{9}^{V}   \nonumber\\
  &-  \,C_{10}^{V} \,C_{9}^{\prime U} -  \,C_{10}^{V} \,C_{9}^{\prime V} + 
  \,C_{10}^{V} \,C_{9}^{U} + \,C_{10}^{V} \,C_{9}^{V} \}+ 2.7\{ \,C_{10}^{\prime U} +  \,C_{10}^{\prime V}\}.
\end{align}

\begin{align}
10&\times\mathcal{A}_{FB}^{[14,s_{max}]}(B_c\to D_s^{\ast} \,\tau^+\tau^-)=\nonumber\\
   &6.8  + 0.2 \{\,C_{10}^{\prime U} \,C_{9}^{\prime U} -  \,C_{10}^{\prime U} \,C_{9}^{U}  -  \,C_{10}^{U} \,C_{9}^{\prime U}+  \,C_{10}^{U} \,C_{9}^{U}+  \,(C_{9}^{\prime U})^2 + \,(C_{9}^{U})^2\} - 0.4\{ \,C_{10}^{\prime U}\nonumber\\
   &+  \,C_{10}^{U} -\,C_{9}^{\prime U} \,C_{9}^{U} + \,C_{9}^{U}\}  - 0.8 \,C_{9}^{\prime U}. \label{tau flexp}
\end{align}

In these expressions, $s_{min}\equiv4m_{\ell}^2 $ and $s_{max}\equiv(M_{B_{c}}-M_{D^{\ast}_{s}})^2$ correspond, respectively, to the maximum and minimum momentum transfer to the final state lepton.

\subsection{Analysis of the Physical Observables in $B_c\to D^{(\ast)}_s\,\ell^+\,\ell^-$ process}

We now go on to present the results of our analysis for the decay $B_c\to D^{(\ast)}_s\,\ell^+\,\ell^-$. We will first discuss the three scenarios S1-S3 (part of F-I) that are one dimensional (1D) in the sense of allowing one extra degree of freedom in terms of one WC. We then 
go on to discuss the $D>1$ dimensional scenarios of F-II.

\subsubsection{1D NP scenarios}
\begin{figure}[H]
    \centering
    \begin{subfigure}[b]{0.32\textwidth}
        \includegraphics[width=\textwidth]{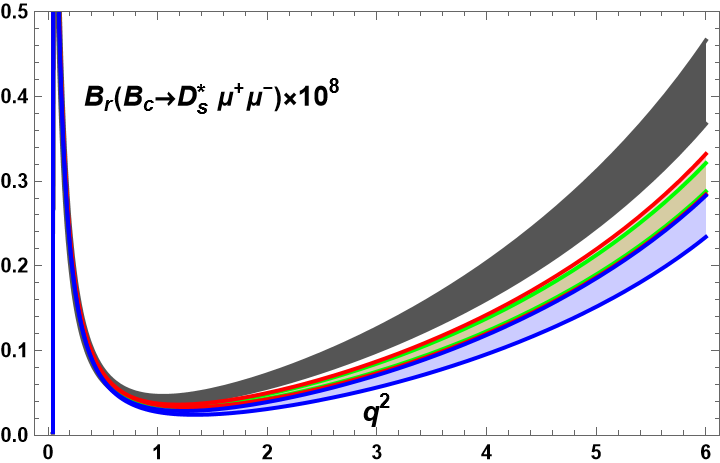}
        \caption{}
        \label{low q BR 1D}
    \end{subfigure} 
    \begin{subfigure}[b]{0.32\textwidth}
        \includegraphics[width=\textwidth]{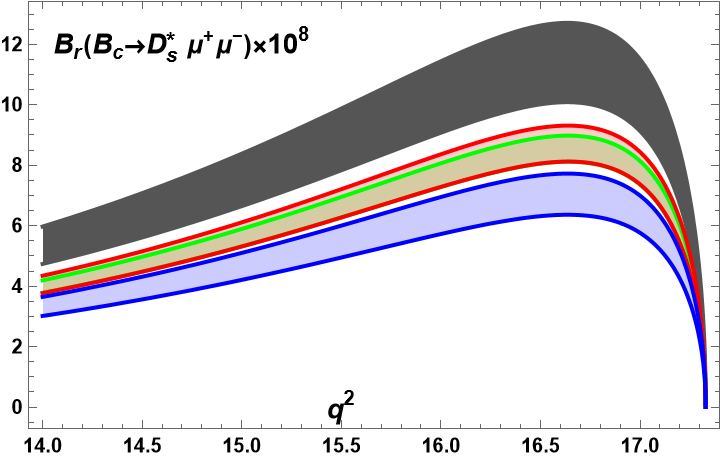}
        \caption{}
        \label{high q BR 1D}
    \end{subfigure}
    \begin{subfigure}[b]{0.32\textwidth}
        \includegraphics[width=\textwidth]{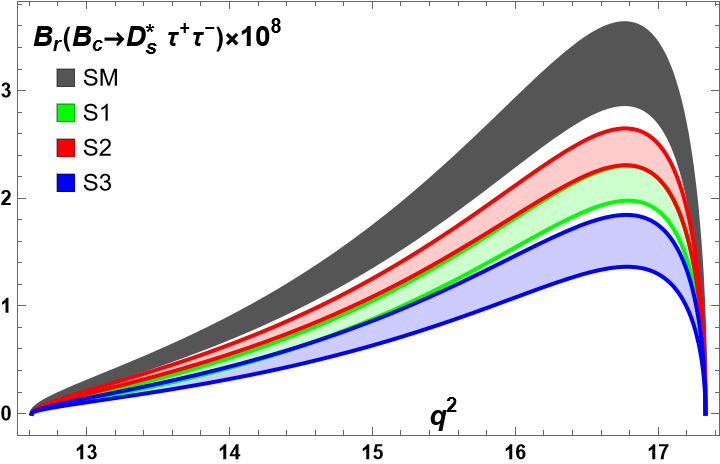}
        \caption{}
        \label{BR 1D}
    \end{subfigure}
    \begin{subfigure}[b]{0.32\textwidth}
        \includegraphics[width=\textwidth]{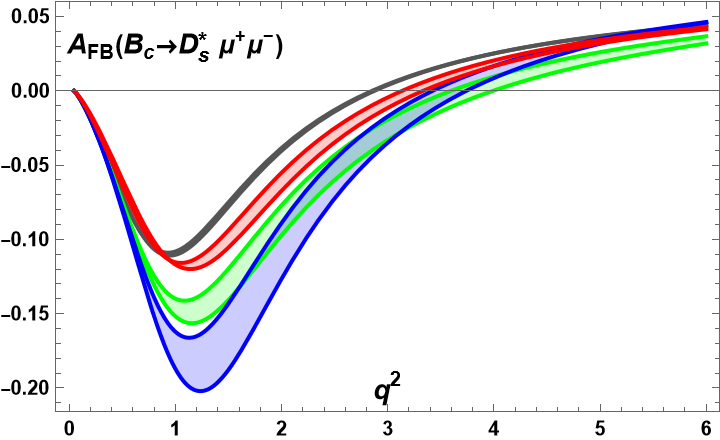}
        \caption{}
        \label{low q AFB 1D}
    \end{subfigure} 
    \begin{subfigure}[b]{0.32\textwidth}
        \includegraphics[width=\textwidth]{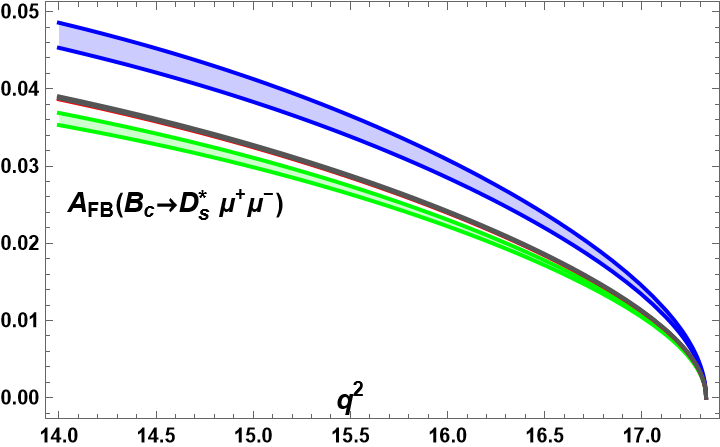}
        \caption{}
        \label{high q AFB 1D}
    \end{subfigure}
    \begin{subfigure}[b]{0.32\textwidth}
        \includegraphics[width=\textwidth]{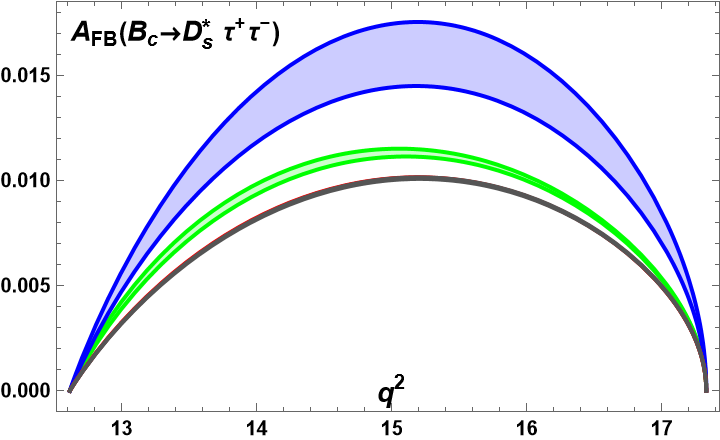}
        \caption{}
        \label{AFB 1D}
    \end{subfigure}
    \begin{subfigure}[b]{0.32\textwidth}
        \includegraphics[width=\textwidth]{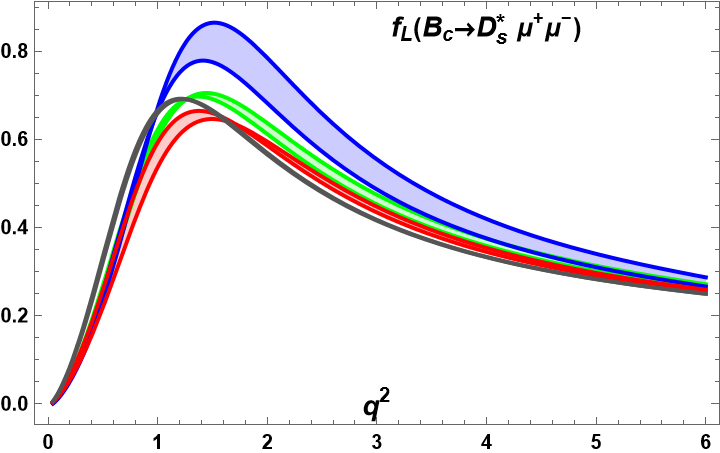}
        \caption{}
        \label{low q fL 1D}
    \end{subfigure} 
    \begin{subfigure}[b]{0.32\textwidth}
        \includegraphics[width=\textwidth]{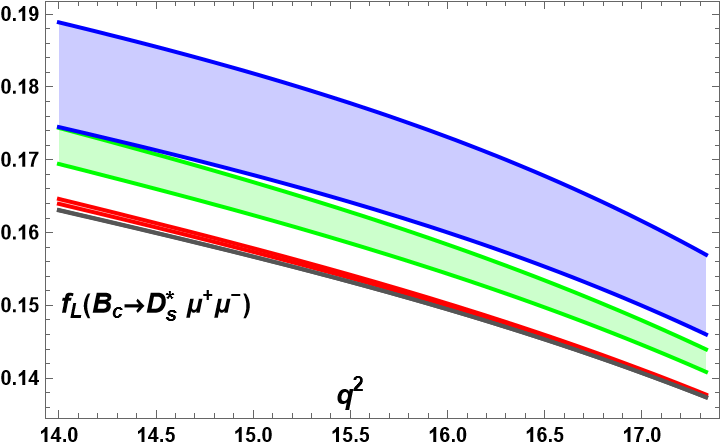}
        \caption{}
        \label{high q fL 1D}
    \end{subfigure}
    \begin{subfigure}[b]{0.32\textwidth}
        \includegraphics[width=\textwidth]{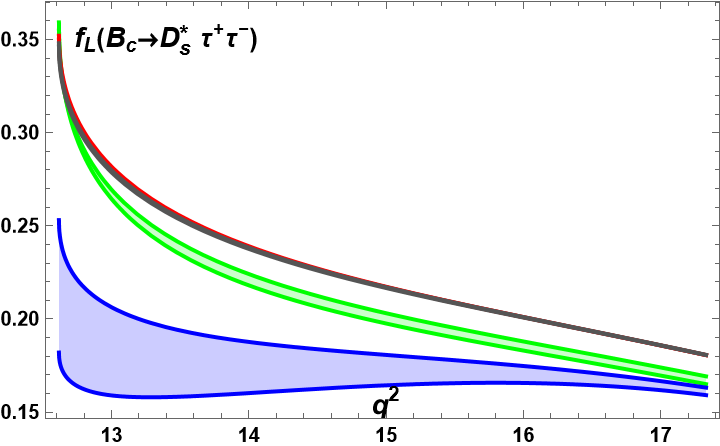}
        \caption{}
        \label{fL 1D fit}
    \end{subfigure}
    \caption{(a-c) The branching ratio, $\mathcal{B}_r(B_c\to D_s^*\ell^+\ell^-):\,\ell=\mu,\tau$, (d-f) the forward-backward asymmetry, $A_{FB}$, and (g-i) the helicity fraction, $f_L$, against the momentum transfer squared, $q^2$. The gray band is for the SM while the green, red, and blue bands correspond to the 1D NP scenarios: S1, S2, and S3, respectively. The width of these bands represents the uncertainty in the SM values due to the form factors as well as the $1\sigma$ allowed ranges for the WCs.} 
    \label{1D scenarios BR}
\end{figure}

In this section, we discuss the impact of 1D NP scenarios on the values of the various physical observables. In Fig.~(\ref{1D scenarios BR}), we present the result for the branching 
ratio ($\mathcal{B}_r$), forward-backward asymmetry ($A_{FB}$), and longitudinal helicity fraction ($f_L$) for the decay $B_c\to D_s^*\ell^+\ell^-$, $l=\mu, \tau$. The color coding of Fig.~(\ref{1D scenarios BR}) 
is: the charcoal gray color represents the SM result, and the green, red, and blue bands correspond 
to S1, S2, and S3, respectively. The widths of these bands is due to form factor uncertainties as 
well as the $1\sigma$ allowed range for the WCs.

In Figs.~(\ref{low q BR 1D} - \ref{BR 1D}), we
 plot the branching ratio of $B_c\to D_s^*\ell^+\ell^-$ ($\ell=\mu,\tau$) against the momentum transfer squared, $q^2$. These plots show that the branching ratio is an increasing function of $q^2$ where one can also see that
for the case of $\mu$, NP effects become significant in the low $q^2$ region, after $q^2\gtrsim2$~GeV$^2$. Continuing with this trend, for both the $\mu$ and $\tau$ cases, NP effects are prominent throughout the high $q^2$ region. However, in the case of $\mu$ for both low and high $q^2$ regions, the effects of NP scenarios S1 and S2 overlap with each other. Therefore, the branching ratio of $B_c\to D_s^*\mu^+\mu^-$ is not a suitable observable to distinguish scenarios S1 and S2. In contrast, for the case of $\tau$, the effects of scenarios S1, S2, and S3 are fairly distinct from each other, particularly for $15~\rm{GeV}^{2} \lesssim q^2 \lesssim 17~\rm{GeV}^{2}.$ Therefore, a precise measurement of the branching ratio $B_c\to D_s^*\tau^+\tau^-$ in this bin provides not only a complementary check of NP but also a better observable to distinguish among the considered 1D NP scenarios, especially to differentiate between S1 and S2.

In Figs.~(\ref{low q AFB 1D} - \ref{AFB 1D}), we show the forward-backward asymmetry ($A_{FB}$) as a function of $q^2$. It can be seen that for the case of $\mu$, NP effects are quite prominent and distinguishable from each other in the $1~\rm{GeV}^{2}\lesssim q^2\lesssim1.8~\rm{GeV}^{2}$ region. On the other hand, in the high $q^2$ region for both $\mu$ and $\tau$ cases, the value of $A_{FB}$ is affected only by the scenarios S1 and S3, while the effects of S2 are negligible and overlap with the SM. Furthermore, in the low $q^2$ region for the case of $\mu$, all these three scenarios decrease the SM value of $A_{FB}$ while in the high $q^2$ region scenario S1 (S3) decreases (increases) and for the case of $\tau$, both S1 and S3 increase the value of $A_{FB}$. In particular, the maximum increment by the scenario S3 (S1) in the value of $A_{FB}$ for $B\to D^*\tau^+\tau^-$ is $\sim48\%$ ($\sim18\%$) at $q^2\simeq15$ GeV$^2$. It can further be seen that scenario S2 agrees with the SM value of $A_{FB}$ throughout the kinematical region for both the $\mu$ and $\tau$ cases. 

In Figs.(\ref{low q fL 1D} - \ref{fL 1D fit}), we plot the longitudinal helicity fraction($f_L$) as a function of  $q^2$. One can see for the case of $\mu$ in the low $q^2$ region, the effects of S3 are quite prominent while the effects of S1 and S2 are rather mild. Here, we observe that in the presence of S3, the maximum SM value of $f_L=0.66$ at $q^2\simeq1.1$ GeV$^2$ is not only changed to 0.86 at $q^2\simeq1.5$ GeV$^2$ but is affected throughout the kinematical range. However, in the high $q^2$ region, both S1 and S3 affect the value of $f_L$ while the effects of S2 are still insignificant. Similarly, for the case of $\tau$, the effects of S1 and S3 are significant throughout the $q^2$ region, especially in the $13\lesssim q^2\lesssim15$ GeV$^2$ region where the effects are not only prominent but also fairly distinguishable from each other. It is also important to note that at high $q^2$, the effects of S1 and S3 increase the $f_L$ for the case of muons while decrease $f_L$ for the case of tauons throughout.

\begin{figure}[H]
    \centering
    \begin{subfigure}[b]{0.32\textwidth}
        \includegraphics[width=\textwidth]{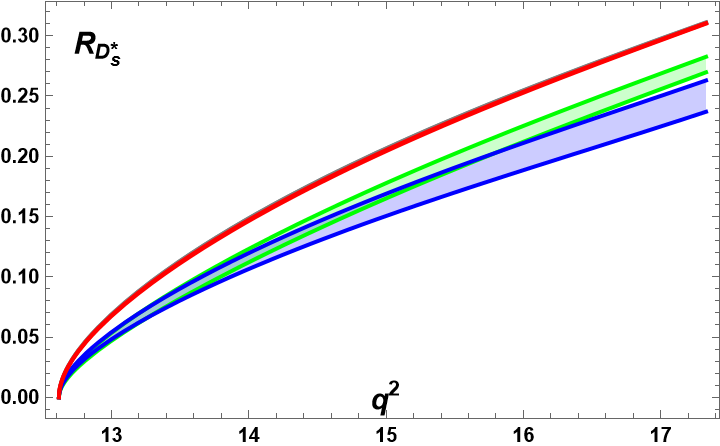}
        \caption{}
        \label{RD 1D}
    \end{subfigure}
       \begin{subfigure}[b]{0.32\textwidth}
        \includegraphics[width=\textwidth]{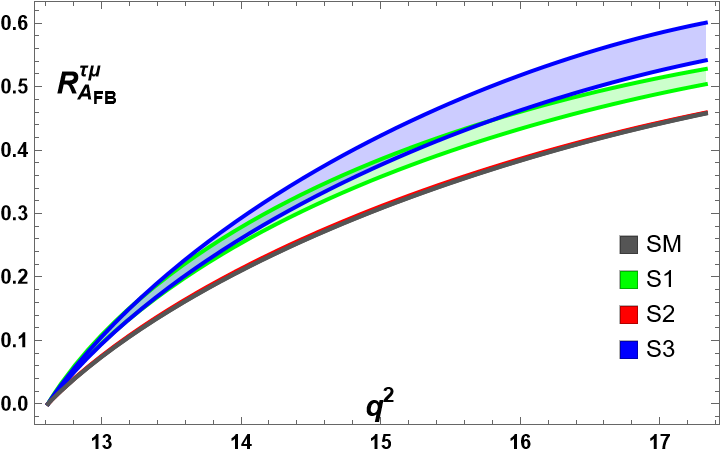}
        \caption{}
        \label{LFR AFB 1D}
    \end{subfigure}
    \begin{subfigure}[b]{0.32\textwidth}
        \includegraphics[width=\textwidth]{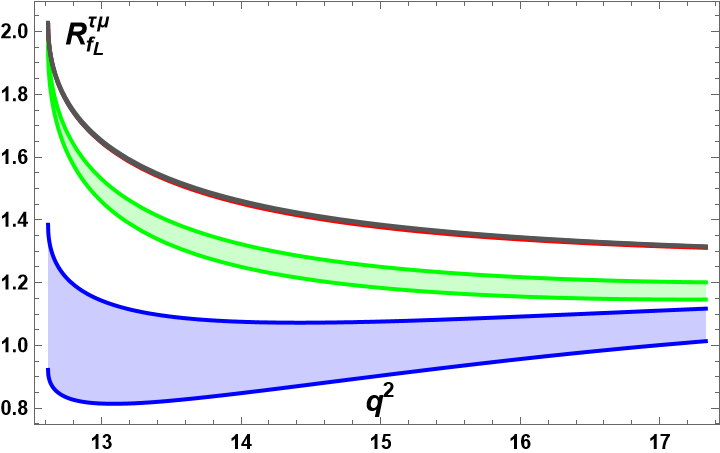}
        \caption{}
        \label{LFR HF 1D}
    \end{subfigure}
    \caption{(a) The lepton flavor universality ratio, $R_{D_s^*}$, (b) the lepton flavor ratio, $R_{A_{FB}}^{\tau\mu}$, and (c) the lepton flavor ratio, $R_{f_{L}}^{\tau\mu}$, as a function of $q^2$ in and beyond the SM scenarios.}
    \label{1D scenarios}
\end{figure}

In Fig.(\ref{RD 1D}), the lepton flavor universality ratio, $R_{D_s^*}$, is plotted against $q^2$ where it can be seen that scenario S2 overlaps with the SM curve while scenarios S1 and S3 lower the $R_{D_s^*}$ values compared with the SM. In particular, one can notice that in scenario S3, the SM value of $R_{D_s^*}=0.31$ at $\text{s}_{\text{max}}$ is reduced to 0.24. Moreover, as a complementary check on LFU, we have calculated the ratio between the $A_{FB}$ ($f_L$) when the final state leptons are tauons to the $A_{FB}$ ($f_L$) when the final state leptons are muons; mathematically $R_{A_{FB}}^{\tau\mu}\equiv\frac{A_{FB}(B_c\to D_s^*\tau^+\tau^-)}{A_{FB}(B_c\to D_s^*\mu^+\mu^-)}$ $\left(R_{f_{L}}^{\tau\mu}\equiv\frac{f_{L}(B_c\to D_s^*\tau^+\tau^-)}{f_{L}(B_c\to D_s^*\mu^+\mu^-)}\right)$. These ratios, in the presence of 1D NP, are plotted as a function of $q^2$ in Figs.(\ref{LFR AFB 1D}) and (\ref{LFR HF 1D}) which clearly show a distinct advantage of considering $R^{\tau\mu}_{f_L}$.

 In conclusion, the NP effects on $\mathcal{B}_r$ are significant in high $q^2$ for both leptons, distinguishing S1 and S2 is challenging in the muon case. $A_{FB}$ shows prominent NP deviations, especially in the low $q^2$ region for muon and high $q^2$ for taun, where S1 and S3 induce opposite effects. The helicity fraction $f_L$ is significantly altered in NP scenarios, particularly S3, which increases $f_L$ for muon while decreasing it for taun. The lepton flavor universality ratio $R_{D_s^*}$ is lowered in S1 and S3, while S2 remains SM like. Additional ratios $R_{A_{FB}}^{\tau\mu}$ and $R_{f_L}^{\tau\mu}$ further probe LFU violation, with $R_{f_L}^{\tau\mu}$ showing clearer NP sensitivity.

\begin{figure}[H]
	\centering
  \begin{subfigure}[b]{0.44\textwidth}
    	 \includegraphics[width=\textwidth]{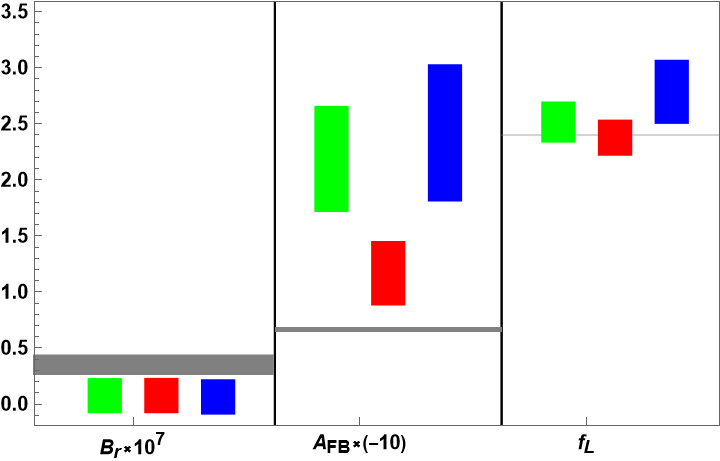}
        \caption{}
        \label{fit-1 bar muon low}
      \end{subfigure}
       \begin{subfigure}[b]{0.44\textwidth}
    	 \includegraphics[width=\textwidth]{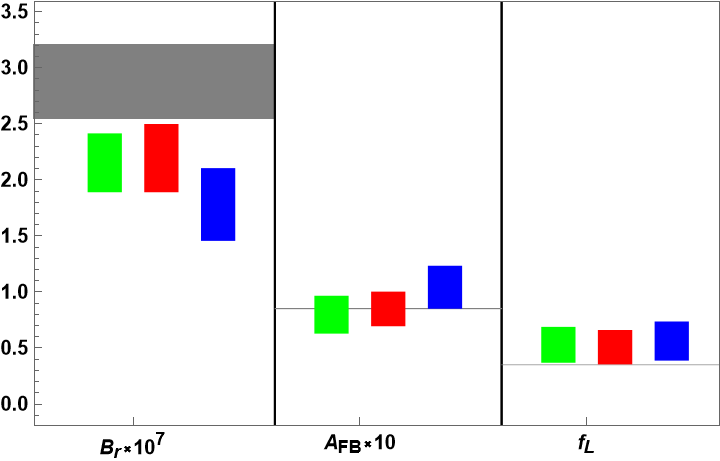}
        \caption{}
        \label{fit-1 bar muon high}
      \end{subfigure}
 \begin{subfigure}[b]{0.44\textwidth}
    	 \includegraphics[width=\textwidth]{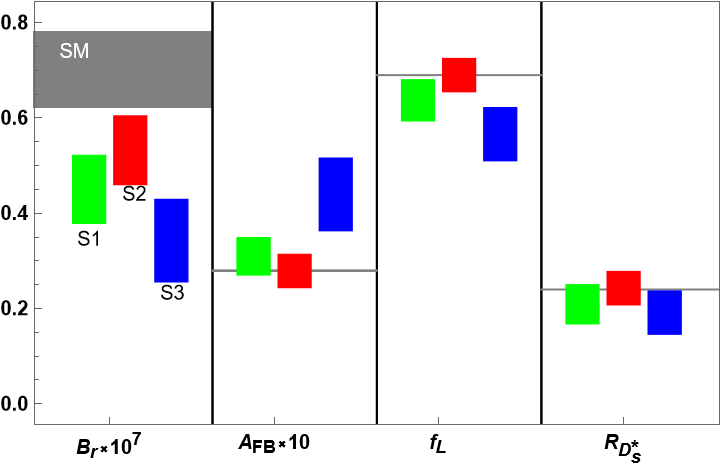}
        \caption{}
        \label{fit-1 bar}
      \end{subfigure}
      \begin{subfigure}[b]{0.42\textwidth}
        \includegraphics[width=\textwidth]{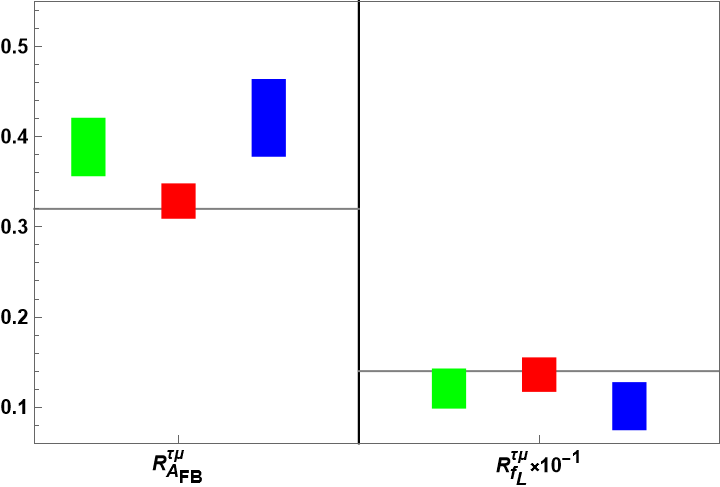}
        \caption{}
        \label{LFR bar 1D}
    \end{subfigure}
	\caption{The variation in the magnitudes of $\mathcal{B}_r$, $A_{FB}$ and $f_L$ due to the presence of NP in the (a) $[s_{min},6]$ bin, (b) in the $[14, s_{max}]$ bin, for the case of muon, and (c) for the case of tauon in $[14, s_{max}]$ bin and $R_{D_s^*}$ magnitude variation while (d) represents the variation in the magnitudes of $R_{A_{FB}}^{\tau\mu}$ and $R_{f_L}^{\tau\mu}$.}  
	\label{bar plots fit-1}
\end{figure}

It is important to remind ourselves that experimental results are typically 
reported after integration over various $q^2$ bins. With this view, we present in Fig.~(\ref{bar plots fit-1}) the 
results of the various physical observables discussed integrated over the $q^2$ regions of interest. Figs.~(\ref{fit-1 bar muon low}) and (\ref{fit-1 bar muon high}) show the integrated results corresponding to the $\mu$ case for the 
$\mathcal{B}_r$, $A_{FB}$, and $f_L$ in the low and high $q^2$ regions, while Fig.~(\ref{fit-1 bar}) presents the same for the $\tau$ and ratios $R_{D^*_S}$. In the same way, we reproduce in Fig.~(\ref{LFR bar 1D}) the results of the ratios $R^{\tau\mu}_{A_{FB}}$, and $R^{\tau\mu}_{f_{L}}$ integrated over the high $q^2$ region. From these bar plots, one can easily and quantitatively observe the deviation from the SM values of these observables by considering 1D NP scenarios. Therefore, the precise measurements of these observables, in low and high $q^2$ bins, provides 
a window of opportunity into the status of NP. In addition, while it may not always be possible, we might be able to distinguish between the 1D NP scenarios as well. 

In conclusion, the intricate dependence of the observables on specific NP couplings. The relative impacts of vector and axial-vector interactions manifest differently across $\mathcal{B}_r$, $A_{FB}$, and $f_L$, making their combined analysis crucial for identifying the underlying NP structure. Precise measurements of these observables, particularly in the tauon sector, provide a powerful probe of LFU violation and allow for differentiation among competing NP scenarios.

\begin{figure}[H]
    \centering
    \begin{subfigure}[b]{0.32\textwidth}
        \includegraphics[width=\textwidth]{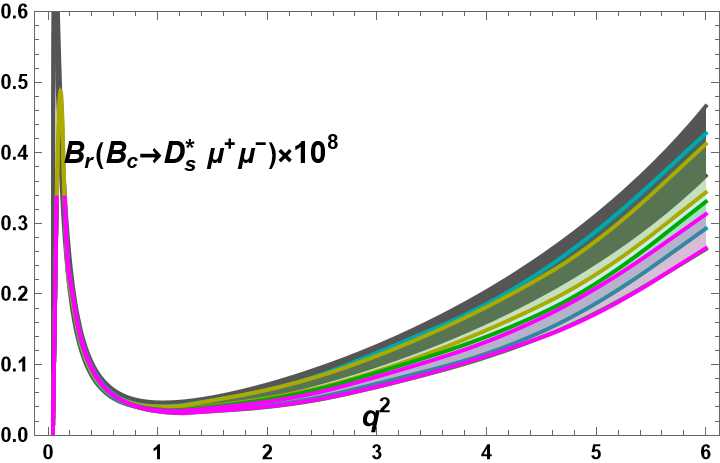}
        \caption{}
        \label{low q BR MI}
    \end{subfigure} 
    \begin{subfigure}[b]{0.32\textwidth}
        \includegraphics[width=\textwidth]{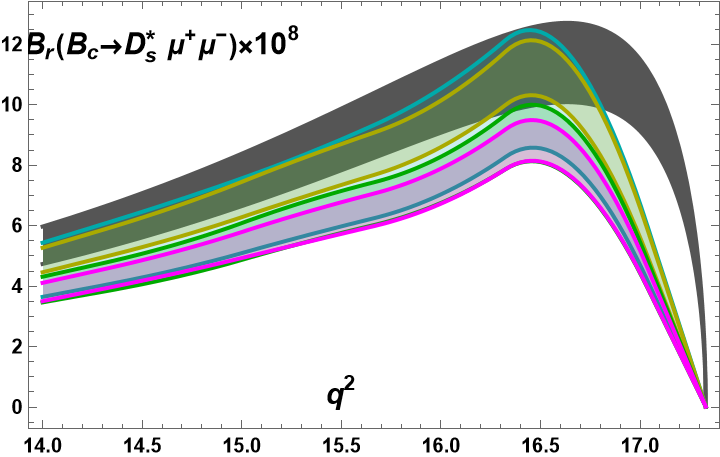}
        \caption{}
        \label{high q BR MI}
    \end{subfigure}
    \begin{subfigure}[b]{0.32\textwidth}
        \includegraphics[width=\textwidth]{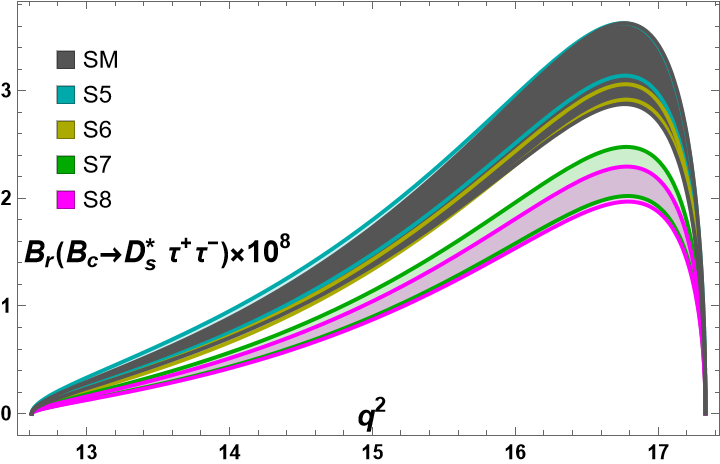}
        \caption{}
        \label{BR MI taun}
    \end{subfigure}
    \begin{subfigure}[b]{0.32\textwidth}
        \includegraphics[width=\textwidth]{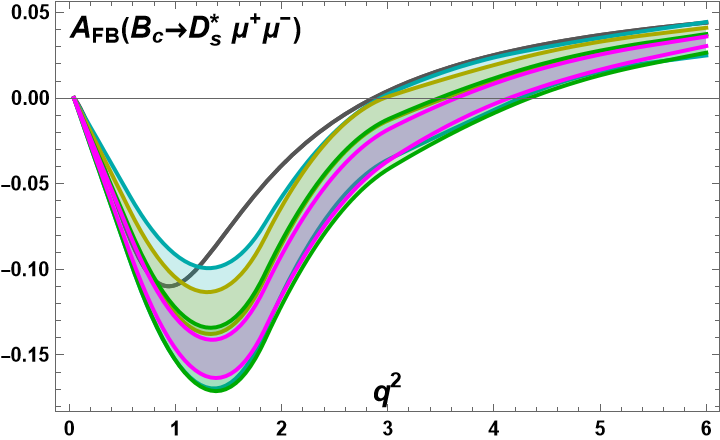}
        \caption{}
        \label{low q AFB MI}
    \end{subfigure} 
    \begin{subfigure}[b]{0.32\textwidth}
        \includegraphics[width=\textwidth]{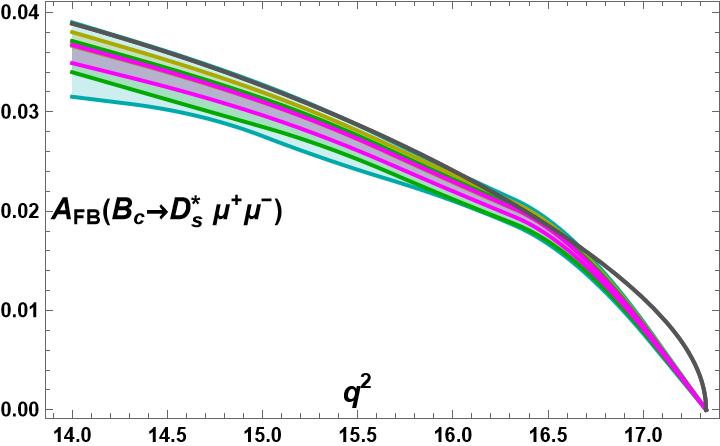}
        \caption{}
        \label{high q AFB MI}
    \end{subfigure}
    \begin{subfigure}[b]{0.32\textwidth}
        \includegraphics[width=\textwidth]{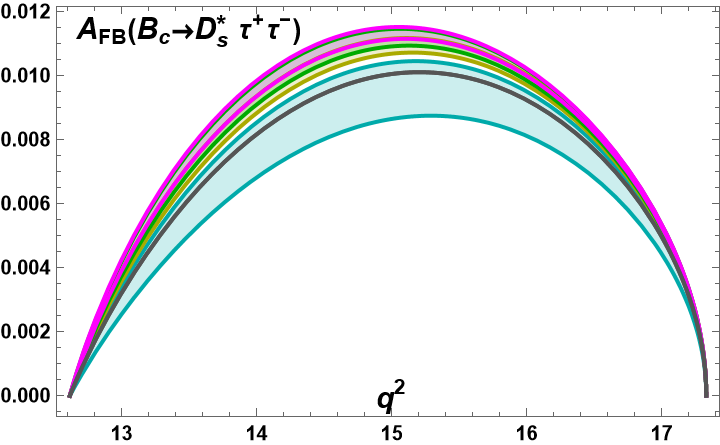}
        \caption{}
        \label{AFB MI taun}
    \end{subfigure}
        \begin{subfigure}[b]{0.32\textwidth}
        \includegraphics[width=\textwidth]{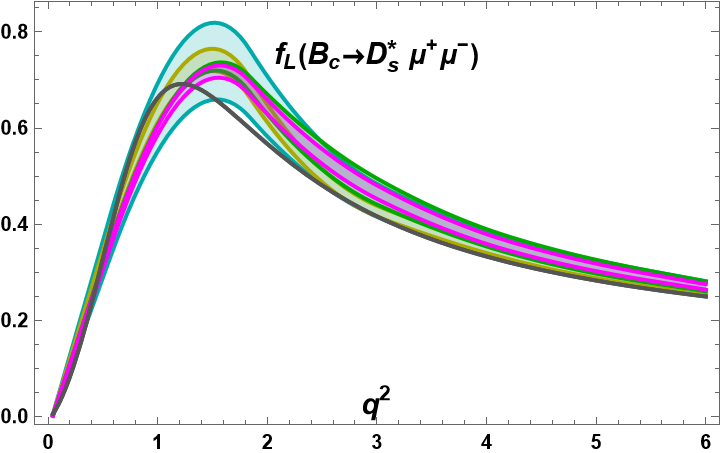}
        \caption{}
        \label{low q fl MI}
    \end{subfigure} 
    \begin{subfigure}[b]{0.32\textwidth}
        \includegraphics[width=\textwidth]{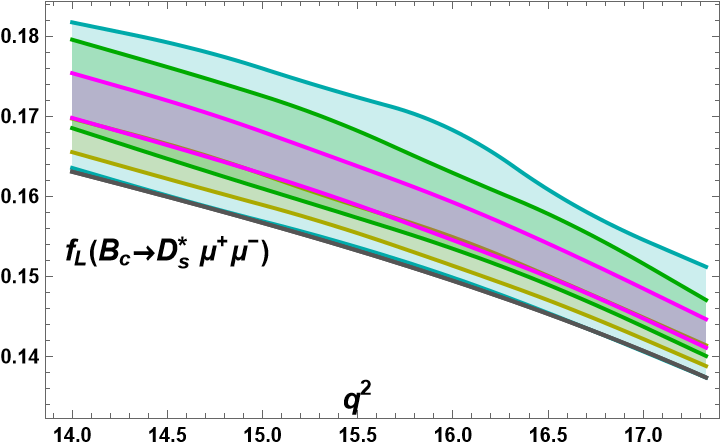}
        \caption{}
        \label{high q fl MI}
    \end{subfigure}
    \begin{subfigure}[b]{0.32\textwidth}
        \includegraphics[width=\textwidth]{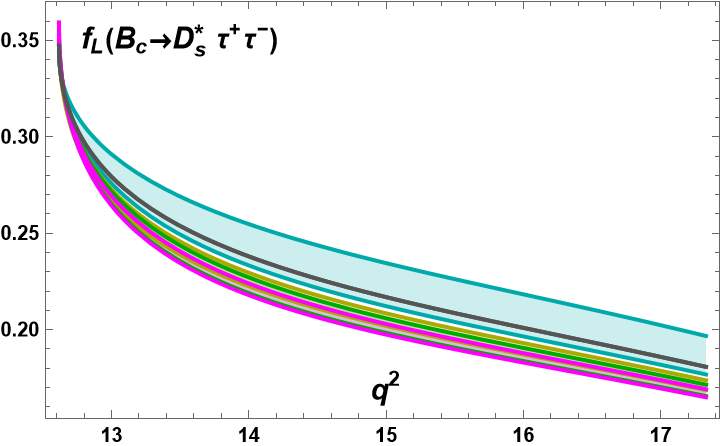}
        \caption{}
        \label{fl MI taun}
    \end{subfigure}
    \caption{(a-c) The $\mathcal{B}_r$, (d-f) the $A_{FB}$ and (g-i) the $f_L$ as a function of $q^2$. The gray curve is for the SM, where the width represents the uncertainty in the SM values due to the form factors. The darker cyan, darker yellow, darker green, and magenta bands correspond to the D$>$1 NP scenarios: S5, S6, S7, and S8, respectively, where the width of the bands show the $1\sigma$ range of the parametric space.
    }
    \label{BR MI}
\end{figure}

\subsubsection{D$>$1 NP scenarios}
We now turn our attention towards the higher dimensional scenarios of F-II. We begin 
by discussing scenarios S5, S6, S7, and S8 which are depicted in darker cyan, darker 
yellow, darker green and magenta colors in Fig.~(\ref{BR MI}) and compared with the 
SM results shown in gray. Fig.~(\ref{BR MI}) is rather similar in content to Fig.~(\ref{1D scenarios BR}) with the only difference between the two being that they discuss different NP scenarios.

In Figs. (\ref{low q BR MI}-\ref{BR MI taun}), we have plotted the branching ratio, $\mathcal{B}_r(B_c\to D_s^*\ell^+\ell^-)$ where ($\ell=\mu,\tau$) against $q^2$ where one can see that for the case of $\mu$, the NP effects become significant in the $q^2\geq2$ GeV$^2$ for the scenarios S7 and S8 while S5 and S6 almost overlap with the SM. Although the effects of S7 and S8 are prominent and visibly decrease the $\mathcal{B}_r$ from its SM value, however, have significant overlap with each other. Similar is the case for the high $q^2$ bin both for the muon and tauon cases. We do note in passing that as far as the $\mathcal{B}_r$ is concerned, the effects of S7 and S8 are most prominent in the tauon case.

In Figs.~(\ref{low q AFB MI}-\ref{AFB MI taun}), we have plotted $A_{FB}$ versus $q^2$ where we observe that all the four scenarios S5-S8 affect its value and at low $q^2$ $\mu$, not only change the position of its minimum value, but also relocate the position of its zero crossing. In the case of high $q^2$ ($\mu$,) all scenarios are destructive and, consequently, lower the $A_{FB}$ value, while for the tauon case, only S5 lowers the value with the effects of S6, S7, and S8 being constructive. 

In Figs.~(\ref{low q fl MI}-\ref{fl MI taun}), we have plotted $f_{L}$ as a function of $q^2$. The trend of NP in this observable both for muon and tauon is opposite to the trend present in $A_{FB}$. However, for the muon, the NP effects are comparatively less prominent than $A_{FB}$ at the low $q^2$ bin, while in the high $q^2$ region, these effects are vice versa.

\begin{figure}[H]
    \centering
    \begin{subfigure}[b]{0.32\textwidth}
        \includegraphics[width=\textwidth]{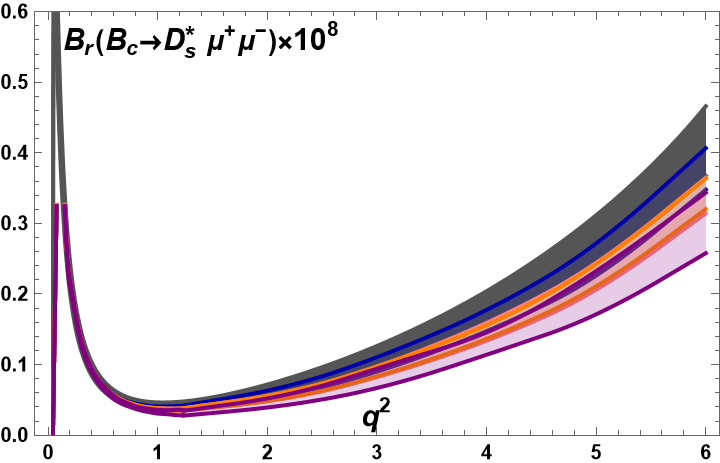}
        \caption{}
        \label{low q BR zprime}
    \end{subfigure} 
    \begin{subfigure}[b]{0.32\textwidth}
        \includegraphics[width=\textwidth]{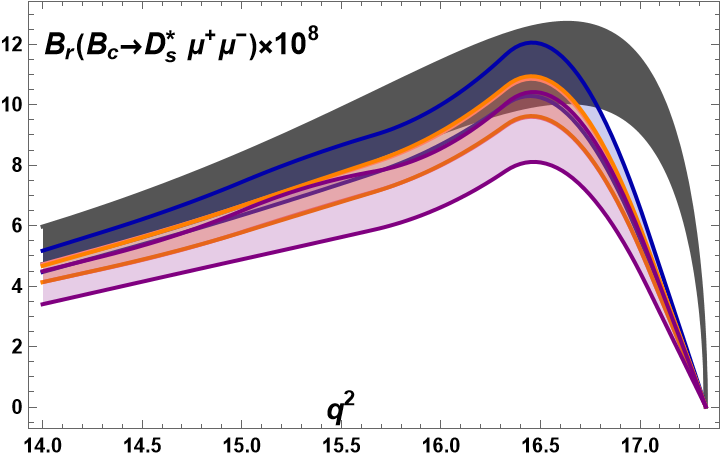}
        \caption{}
        \label{high q BR zprime}
    \end{subfigure}
    \begin{subfigure}[b]{0.32\textwidth}
        \includegraphics[width=\textwidth]{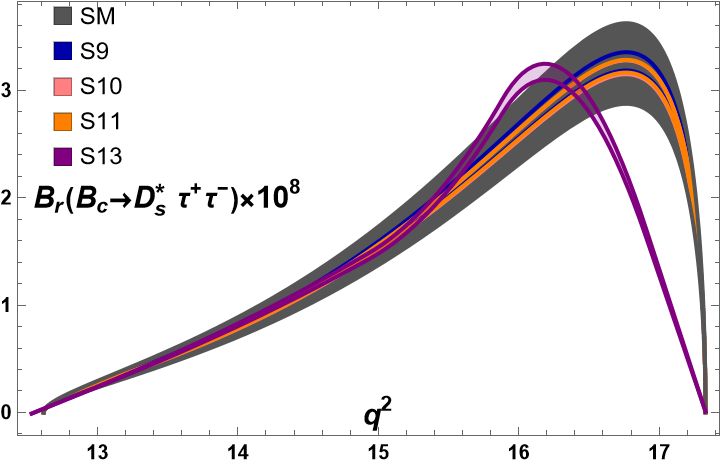}
        \caption{}
        \label{BR zprime taun}
    \end{subfigure}
     \begin{subfigure}[b]{0.32\textwidth}
        \includegraphics[width=\textwidth]{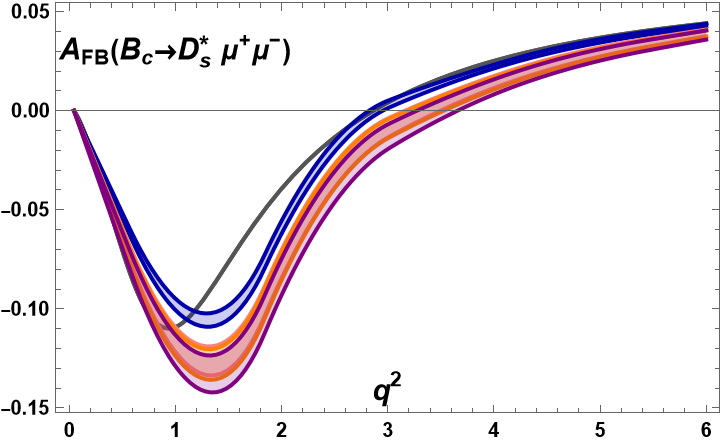}
        \caption{}
        \label{low q afb zprime}
    \end{subfigure} 
    \begin{subfigure}[b]{0.32\textwidth}
        \includegraphics[width=\textwidth]{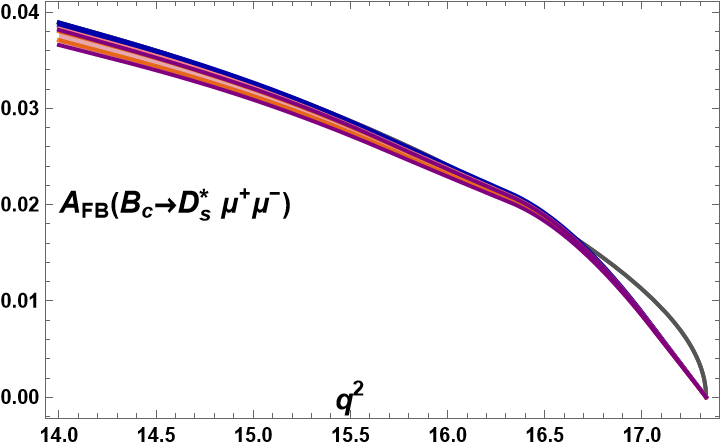}
        \caption{}
        \label{high q afb zprime}
    \end{subfigure}
    \begin{subfigure}[b]{0.32\textwidth}
        \includegraphics[width=\textwidth]{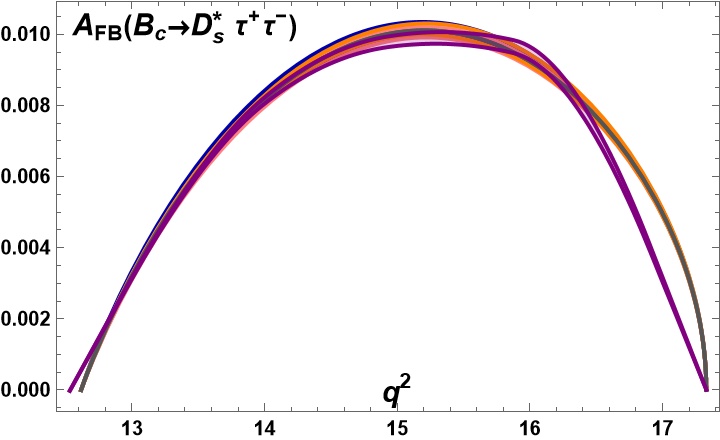}
        \caption{}
        \label{afb zprime taun}
    \end{subfigure}
     \begin{subfigure}[b]{0.32\textwidth}
        \includegraphics[width=\textwidth]{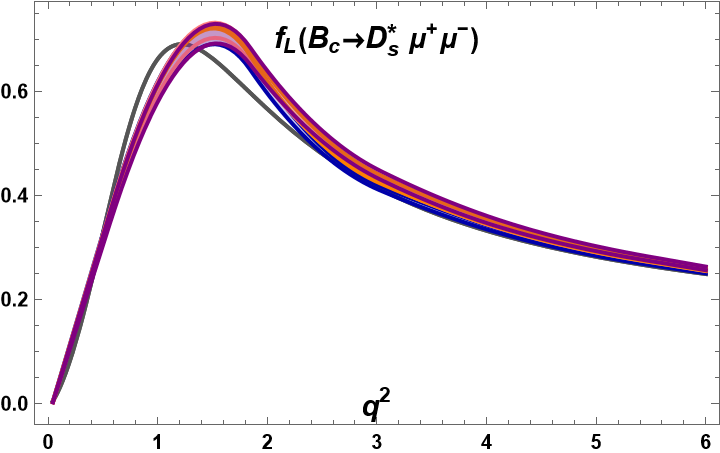}
        \caption{}
        \label{low q fl zprime}
    \end{subfigure} 
    \begin{subfigure}[b]{0.32\textwidth}
        \includegraphics[width=\textwidth]{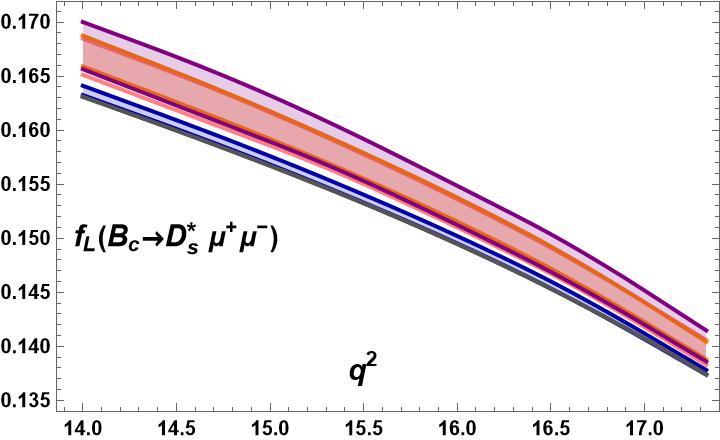}
        \caption{}
        \label{high q fl zprime}
    \end{subfigure}
    \begin{subfigure}[b]{0.32\textwidth}
        \includegraphics[width=\textwidth]{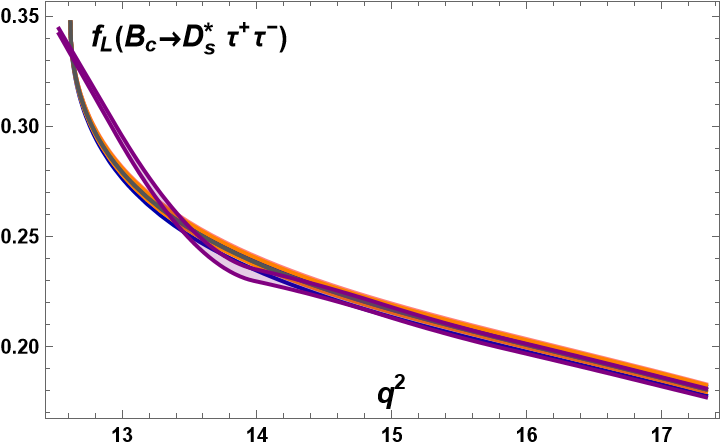}
        \caption{}
        \label{fl zprime taun}
    \end{subfigure}
    \caption{ (a-c) The $\mathcal{B}_r$, (d-f) the $A_{FB}$ and (g-i) the $f_L$ as a function of $q^2$. The gray curve is for the SM, where the width represents the uncertainty in the SM values due to the form factors. The darker blue, pink, orange, and purple bands correspond to the D$>$1 NP scenarios: S9, S10, S11, and S13, respectively, where the width of the bands show the $1\sigma$ range of the parametric space.}
    \label{zprime}
\end{figure}

Fig.~(\ref{zprime}), we have plotted the considered observables against $q^2$ in the presence of those NP scenarios which are motivated by the 2HDM (S9) and $Z^\prime$ models (S10, S11, and S13). We show the results for S9, 10, S11, and S13 scenarios in darker blue, pink, orange and purple, respectively. The NP influence in the branching ratio, for the case of muon (see Figs.~(\ref{low q BR zprime}) and (\ref{high q BR zprime}), are almost the same as the model independent scenarios that are described in Figs. (\ref{low q BR MI}) and (\ref{high q BR MI}). However, for the case of tauon, the effects of scenario S13 are quite distinguishable from the other scenarios as it changes the position of the maximum value of the branching ratio from the high $q^2\simeq17$ GeV$^2$ towards the low value of $q^2\simeq16$ GeV$^2$ (see Fig.~\ref{BR MI taun}).

Similarly, in Figs.~(\ref{low q afb zprime}-\ref{afb zprime taun}),  we have plotted the $A_{FB}$ as a function of $q^2$ where one can see that the NP effects are only prominent for the case of muon in low $q^2$ bin. It is also important to see here that the effects of scenario S9 are distinguished from the other scenarios (see blue band in Fig.~\ref{low q afb zprime}).

In Figs.~(\ref{low q fl zprime}-\ref{fl zprime taun}), we show the variation of $f_{L}$ as a function of $q^2$. From these plots, it can be seen that NP effects are more significant for the case of muon in the high $q^2$ region, except for S9 which mostly overlaps with the SM in this region. NP effects for 
these scenarios in the case of $f_L$ seem rather muted for the low $q^2$ region of the 
$\mu$ and the $\tau$, but the high $q^2$ region for the $\mu$ shows some promise in terms 
of differentiability from the SM values.

\begin{figure}[H]
\centering
    \begin{subfigure}[b]{0.32\textwidth}
        \includegraphics[width=\textwidth]{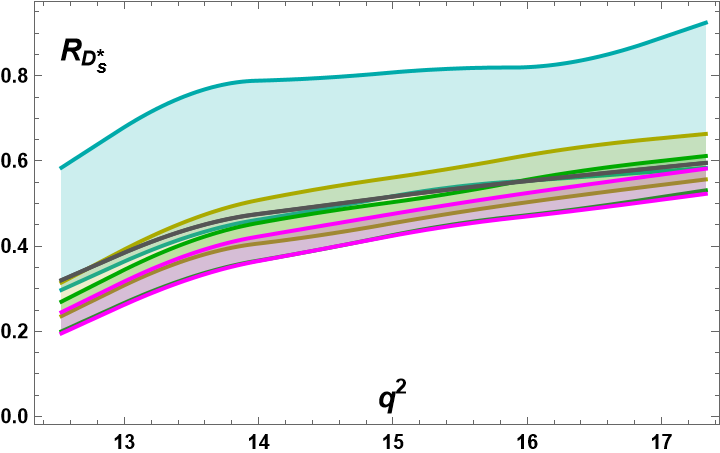}
        \caption{}
        \label{RD MI 2D}
    \end{subfigure}
    \begin{subfigure}[b]{0.32\textwidth}
        \includegraphics[width=\textwidth]{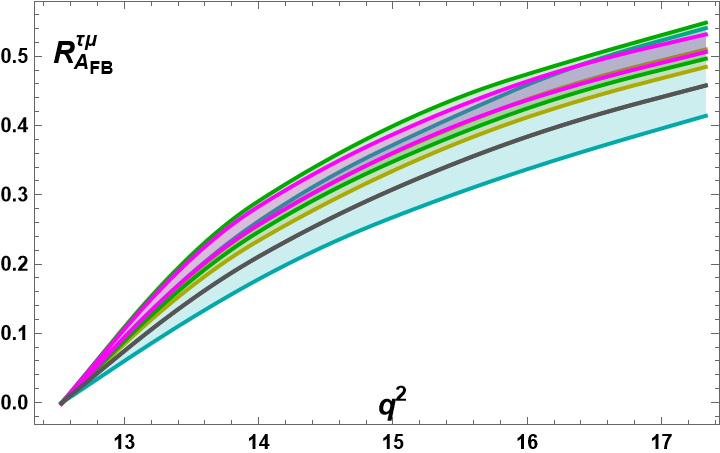}
          \caption{}
        \label{LFR AFB MI 2D}
    \end{subfigure}
     \begin{subfigure}[b]{0.32\textwidth}
        \includegraphics[width=\textwidth]{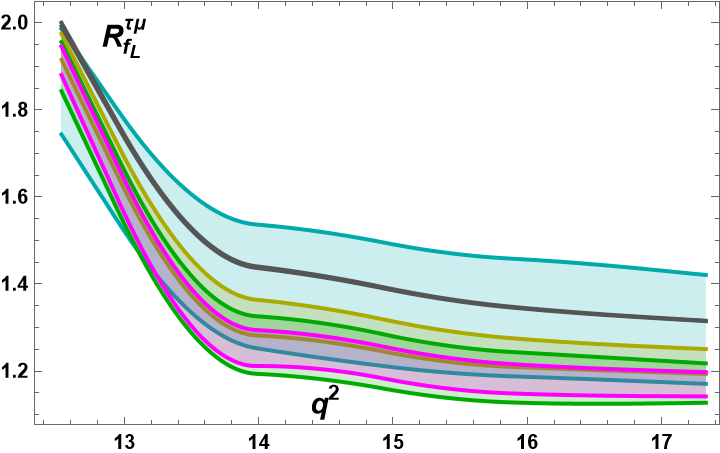}
          \caption{}
        \label{LFR HF MI 2D}
    \end{subfigure}
    \begin{subfigure}[b]{0.32\textwidth}
        \includegraphics[width=\textwidth]{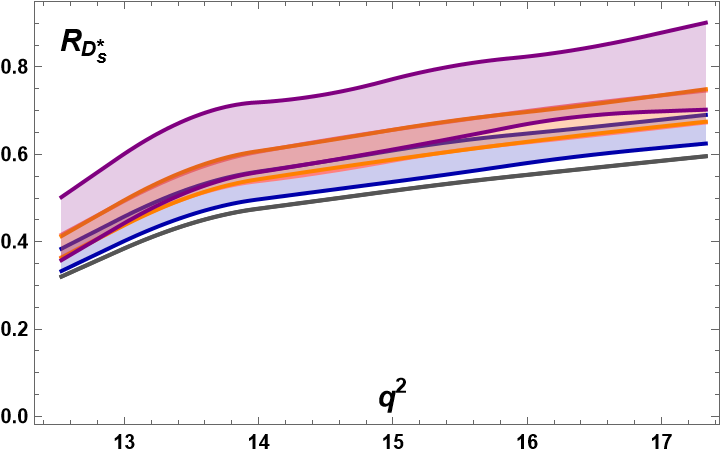}
        \caption{}
        \label{RD zprime 2D}
    \end{subfigure}
    \begin{subfigure}[b]{0.32\textwidth}
        \includegraphics[width=\textwidth]{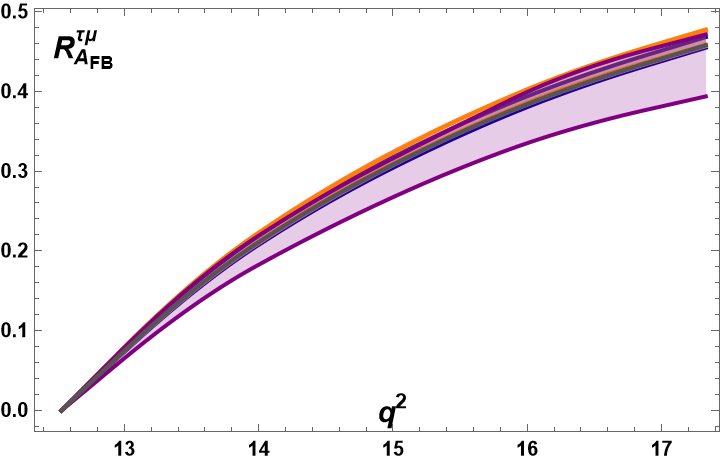}
          \caption{}
        \label{LFR AFB zprime 2D}
    \end{subfigure}
    \begin{subfigure}[b]{0.32\textwidth}
        \includegraphics[width=\textwidth]{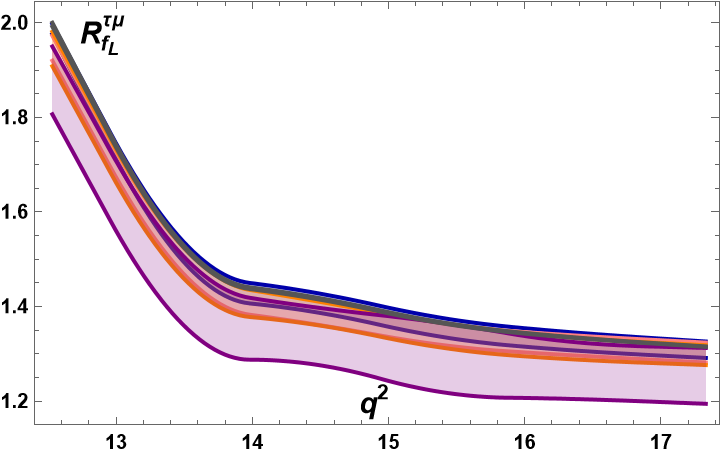}
          \caption{}
        \label{LFR HF zprime 2D}
    \end{subfigure}
    \caption{ The $R_{D_s^*}$, the $R_{A_{FB}}^{\tau\mu}$, and the $R_{f_{L}}^{\tau\mu}$ as a function of $q^2$ are drawn in (a), (b) and (c), respectively, in the presence of S5, S6, S7 and S8 while (d), (e) and (f) depict the variation of these observables in the presence of S9, S10, S11 and S13.}
    \label{2D scenarios}
\end{figure}

The LFU ratio $R_{D^*}$ and the ratios $R_{A_{FB}}^{\tau\mu}$, $R_{f_L}^{\tau\mu}$ are plotted in Fig.~(\ref{2D scenarios}) where one can see that the values of these ratios are sensitive to all the considered higher dimensional scenarios. In particular, however, the scenario, S5 and S13 have more prominent and distinguishable effects than the other scenarios.

In conclusion, the impact of $D>1$ NP scenarios (S5–S8) on the decay $B_c \to D_s^* \ell^+ \ell^-$ is analyzed through $\mathcal{B}_r$, $A_{FB}$, and $f_L$. S7 and S8 significantly reduce $\mathcal{B}_r$ for $\mu$ at $q^2 \geq 2$ GeV$^2$, with stronger effects in the $\tau$ channel. In $A_{FB}$, all scenarios shift the minimum and zero crossing for $\mu$, while S5 lowers and S6–S8 enhance it for $\tau$. NP effects in $f_L$ follow an opposite trend to $A_{FB}$, becoming more pronounced at high $q^2$ for $\mu$. Additional NP scenarios from 2HDM (S9) and $Z'$ models (S10, S11, S13) show similar $\mathcal{B}_r$ effects, with S13 shifting the peak for $\tau$, while S9 distinctly modifies $A_{FB}$ at low $q^2$. LFU ratios $R_{D_s^*}$, $R_{A_{FB}}^{\tau\mu}$, and $R_{f_L}^{\tau\mu}$ reveal sensitivity to all scenarios, with S5 and S13 showing the most prominent deviations.

\begin{figure}[H]
    \centering
    \begin{subfigure}[b]{0.32\textwidth}
        \includegraphics[width=\textwidth]{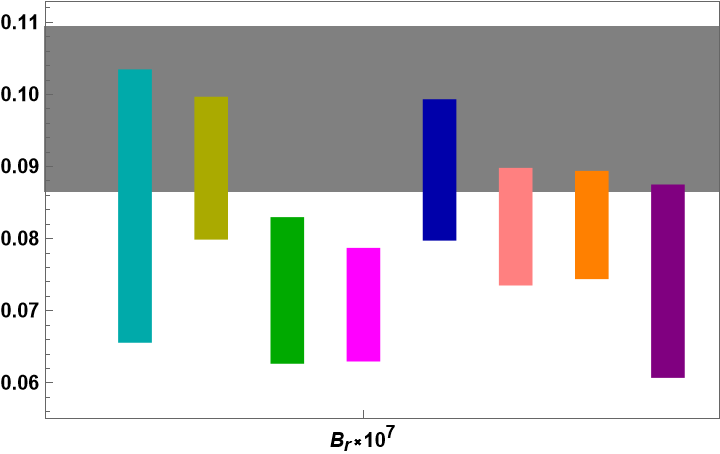}
        \caption{}
        \label{BR bar low 2D}
    \end{subfigure}
        \begin{subfigure}[b]{0.32\textwidth}
        \includegraphics[width=\textwidth]{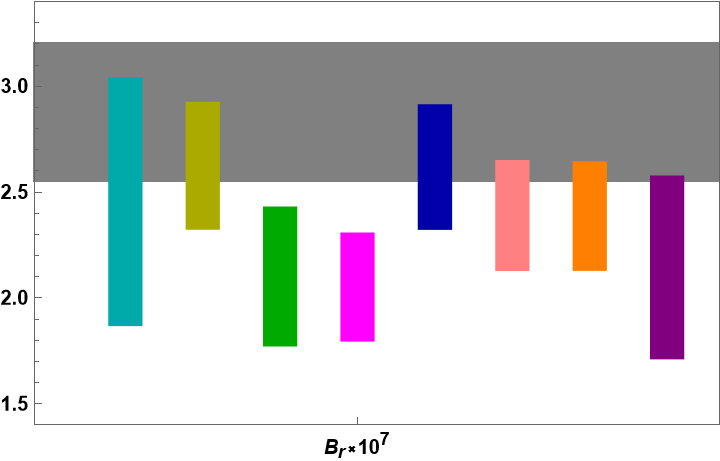}
        \caption{}
        \label{Br bar high 2D}
    \end{subfigure} 
     \begin{subfigure}[b]{0.32\textwidth}
        \includegraphics[width=\textwidth]{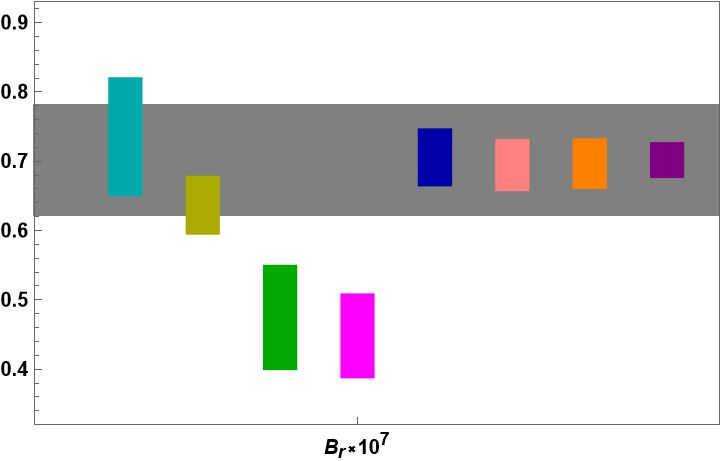}
        \caption{}
        \label{BR bar 2D}
    \end{subfigure}
    \begin{subfigure}[b]{0.32\textwidth}
        \includegraphics[width=\textwidth]{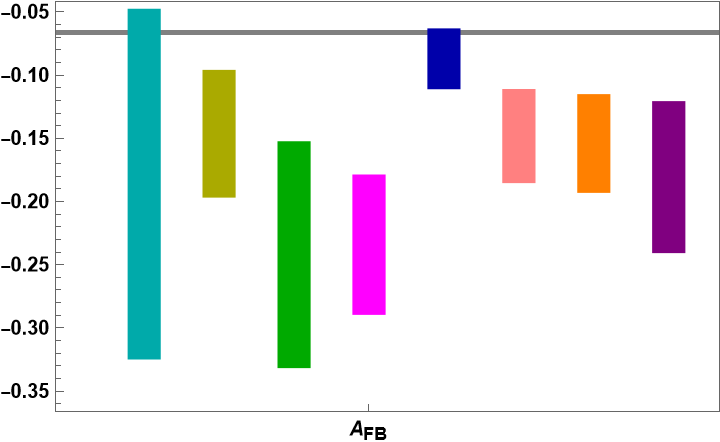}
        \caption{}
        \label{AFB bar low 2D}
    \end{subfigure}
        \begin{subfigure}[b]{0.32\textwidth}
        \includegraphics[width=\textwidth]{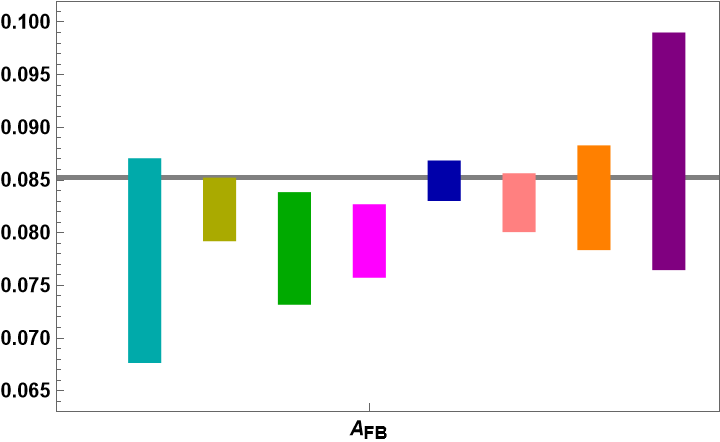}
        \caption{}
        \label{ AFB bar high 2D}
    \end{subfigure}
    \begin{subfigure}[b]{0.32\textwidth}
        \includegraphics[width=\textwidth]{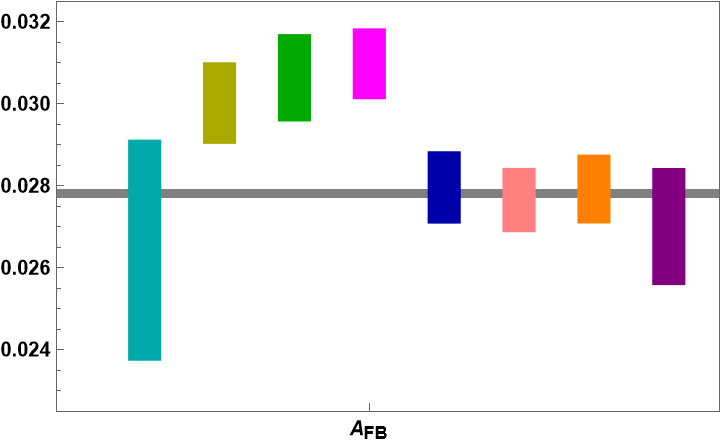}
        \caption{}
        \label{AFB bar 2D}
    \end{subfigure}
    \begin{subfigure}[b]{0.32\textwidth}
        \includegraphics[width=\textwidth]{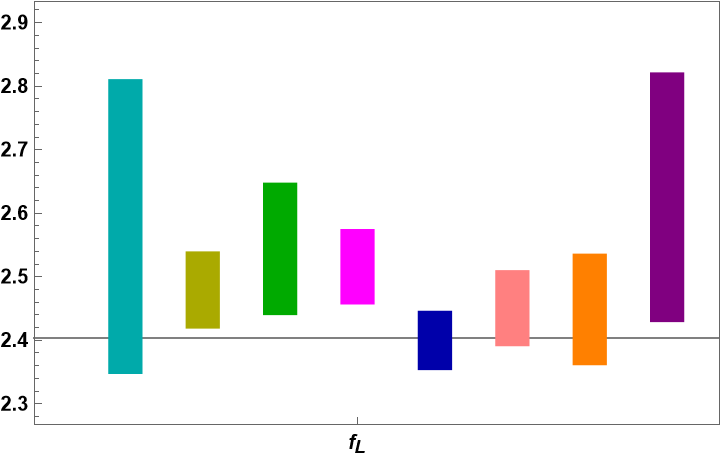}
        \caption{}
        \label{FL bar low 2D}
    \end{subfigure}
    \begin{subfigure}[b]{0.32\textwidth}
        \includegraphics[width=\textwidth]{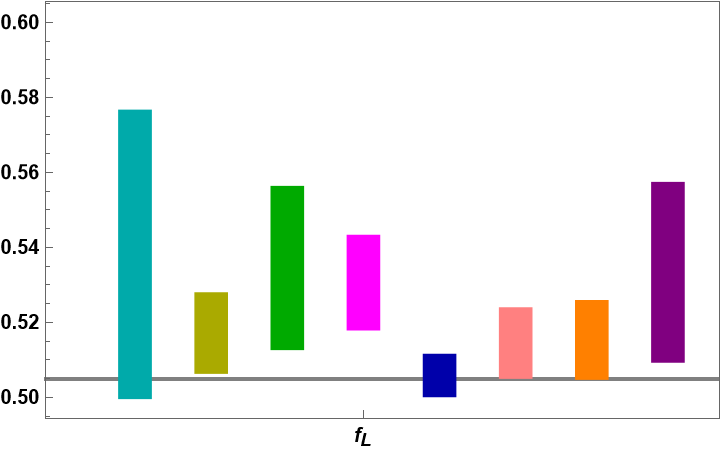}
        \caption{}
        \label{HF bar high 2D }
    \end{subfigure}
    \begin{subfigure}[b]{0.32\textwidth}
        \includegraphics[width=\textwidth]{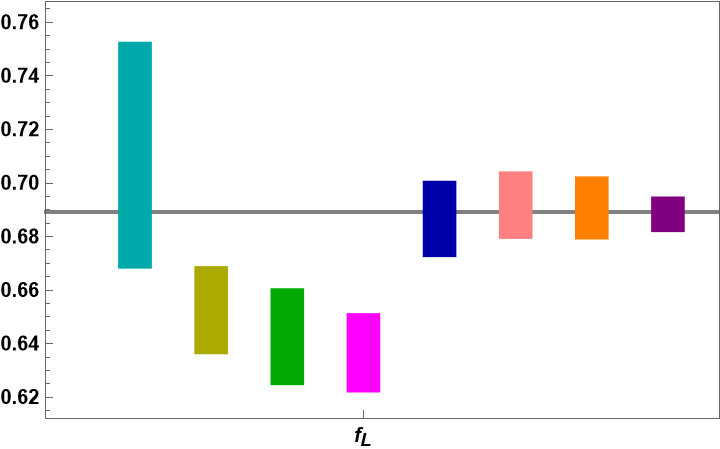}
        \caption{}
        \label{FL bar 2D}
    \end{subfigure}
    \begin{subfigure}[b]{0.32\textwidth}
        \includegraphics[width=\textwidth]{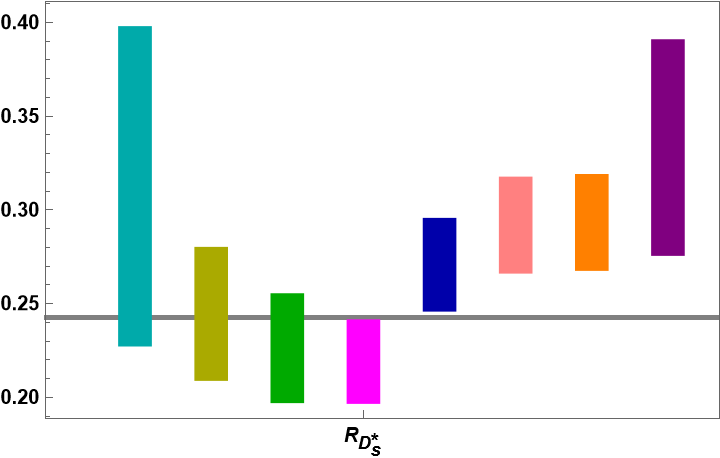}
        \caption{}
        \label{RD bar 2D}
    \end{subfigure}
    \begin{subfigure}[b]{0.32\textwidth}
        \includegraphics[width=\textwidth]{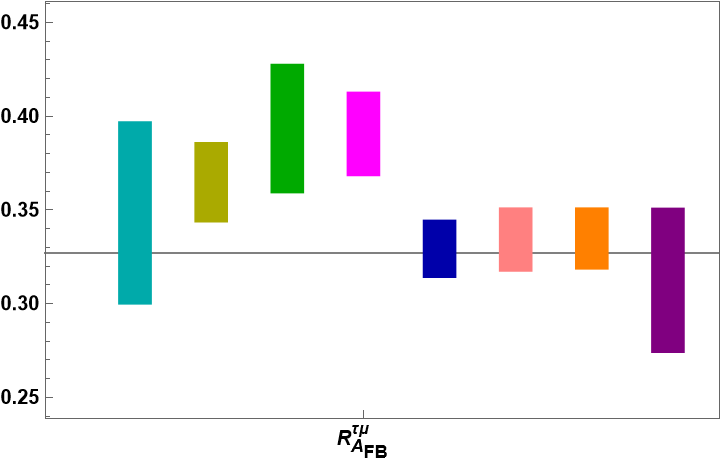}
        \caption{}
        \label{LFR AFB bar 2D}
    \end{subfigure}
    \begin{subfigure}[b]{0.32\textwidth}
        \includegraphics[width=\textwidth]{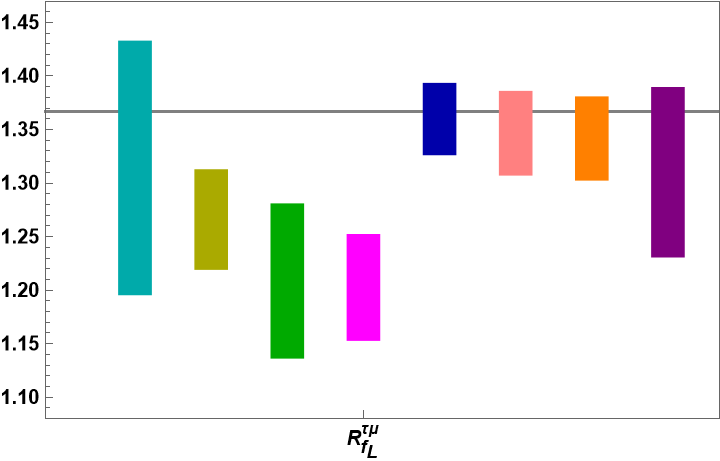}
        \caption{}
        \label{LFR HF bar 2D }
    \end{subfigure}
    \caption{The variation in the magnitudes of $\mathcal{B}_r$, $A_{FB}$, $f_L$ and $R_{D_s^*}$ due to the presence of $D>1$ NP are drawn in (a), (d), (g) and (j), respectively, in the $[s_{min},6]$ bin while (b), (e), (h) and (k) are in the $[14, s_{max}]$ bin for the case of muon, and (c), (f), (i) and (l) correspond to the case of tauon in $[14, s_{max}]$.}
    \label{2D bar scenarios}
\end{figure}

 In Fig.~(\ref{2D bar scenarios}) we plot the results 
of the physical observables for F-II after integrating over the various $q^2$ bins much like we did in Fig.~(\ref{bar plots fit-1}) for the 1D cases. The first two columns in the first three
rows of Fig. (\ref{2D bar scenarios}) correspond to the $\mu$ case in the low and high $q^2$ regions, while the third column represents the case when the final state leptons are
tauons. In the last row, we show the results for the ratios $R_{D^*_s}$, 
$R^{\tau\mu}_{A_{FB}}$ and $R^{\tau\mu}_{f_{L}}$ integrated over the high $q^2$ bin.
From these bar plots, one can easily and quantitatively observe the correction 
to the SM values for the considered observables due to D>1 NP scenarios. Therefore, the precise measurements of these observables, in low and high $q^2$ bins, not only provides a complementary check 
to explore the status of NP but also helps to discriminate among the D>1 NP scenarios.

\begin{figure}[H]
    \centering
    \begin{subfigure}[b]{0.32\textwidth}
        \includegraphics[width=\textwidth]{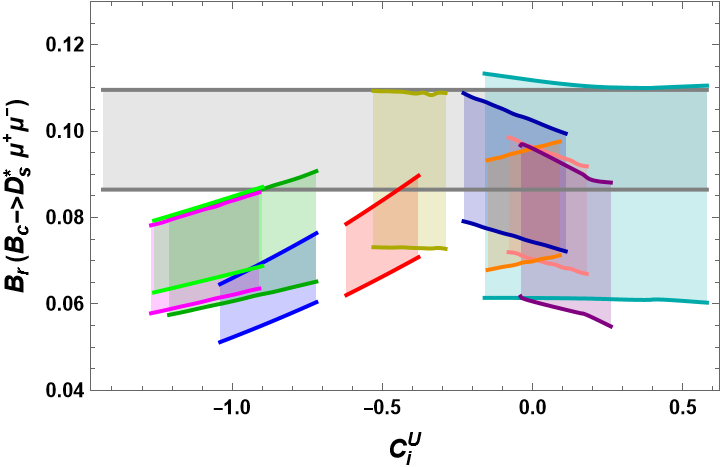}
        \caption{}
        \label{WC muon low q BR}
    \end{subfigure}
    \begin{subfigure}[b]{0.32\textwidth}
        \includegraphics[width=\textwidth]{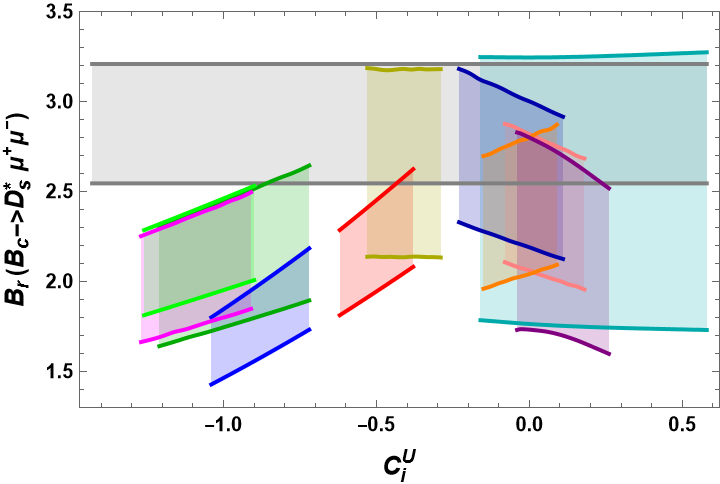}
        \caption{}
        \label{WC muon high q BR}
    \end{subfigure}
    \begin{subfigure}[b]{0.32\textwidth}
        \includegraphics[width=\textwidth]{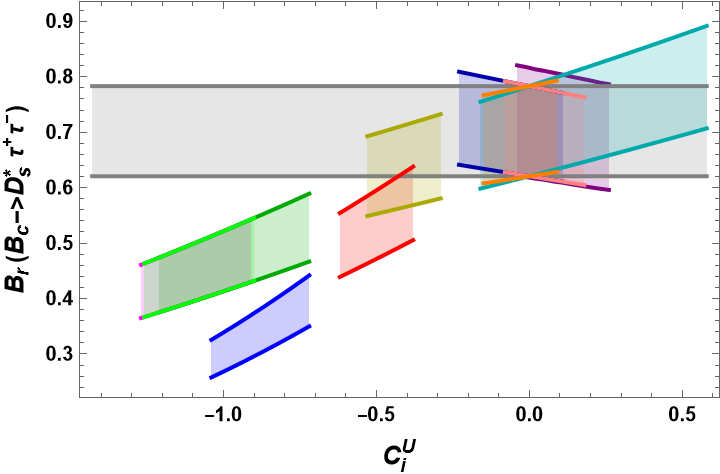}
        \caption{}
        \label{WC BR}
    \end{subfigure}
    \begin{subfigure}[b]{0.32\textwidth}
        \includegraphics[width=\textwidth]{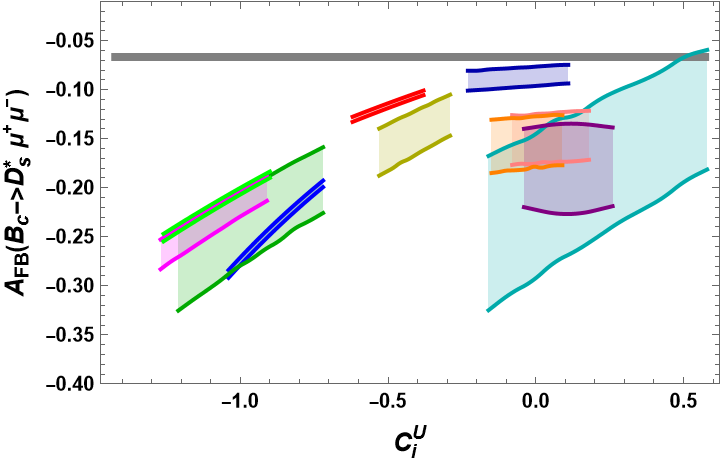}
        \caption{}
        \label{WC muon low q AFB}
    \end{subfigure}
        \begin{subfigure}[b]{0.32\textwidth}
        \includegraphics[width=\textwidth]{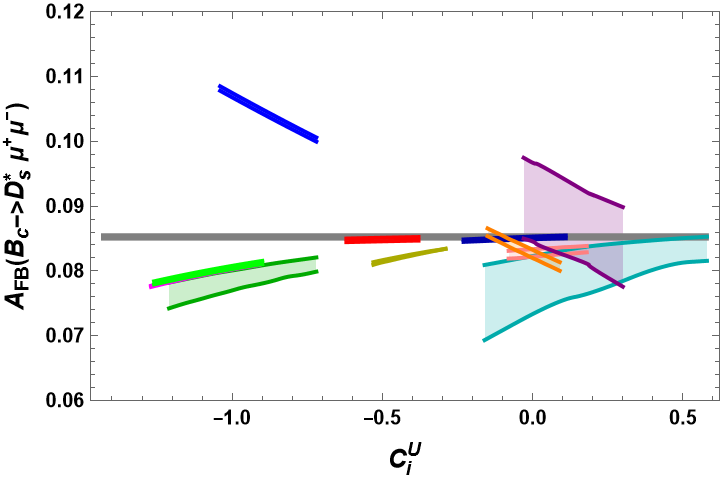}
        \caption{}
        \label{WC muon high q AFB}
    \end{subfigure}
     \begin{subfigure}[b]{0.32\textwidth}
        \includegraphics[width=\textwidth]{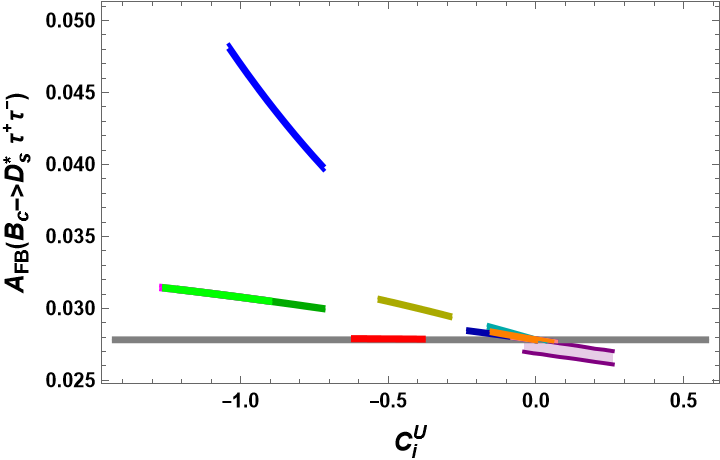}
        \caption{}
        \label{WC AFB}
    \end{subfigure}
    \begin{subfigure}[b]{0.32\textwidth}
        \includegraphics[width=\textwidth]{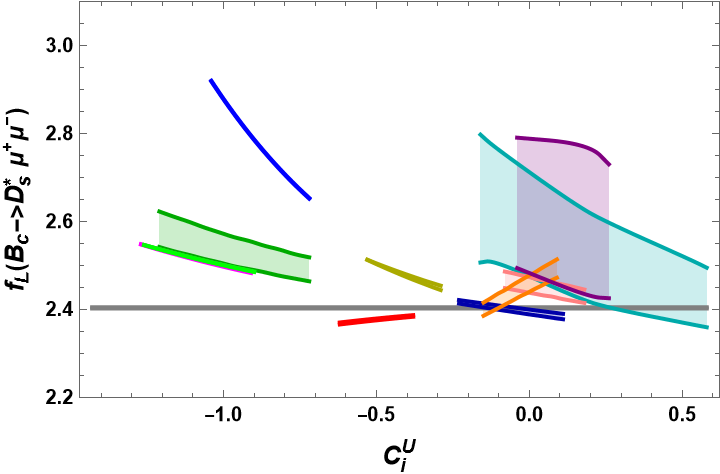}
        \caption{}
        \label{WC muon low q fL}
    \end{subfigure}
     \begin{subfigure}[b]{0.32\textwidth}
        \includegraphics[width=\textwidth]{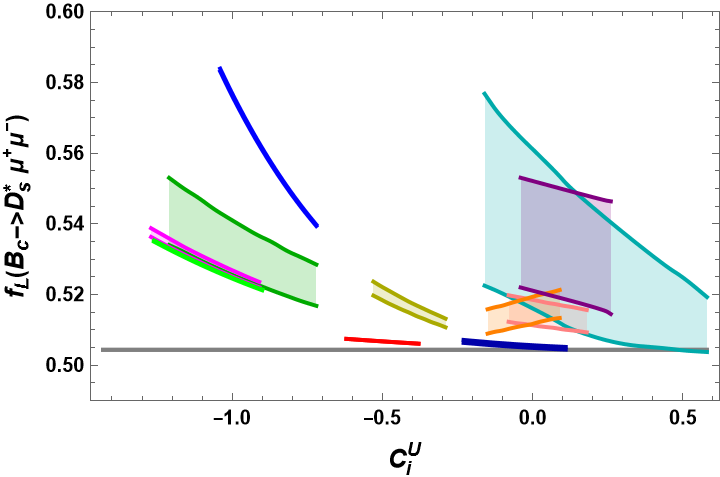}
        \caption{}
        \label{WC muon high q fL}
    \end{subfigure}
        \begin{subfigure}[b]{0.32\textwidth}
        \includegraphics[width=\textwidth]{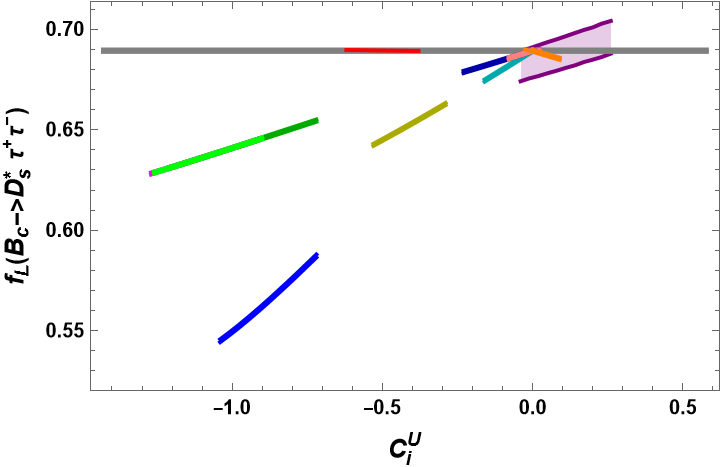}
        \caption{}
        \label{WC fL}
    \end{subfigure}
    \begin{subfigure}[b]{0.32\textwidth}
        \includegraphics[width=\textwidth]{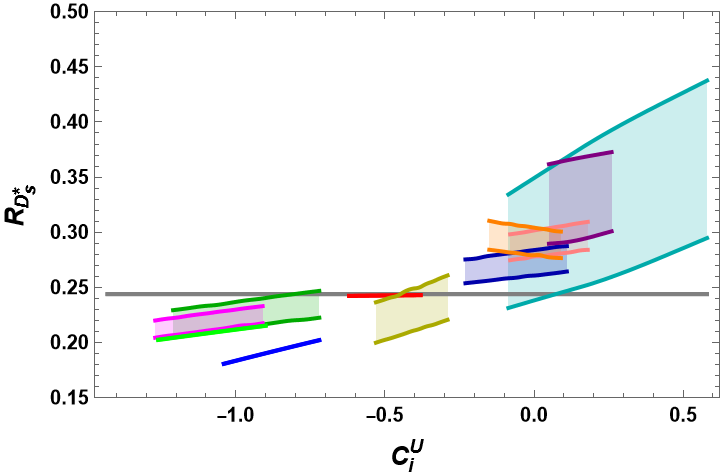}
        \caption{}
        \label{WC RD}
    \end{subfigure}
    \begin{subfigure}[b]{0.32\textwidth}
        \includegraphics[width=\textwidth]{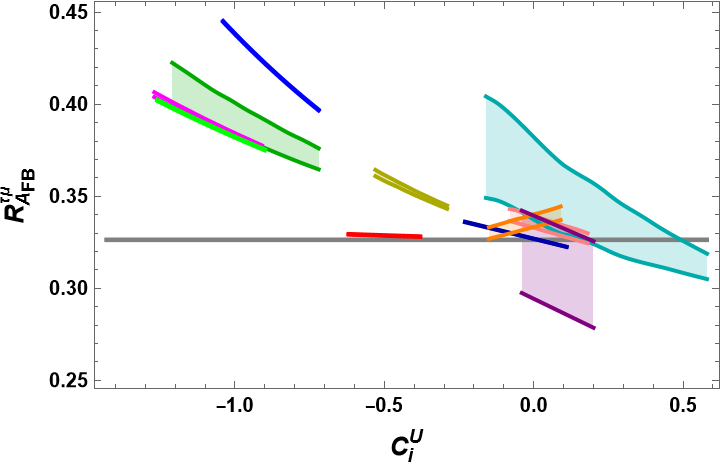}
        \caption{}
        \label{WC RAFB}
    \end{subfigure}
     \begin{subfigure}[b]{0.32\textwidth}
        \includegraphics[width=\textwidth]{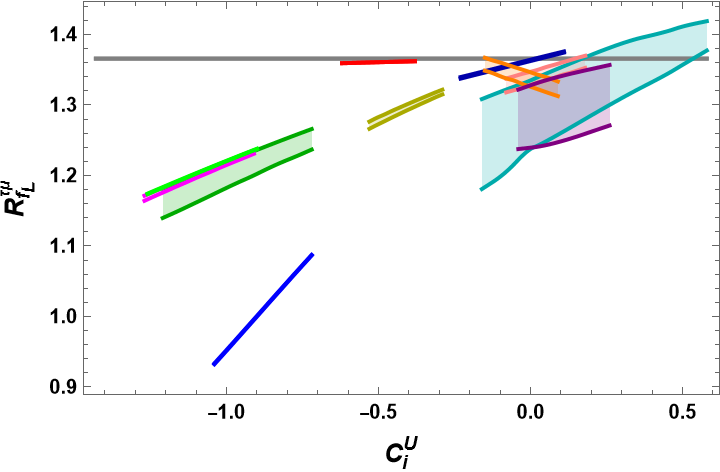}
        \caption{}
        \label{WC RfL}
    \end{subfigure}
  \caption{The variation in the magnitudes of $\mathcal{B}_r$, $A_{FB}$, $f_L$ and $R_{D_s^*}$ as a function of NP WCs are drawn in (a), (d), (g) and (j), respectivle, in the $[s_{min},6]$ bin while (b), (e), (h) and (k) are in the $[14, s_{max}]$ bin for the case of muon, and (c), (f), (i) and (l) correspond to the case of tauon in $[14, s_{max}]$.  The $\mathcal{B}_r$, $A_{FB}$, $f_L$ and $R_{D_s^*}$ as a function of NP WCs.}
    \label{WC}
\end{figure}

Moreover, to see the explicit dependence of the observables on the allowed parametric space of the NP WCs, in Fig.~(\ref{WC}), we have plotted the observables as a function of NP WCs, $C_i^{(\prime)j}$, by integrating over the low and high $q^2$ bins where $i=9,10,9\mu,10\mu$ and $j=U,V$. In order to generate this figure, we have used the expressions defined in Eqs.~(\ref{tua br expr}-\ref{tau flexp}) where we randomly change the values of $C_{9\mu,10\mu}^{(\prime)V}$ and make the piece-wise variation in the values of $C_{9,10}^{(\prime)U}$ in their $1\sigma$ ranges that are listed in 
Tables~(\ref{framework-I}-\ref{framework-II1}). From these plots, one can easily see how the values of the observables change when we change the values of NP WCs in
their allowed $1\sigma$ range. Consequently, the precise measurement of these observables may play a pivotal role in not only further constraining the parametric space of various NP scenarios but may also provide an insight to accommodate the discrepancies present in the $b\to s\ell^+\ell^-$ data.

\begin{figure}[H]
    \centering
    \begin{subfigure}[b]{0.32\textwidth}
        \includegraphics[width=\textwidth]{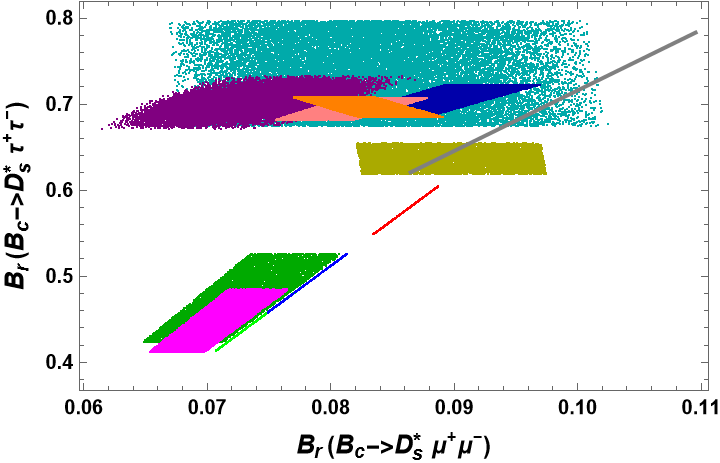}
        \caption{}
        \label{cor7_2}
    \end{subfigure}
    \begin{subfigure}[b]{0.32\textwidth}
        \includegraphics[width=\textwidth]{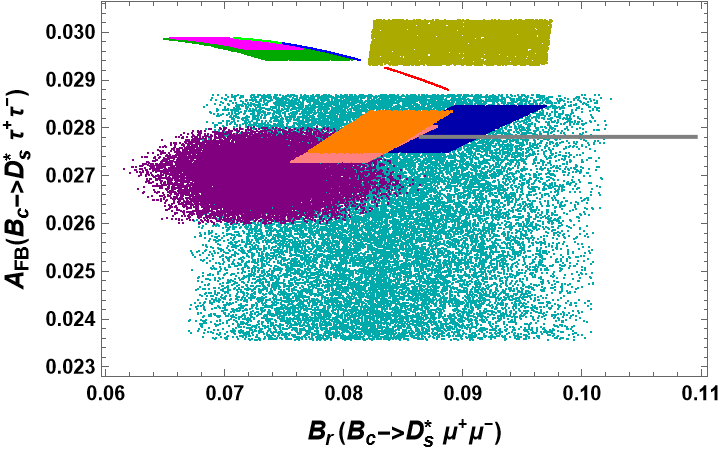}
        \caption{}
        \label{cor7_5}
    \end{subfigure}
    \begin{subfigure}[b]{0.32\textwidth}
        \includegraphics[width=\textwidth]{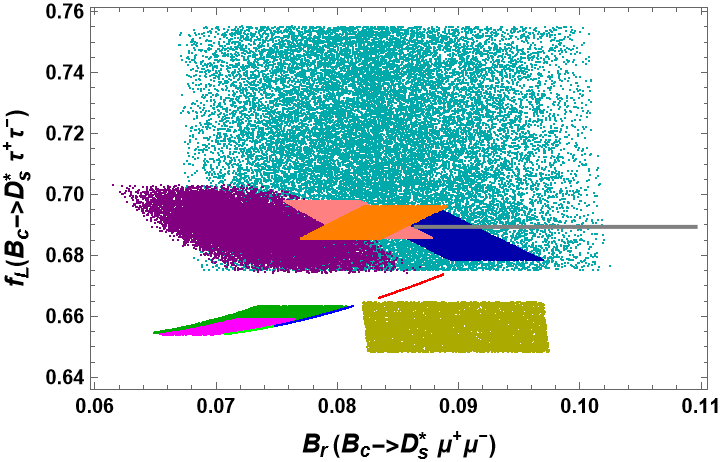}
        \caption{}
        \label{cor7_6}
    \end{subfigure}
    \begin{subfigure}[b]{0.32\textwidth}
        \includegraphics[width=\textwidth]{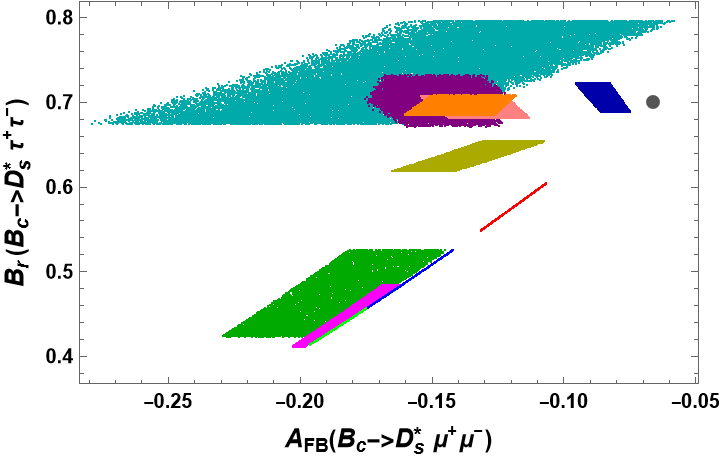}
        \caption{}
        \label{cor8_2}
    \end{subfigure}
        \begin{subfigure}[b]{0.32\textwidth}
        \includegraphics[width=\textwidth]{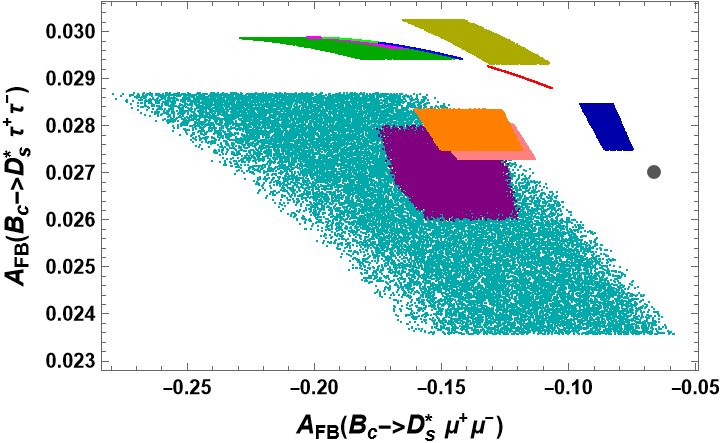}
        \caption{}
        \label{cor8_5}
    \end{subfigure}
     \begin{subfigure}[b]{0.32\textwidth}
        \includegraphics[width=\textwidth]{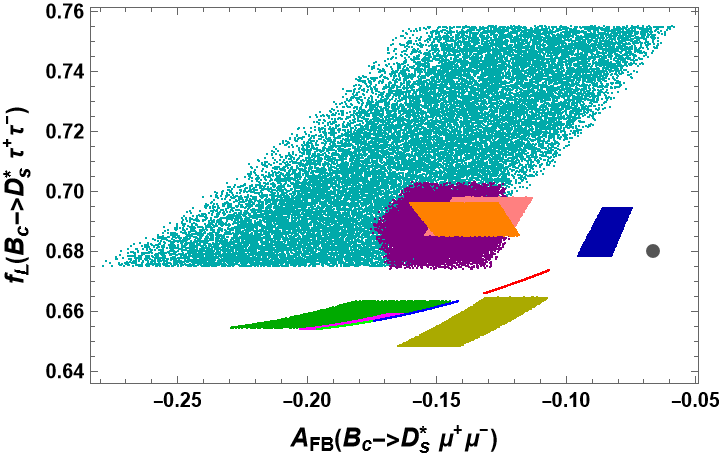}
        \caption{}
        \label{cor8_6}
    \end{subfigure}
    \begin{subfigure}[b]{0.32\textwidth}
        \includegraphics[width=\textwidth]{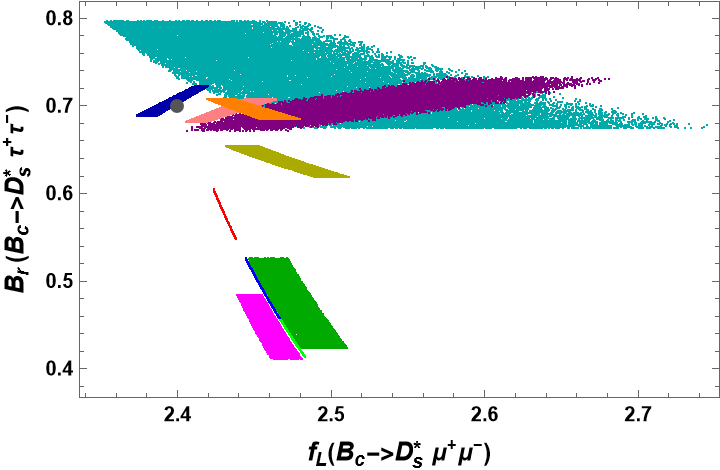}
        \caption{}
        \label{cor9_2}
    \end{subfigure}
     \begin{subfigure}[b]{0.32\textwidth}
        \includegraphics[width=\textwidth]{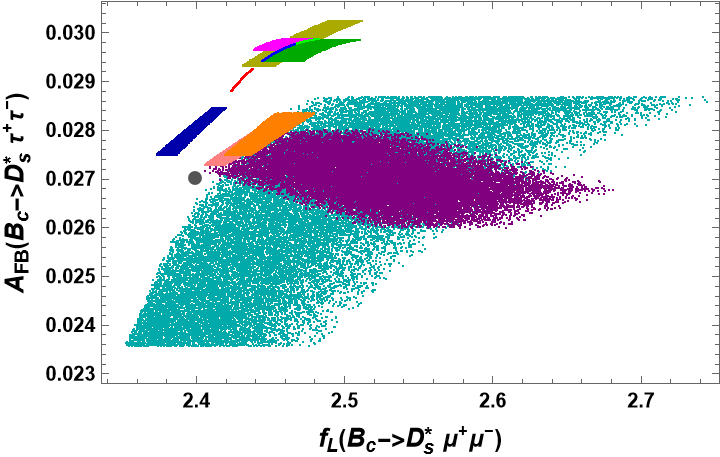}
        \caption{}
        \label{cor9_5}
    \end{subfigure}
        \begin{subfigure}[b]{0.32\textwidth}
        \includegraphics[width=\textwidth]{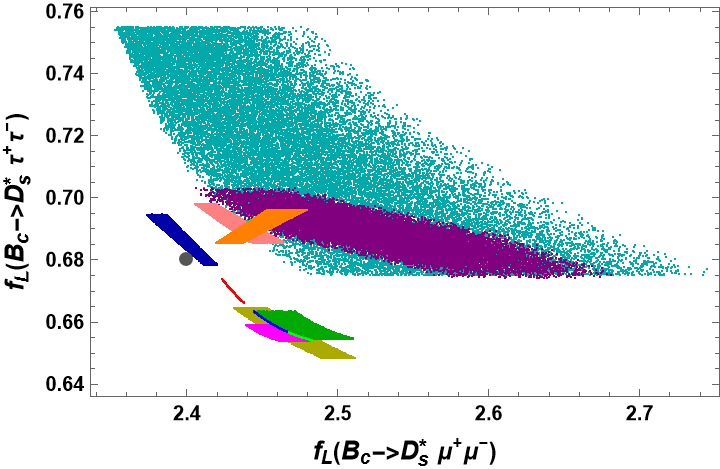}
        \caption{}
        \label{cor9_6}
    \end{subfigure}
\caption{Correlation among the different observables of $B\to D^*_s\mu^+\mu^-$ in the low bin $q^2$ and $B\to D^*_s\tau^+\tau^-$.}
    \label{corr}
\end{figure}

Finally, we have computed the correlations between the different observables of the $B \to D^*_s \mu^+ \mu^-$ decay channel in the low $q^2$ bin and the corresponding observables of the $B \to D^*_s \tau^+ \tau^-$ decay channel, as shown in Fig. (\ref{corr}). These plots allow us to establish explicit relationships between the measurements in the two decay channels. In particular, while measuring the observables in the $B \to D^*_s \mu^+ \mu^-$ channel within the low $q^2$ region, we can predict the significances of the related observables in the $B \to D^*_s \tau^+ \tau^-$ channel. 

Such correlations are valuable for understanding the phenomenology of these decays and provide a powerful tool for testing NP models. By comparing the measured observables indicated from various NP scenarios, these correlations can assist in evaluating the validity of different theoretical frameworks. Furthermore, they can help distinguish between different NP models, as deviations in the correlations could indicate NP effects that differ across the two decay modes. Thus, this technique is important for both the exploration of NP and the potential identification of its nature.

In the interest of completeness, we show the numerical values of the observables for $B_c\to D_s^{*}\mu^+(\tau^+)\mu^-(\tau^-)$, in the SM and in the presence of 1D and D>1 NP scenarios by using their allowed $1\sigma$ parametric space, duly integrated over the low and the high $q^2$ bins. These values for the case of muons are tabulated in Tables~\ref{table Dsstar muon 
low} (low $q^2$) and \ref{table Ds muon high} (high $q^2$), and for the case of tauons in Table~\ref{table bin Dstar tuaon}.

\begin{table}[H]
\centering
\setlength{\tabcolsep}{3pt} 
\renewcommand{\arraystretch}{1.5}
\resizebox{1\textwidth}{!}{
\begin{tabular}{|>{\bfseries}c|c|c|c||>{\bfseries}c|c|c|c|}
\hline
 & $10^7 \times \mathcal{B}_r$ & $A_{FB}$& $f_L$ && $10^7 \times  \mathcal{B}_r$& $A_{FB}$&$f_L$\\
\hline
\hline
SM&$0.097^{+0.01}_{-0.01}$ &$\text{-}0.066^{+0.002}_{-0.002}$ &$2.402^{+0.001}_{-0.001}$&S7 &$(0.064-0.080)$ &$(\text{-}0.16-\text{-}0.31)$&$(2.46-2.62)$ \\
S1 &$(0.070-0.077)$ &$(\text{-}0.18-\text{-}0.25)$ &$(2.48-2.54)$ &S8 &$(0.065-0.076)$ & $(\text{-}0.19-\text{-}0.27)$&$(2.48-2.54)$ \\
 S2&$(0.070-0.079)$ &$(\text{-}0.10-\text{-}0.13)$ &$(2.36-2.38)$&S9 &$(0.082-0.096)$&$(\text{-}0.07-\text{-}0.09)$&$(2.38-2.41)$ \\
 S3&$(0.057-0.068)$ &$(\text{-}0.19-\text{-}0.28)$ &$(2.65-2.91)$&S10 &$(0.075-0.087) $ &$(\text{-}0.12-\text{-}0.17)$&$(2.41-2.48)$\\
S5&$(0.067-0.101)$ &$(\text{-}0.06-\text{-}0.31)$ &$(2.36-2.78)$&S11 &$(0.076-0.087)$ &$(\text{-}0.12-\text{-}0.18)$&$(2.38-2.50)$  \\
 S6&$(0.082-0.097)$ &$(\text{-}0.10-\text{-}0.18)$ &$(2.44-2.51)$&S13 &$(0.062-0.085) $&$(\text{-}0.13-\text{-}0.23)$&$(2.43-2.82)$ \\
\hline\hline
\end{tabular}}
\caption{Observable/$q^2$ bin~(in ${\rm GeV^2}$) values [$s_{min}$,6] of different observables of $B_c \to D_s^*\mu^+\mu^-$ 
decays.}
\label{table Dsstar muon low}
\end{table}

\begin{table}[H]
\centering
\setlength{\tabcolsep}{3pt}
\renewcommand{\arraystretch}{1.5} 
\resizebox{1\textwidth}{!}{
\begin{tabular}{|>{\bfseries}c|c|c|c||>{\bfseries}c|c|c|c|}
\hline
 & $10^7 \times \mathcal B_r$ & $A_{FB}$& $f_L$ && $10^7 \times \mathcal B_r$& $A_{FB}$&$f_L$\\
\hline
\hline
SM&$2.86^{+0.34}_{-0.32}$ &$0.085^{+0.0004}_{-0.0004}$ &$0.504^{+0.0001}_{-0.0001}$&S7 &$(1.85-2.35)$ &$(0.074-0.082)$&$(0.51-0.55)$ \\
S1 &$(2.04-2.26)$ &$( 0.078-0.081)$ &$(0.52-0.53)$ &S8 &$(1.87-2.22)$ & $(0.077-0.081)$&$(0.52-0.53)$ \\
 S2&$(2.04-2.34)$ &$(0.084-0.085)$ &$(0.50-0.51)$&S9 &$(2.40-2.83)$&$(0.084-0.085)$&$(0.50-0.51)$ \\
 S3&$(1.60-1.95)$ &$(0.100-0.108)$ &$(0.53-0.58)$&S10 &$(2.20-2.57) $ &$(0.081-0.084)$&$(0.51-0.52)$\\
S5&$(1.95-2.96)$ &$(0.069-0.085)$ &$(0.50-0.57)$&S11 &$(2.20-2.56)$ &$(0.079-0.086)$&$(0.51-0.53)$  \\
 S6&$(2.40-2.84)$ &$(0.080-0.083)$ &$(0.51-0.52)$&S13 &$(1.78-2.51) $&$(0.077-0.097)$&$(0.51-0.55)$ \\
\hline\hline
\end{tabular}}
\caption{Observable/$q^2$ bin~(in ${\rm GeV^2}$) values [14,$s_{max}$] of different observables of $B_c \to D_s^*\mu^+\mu^-$ 
decays.}
\label{table Ds muon high}
\end{table}

\begin{table}[H]
\centering
\setlength{\tabcolsep}{3pt} 
\renewcommand{\arraystretch}{1.5} 
\resizebox{1\textwidth}{!}{
\begin{tabular}{|>{\bfseries}c|c|c|c|c|c|c|}
\hline
 & $10^7 \times \mathcal B_r$ & $R_{D_s^*}$ &$A_{FB}$& $f_L$& $R_{A_{FB}}^{\tau\mu}$& $R_{f_{L}}^{\tau\mu}$\\
\hline
\hline
SM&$0.699^{+0.08}_{-0.07}$ &$0.243^{+0.01}_{-0.01}$ &$0.027^{+0.00}_{-0.00}$ &$0.689^{+0.00}_{-0.00}$ &$0.326^{+0.01}_{-0.01} $ &$1.366^{+0.01}_{-0.09}$ \\
S1 &$(0.41-0.49)$ &$(0.20-0.22)$ &$(0.030-0.031)$ &$(0.62-0.64) $ & $(0.35-0.40)$&$(1.23-1.17)$ \\
 S2&$(0.49-0.57)$ &$(0.24-0.25)$ &$(0.027-0.028)$ &$(0.68-0.69)$&$(0.32-0.33)$ &$(1.36-1.37)$ \\
 S3&$(0.29-0.39)$ &$(0.18-0.20)$ &$(0.039-0.048)$ &$(0.54-0.58) $ &$(0.39-0.44)$ & $(0.93-1.08) $\\
S5&$(0.68-0.79)$ &$(0.24-0.28)$ &$(0.024-0.028)$ &$(0.67-0.74)$ &$(0.30-0.38)$ &$(1.21-1.41)$ \\
 S6&$(0.63-0.67)$ &$(0.22-0.23)$ &$(0.029-0.030)$ &$(0.64-0.66) $&$(0.35-0.37)$ &  $(1.23-1.29)$\\
 S7&$(0.45-0.56)$ &$(0.21-0.23)$ &$(0.029-0.031)$ &$(0.63-0.65)$ & $(0.36-0.41)$&$(1.15-1.26)$\\
 S8&$(0.44-0.51)$ &$(0.21-0.22)$ &$(0.030-0.031)$ &$(0.62-0.64)$ &$(0.37-0.40)$ &$(1.16-1.23)$ \\
S9&$(0.67-0.71)$ &$(0.24-0.25)$ &$(0.027-0.028)$ &$(0.67-0.69)$  &$(0.32-0.33)$ &$(1.34-1.37)$\\
 S10&$(0.67-0.69)$ &$(0.25-0.26)$ &$(0.027-0.028)$ &$(0.68-0.69)$ &$(0.32-0.34)$ &$(1.32-1.37)$ \\
 S11&$(0.67-0.70)$ &$(0.24-0.25)$ &$(0.027-0.28)$ &$(0.68-0.69)$&$(0.33-0.34)$ & $(1.31-1.36)$\\
 S13&$(0.69-0.70)$ &$(0.25-0.26)$ &$(0.025-0.028)$ &$(0.68-0.68)$&$(0.28-0.34)$ & $(1.24-1.37)$ \\
\hline\hline
\end{tabular}}
\caption{Observable/$q^2$ bin~(in ${\rm GeV^2}$) values [14,$s_{max}$] of different observables of $B_c \to D^{\ast}_s\tau^+\tau^-$ 
decays.}
\label{table bin Dstar tuaon}
\end{table}

\subsubsection{$B_c\to D_s\,\ell^+\ell^-$ in the presence of 1D and D $>1$ NP scenarios \label{BctoDs}}

\begin{figure}[H]
    \centering
    \begin{subfigure}[b]{0.32\textwidth}
        \includegraphics[width=\textwidth]{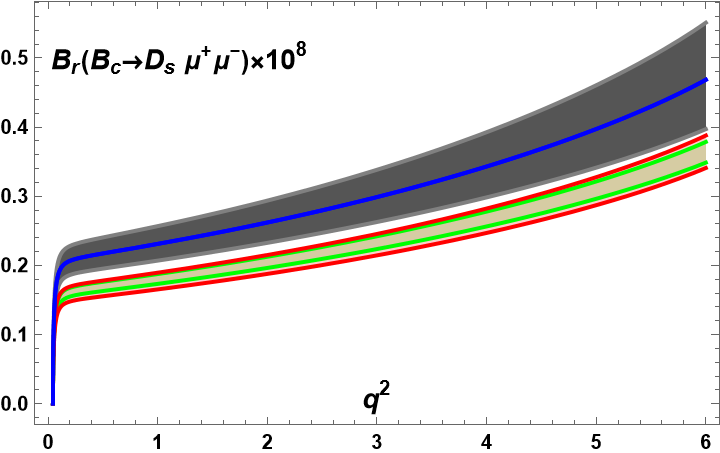}
        \caption{}
        \label{Br 1Ds lowq}
    \end{subfigure}
     \begin{subfigure}[b]{0.32\textwidth}
        \includegraphics[width=\textwidth]{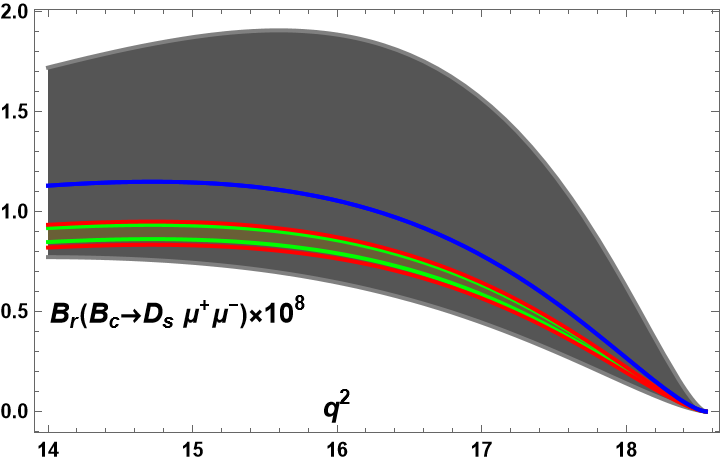}
        \caption{}
        \label{Br 1Ds highq}
    \end{subfigure}
     \begin{subfigure}[b]{0.32\textwidth}
        \includegraphics[width=\textwidth]{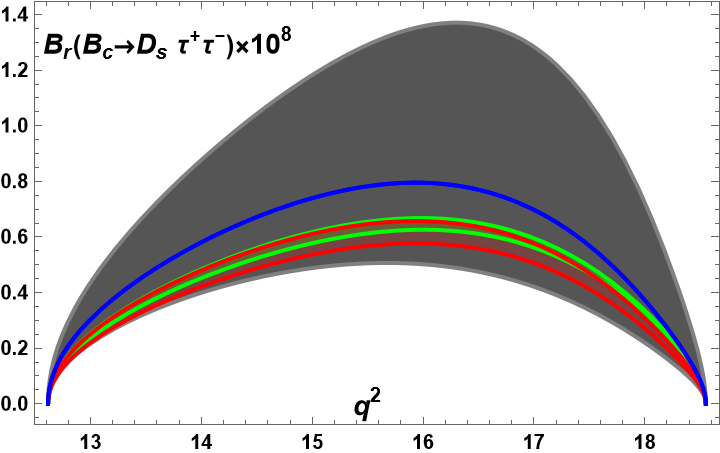}
        \caption{}
        \label{Br RDs 1Dtau}
    \end{subfigure}
    \begin{subfigure}[b]{0.32\textwidth}
        \includegraphics[width=\textwidth]{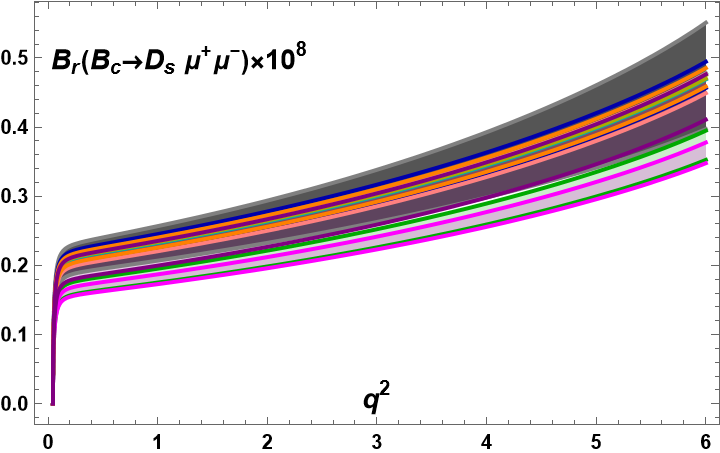}
        \caption{}
        \label{Br 2Ds lowq}
    \end{subfigure}
    \begin{subfigure}[b]{0.32\textwidth}
        \includegraphics[width=\textwidth]{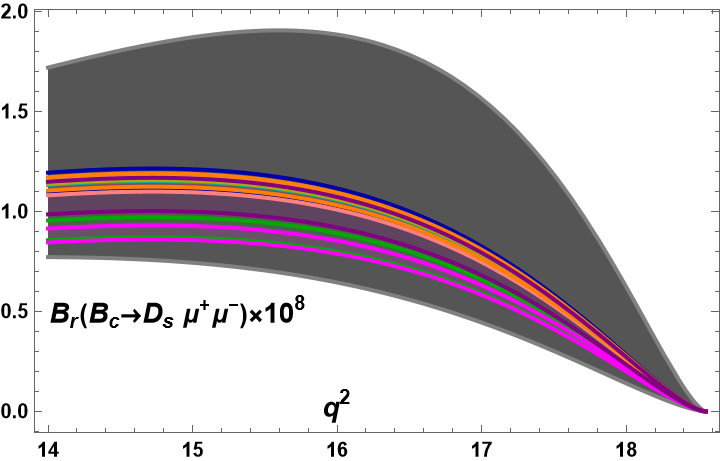}
        \caption{}
        \label{Br 2Ds highq}
    \end{subfigure}
    \begin{subfigure}[b]{0.32\textwidth}
        \includegraphics[width=\textwidth]{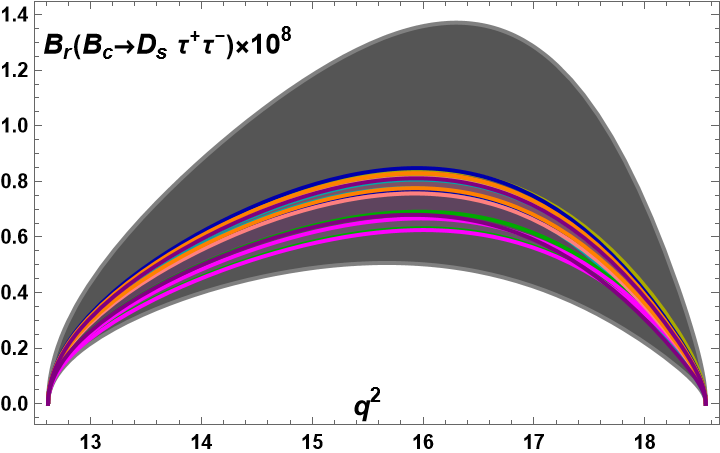}
        \caption{}
        \label{br RDs 2Dtau}
    \end{subfigure}
    \caption{The $\mathcal{B}_r$ as a function of $q^2$ where the plots (a) and (d) represent in D$=1$ and D$>1$ NP scenarios, respectively, for the case of muon in the low $q^2$ region while (b) and (e) represent in the high $q^2$ region. The plots (c) and (f) represent the case when tauons are the final state leptons.}
    \label{Br ratio Ds}
\end{figure}

We observe that the branching ratio of the $B_c\to D_s\,\ell^+\ell^-$ decays are insensitive to most NP scenarios for both the muon and tauon cases as depicted in Fig.~(\ref{Br ratio Ds}), where the color coding is the same as used in Figs.~\ref{1D scenarios BR},~\ref{BR MI}, and~\ref{zprime}. The only scenarios that show some difference from the SM result in terms of the branching ratio are S1, S2, S7, and S8 as can be seen in the low $q^2$ region for the muon case in Figs.~(\ref{Br 1Ds lowq}) and (\ref{Br 2Ds lowq}). 

\begin{figure}[H]
    \centering
    \begin{subfigure}[b]{0.45\textwidth}
        \includegraphics[width=\textwidth]{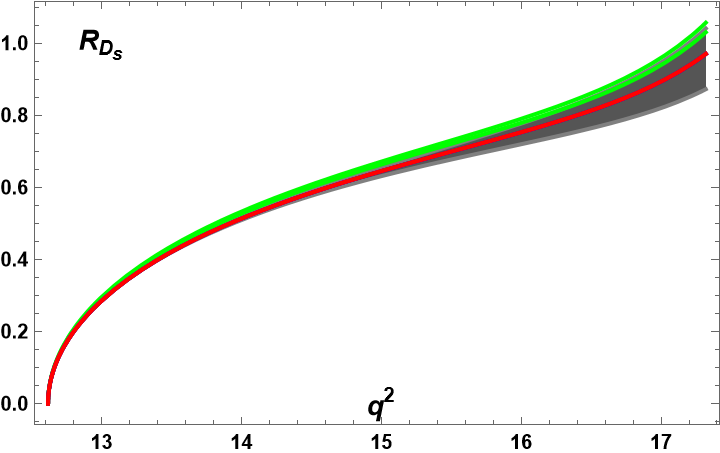}
        \caption{}
        \label{RDs 1D}
    \end{subfigure}
    \begin{subfigure}[b]{0.45\textwidth}
        \includegraphics[width=\textwidth]{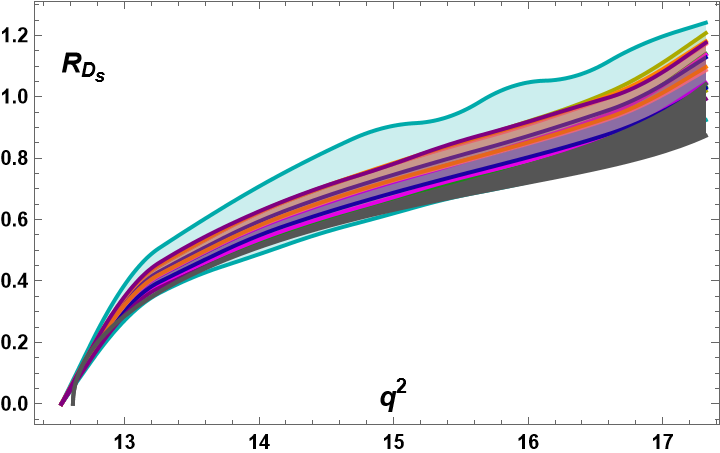}
        \caption{}
        \label{RDs 2D}
    \end{subfigure}
    \begin{subfigure}[b]{0.45\textwidth}
        \includegraphics[width=\textwidth]{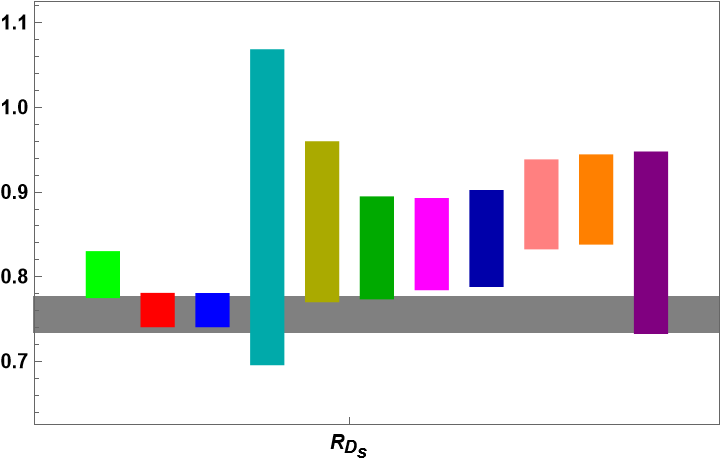}
        \caption{}
        \label{RDs bar}
    \end{subfigure}
     \begin{subfigure}[b]{0.45\textwidth}
        \includegraphics[width=\textwidth]{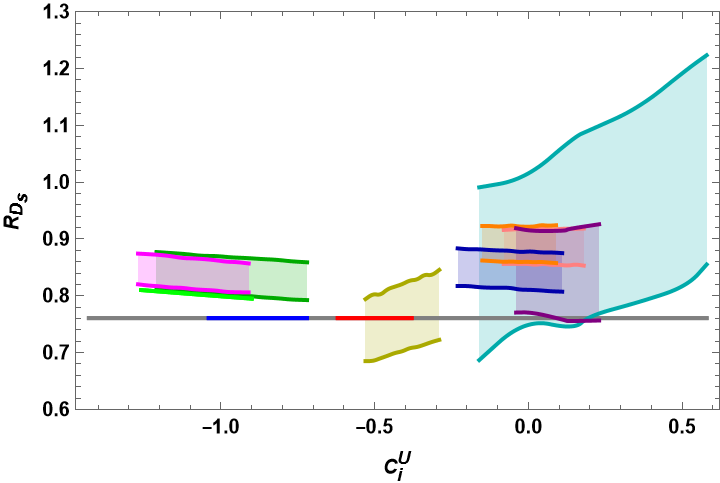}
        \caption{}
        \label{WC RDs}
    \end{subfigure}
    \caption{(a) and (b) depict $R_{D_s}$ as a function of $q^2$ in the presence of D$=1$ and D$>1$ NP scenarios, respectively, while (c) represent the variation in the magnitude of $R_{D_s}$ and (d) correspond to its variation on the allowed $1\sigma$ parametric space of NP WCs.}
    \label{RDs}
\end{figure}

On the other hand, the LFU ratio, $R_{D_s}$ is somewhat sensitive to the considered NP scenarios as can been seen in Fig.~(\ref{RDs}). As for the D=1 NP scenarios, the value of $R_{D_s}$ is influenced only by S1 (see Fig. (\ref{RDs 1D})) which can also be seen explicitly from the variation from the magnitude (integrated over $q^2$) of its SM value by bar plots that are depicted in Fig.~(\ref{RDs bar}). The influence of $D>1$ NP scenarios on $R_{D_s}$ is shown in Fig.~(\ref{RDs 2D}) and the variation in the magnitude of these observables integrated over $q^2$ is shown in Fig.~(\ref{RDs bar}). From these figures, one can notice that the effects of S5 are quite distinct from the other scenarios while the effects of other scenarios overlap with each other. In addition, we have also drawn the $R_{D_s}$ as a function of NP WCs in Fig. (\ref{WC RDs}) where one can explicitly see the effect of $1\sigma$ allowed parametric space of NP WCs on the observables. Finally, the numerical values of branching ratios and the $R_{D_s}$ are also in Table~\ref{table Ds}. While this decay channel may not be uniquely distinctive in separating the various NP scenarios considered, a precise measurement of $R_{D_s}$ can still serve as a complementary check in connection with results from other decays.

\begin{table}[H]
\centering
\setlength{\tabcolsep}{3pt} 
\renewcommand{\arraystretch}{1.5} 
\begin{tabular}{|>{\bfseries}c|c|c||>{\bfseries}c|c|c|}
\hline
 & $10^7 \times \mathcal B_r$ & $R_{D_s}$ && $10^7 \times \mathcal B_r$& $R_{D_s}$\\
\hline
\hline
SM&$1.85^{+0.25}_{-0.25}$ &$0.760^{+0.01}_{-0.02}$ &S7 &$(1.39-1.56)$ &$(0.79-0.87)$ \\
S1 &$(1.38-1.49)$ &$(0.79-0.81)$ &S8 &$(1.38-1.49) $ & $(0.80-0.87)$ \\
 S2&$(1.33-1.52)$ &$(0.75-0.77)$ &S9 &$(1.80-1.96)$&$(0.80-0.88)$ \\
 S3&$(1.85-1.86)$ &$(0.76-0.77)$ &S10 &$(1.80-0.29) $ &$(0.85-0.91)$\\
S5&$1.85-0.29)$ &$(0.71-1.04)$ &S11 &$(1.80-1.92)$ &$(0.85-0.92)$  \\
 S6&$(1.86-1.89)$ &$(0.79-0.93)$ &S13 &$(1.60-1.88) $&$(0.75-0.92)$ \\
\hline\hline
\end{tabular}
\caption{Observable/$q^2$ bin~(in ${\rm GeV^2}$) values [$s_{min}$,6] corresponds to branching ratio $B_c \to D_s\mu^+\mu^-$ decays and for $R_{D_s}$ bin~(in ${\rm GeV^2}$) values [14,$s_{max}$].}
\label{table Ds}
\end{table}

\section{Summary and Conclusion\label{conclusion}}
We investigate the transition $b\to s\ell^+\ell^-$ through $B_c\to D_s^{(\ast)} \,\ell^+\ell^-$ $(\ell=\mu,\tau)$ in different observables by considering some potential NP contributions to $b\to s\ell^+\ell^-$ consisting of universal and non-universal lepton couplings. In our study, we use the helicity formalism for this decay by employing the effective theory approach where the vector and axial vector NP operators are introduced. In this study, we compute the branching ratio $\mathcal{B}_r$, 
the $D^\ast$ helicity fraction $f_L$, the lepton forward-backward asymmetry $A_{FB}$, and the lepton 
flavor universality ratio (LFU) $R^{\tau\mu}_{D_s^*}$. In addition, for a complementary check on LFU, 
we also calculate the ratio of different observables $R_{i}^{\tau\mu}$ where $i=A_{FB}$, $f_L$. In this context, we allow for LFU violating couplings to be present, with the NP universal couplings present for all leptons, while the non-universal coupling available only the muons. Regarding these couplings, we employ the latest global fit to the $b\to s\ell^+\ell^-$ data. 

In order to check the sensitivity of the various NP couplings, we vary them within their 1$\sigma$ ranges as determined. We give predictions of the mentioned observables within the SM and the various NP scenarios. We find that the considered observables are not only sensitive to NP but are also helpful in distinguishing among the different NP scenarios. We have reported our results for 
the observables both as a function of the momentum transfer ($q^2$) and after integrating over the 
low and high $q^2$ bins as typically experiments are more sensitive to these integrated results. In addition, to see the explicit dependence on the couplings, we have calculated the analytical expressions of these observables in terms of NP WCs and plotted them against the NP couplings in their 1$\sigma$ range. These expressions can be very useful for determining the precise values of the universal and non-universal couplings when needed in the future. Finally, we present the numerical results of the observables integrated over various $q^2$ regions in several tables for quick referencing.  

These results can be tested at LHCb, HL-LHC, and FCC-ee, and therefore, the precise measurements of these observables not only deepens our understanding of the $b\to s\ell^+\ell^-$ process, but also provides a complementary check on the status of different NP scenarios.



\begin{thebibliography}{100}
	
	\bibitem{Alguero:2023jeh}
	M.~Alguer\'o, A.~Biswas, B.~Capdevila, S.~Descotes-Genon, J.~Matias and
	M.~Novoa-Brunet, \emph{{To (b)e or not to (b)e: no electrons at LHCb}},
	\href{https://doi.org/10.1140/epjc/s10052-023-11824-0}{\emph{Eur. Phys. J. C}
		{\bfseries 83} (2023) 648}
	[\href{https://arxiv.org/abs/2304.07330}{{\ttfamily arXiv:2304.07330}}].
	
	\bibitem{Albrecht:2021tul}
	J.~Albrecht, D.~van Dyk and C.~Langenbruch, \emph{{Flavour anomalies in heavy
			quark decays}}, \href{https://doi.org/10.1016/j.ppnp.2021.103885}{\emph{Prog.
			Part. Nucl. Phys.} {\bfseries 120} (2021) 103885}
	[\href{https://arxiv.org/abs/2107.04822}{{\ttfamily arXiv:2107.04822}}].
	
	\bibitem{London:2021lfn}
	D.~London and J.~Matias, \emph{{$B$ Flavour Anomalies: 2021 Theoretical Status
			Report}},
	\href{https://doi.org/10.1146/annurev-nucl-102020-090209}{\emph{Ann. Rev.
			Nucl. Part. Sci.} {\bfseries 72} (2022) 37}
	[\href{https://arxiv.org/abs/2110.13270}{{\ttfamily arXiv:2110.13270}}].
	
	\bibitem{CDF:1999uew}
	{\scshape CDF} collaboration, \emph{{Search for the flavor-changing neutral
			current decays $B^+ \to \mu^+ \mu^- K^+$ and $B^0 \to \mu^+ \mu^- K^{*0}$}},
	\href{https://doi.org/10.1103/PhysRevLett.83.3378}{\emph{Phys. Rev. Lett.}
		{\bfseries 83} (1999) 3378}
	[\href{https://arxiv.org/abs/hep-ex/9905004}{{\ttfamily hep-ex/9905004}}].
	
	\bibitem{BaBar:2000jlq}
	{\scshape BaBar} collaboration, \emph{{Search for $B^{+} \to K^{+} \ell^+
			\ell^-$ and $B^0 \to$ K*0 $\ell^+\ell^-$ lepton-}},  in \emph{{30th
			International Conference on High-Energy Physics}}, 7, 2000
	[\href{https://arxiv.org/abs/hep-ex/0008059}{{\ttfamily hep-ex/0008059}}].
	
	\bibitem{Belle:2001oey}
	{\scshape Belle} collaboration, \emph{{Observation of the decay $B \to K
			\ell^{+} \ell^{-}$}},
	\href{https://doi.org/10.1103/PhysRevLett.88.021801}{\emph{Phys. Rev. Lett.}
		{\bfseries 88} (2002) 021801}
	[\href{https://arxiv.org/abs/hep-ex/0109026}{{\ttfamily hep-ex/0109026}}].
	
	\bibitem{BaBar:2003szi}
	{\scshape BaBar} collaboration, \emph{{Evidence for the rare decay $B \to K^*
			\ell^+ \ell^-$ and measurement of the $B \to K \ell^+ \ell^-$ branching
			fraction}}, \href{https://doi.org/10.1103/PhysRevLett.91.221802}{\emph{Phys.
			Rev. Lett.} {\bfseries 91} (2003) 221802}
	[\href{https://arxiv.org/abs/hep-ex/0308042}{{\ttfamily hep-ex/0308042}}].
	
	\bibitem{BaBar:2008jdv}
	{\scshape BaBar} collaboration, \emph{{Direct CP, Lepton Flavor and Isospin
			Asymmetries in the Decays $B \to K^{(*)} \ell^{+} \ell^{-}$}},
	\href{https://doi.org/10.1103/PhysRevLett.102.091803}{\emph{Phys. Rev. Lett.}
		{\bfseries 102} (2009) 091803}
	[\href{https://arxiv.org/abs/0807.4119}{{\ttfamily arXiv:0807.4119}}].
	
	\bibitem{Belle:2003ivt}
	{\scshape Belle} collaboration, \emph{{Observation of B ---\ensuremath{>} K* l+
			l-}}, \href{https://doi.org/10.1103/PhysRevLett.91.261601}{\emph{Phys. Rev.
			Lett.} {\bfseries 91} (2003) 261601}
	[\href{https://arxiv.org/abs/hep-ex/0308044}{{\ttfamily hep-ex/0308044}}].
	
	\bibitem{Belle:2016fev}
	{\scshape Belle} collaboration, \emph{{Lepton-Flavor-Dependent Angular Analysis
			of $B\to K^\ast \ell^+\ell^-$}},
	\href{https://doi.org/10.1103/PhysRevLett.118.111801}{\emph{Phys. Rev. Lett.}
		{\bfseries 118} (2017) 111801}
	[\href{https://arxiv.org/abs/1612.05014}{{\ttfamily arXiv:1612.05014}}].
	
	\bibitem{BELLE:2019xld}
	{\scshape BELLE} collaboration, \emph{{Test of lepton flavor universality and
			search for lepton flavor violation in $B \rightarrow K\ell \ell$ decays}},
	\href{https://doi.org/10.1007/JHEP03(2021)105}{\emph{JHEP} {\bfseries 03}
		(2021) 105} [\href{https://arxiv.org/abs/1908.01848}{{\ttfamily
			arXiv:1908.01848}}].
	
	\bibitem{Belle:2021ecr}
	{\scshape Belle} collaboration, \emph{{Search for the decay
			$B^0$\textrightarrow{}$K^{*0}$\ensuremath{\tau}$^+$\ensuremath{\tau}$^-$ at the Belle
			experiment}}, \href{https://doi.org/10.1103/PhysRevD.108.L011102}{\emph{Phys.
			Rev. D} {\bfseries 108} (2023) L011102}
	[\href{https://arxiv.org/abs/2110.03871}{{\ttfamily arXiv:2110.03871}}].
	
	\bibitem{Belle:2009zue}
	{\scshape Belle} collaboration, \emph{{Measurement of the Differential
			Branching Fraction and Forward-Backward Asymmetry for $B \to
			K^{(*)}\ell^+\ell^-$}},
	\href{https://doi.org/10.1103/PhysRevLett.103.171801}{\emph{Phys. Rev. Lett.}
		{\bfseries 103} (2009) 171801}
	[\href{https://arxiv.org/abs/0904.0770}{{\ttfamily arXiv:0904.0770}}].
	
	\bibitem{Belle:2019oag}
	{\scshape Belle} collaboration, \emph{{Test of Lepton-Flavor Universality in
			${B\to K^\ast\ell^+\ell^-}$ Decays at Belle}},
	\href{https://doi.org/10.1103/PhysRevLett.126.161801}{\emph{Phys. Rev. Lett.}
		{\bfseries 126} (2021) 161801}
	[\href{https://arxiv.org/abs/1904.02440}{{\ttfamily arXiv:1904.02440}}].
	
	\bibitem{BaBar:2012mrf}
	{\scshape BaBar} collaboration, \emph{{Measurement of Branching Fractions and
			Rate Asymmetries in the Rare Decays $B \to K^{(*)} l^+ l^-$}},
	\href{https://doi.org/10.1103/PhysRevD.86.032012}{\emph{Phys. Rev. D}
		{\bfseries 86} (2012) 032012}
	[\href{https://arxiv.org/abs/1204.3933}{{\ttfamily arXiv:1204.3933}}].
	
	\bibitem{CDF:2011buy}
	{\scshape CDF} collaboration, \emph{{Observation of the Baryonic
			Flavor-Changing Neutral Current Decay $\Lambda_{b} \to \Lambda \mu^{+}
			\mu^{-}$}}, \href{https://doi.org/10.1103/PhysRevLett.107.201802}{\emph{Phys.
			Rev. Lett.} {\bfseries 107} (2011) 201802}
	[\href{https://arxiv.org/abs/1107.3753}{{\ttfamily arXiv:1107.3753}}].
	
	\bibitem{CMS:2015bcy}
	{\scshape CMS} collaboration, \emph{{Angular analysis of the decay $B^0 \to
			K^{*0} \mu^+ \mu^-$ from pp collisions at $\sqrt s = 8$ TeV}},
	\href{https://doi.org/10.1016/j.physletb.2015.12.020}{\emph{Phys. Lett. B}
		{\bfseries 753} (2016) 424}
	[\href{https://arxiv.org/abs/1507.08126}{{\ttfamily arXiv:1507.08126}}].
	
	\bibitem{LHCb:2014vgu}
	{\scshape LHCb} collaboration, \emph{{Test of lepton universality using
			$B^{+}\rightarrow K^{+}\ell^{+}\ell^{-}$ decays}},
	\href{https://doi.org/10.1103/PhysRevLett.113.151601}{\emph{Phys. Rev. Lett.}
		{\bfseries 113} (2014) 151601}
	[\href{https://arxiv.org/abs/1406.6482}{{\ttfamily arXiv:1406.6482}}].
	
	\bibitem{LHCb:2016ykl}
	{\scshape LHCb} collaboration, \emph{{Measurements of the S-wave fraction in
			$B^{0}\rightarrow K^{+}\pi^{-}\mu^{+}\mu^{-}$ decays and the
			$B^{0}\rightarrow K^{\ast}(892)^{0}\mu^{+}\mu^{-}$ differential branching
			fraction}}, \href{https://doi.org/10.1007/JHEP11(2016)047}{\emph{JHEP}
		{\bfseries 11} (2016) 047}
	[\href{https://arxiv.org/abs/1606.04731}{{\ttfamily arXiv:1606.04731}}].
	
	\bibitem{LHCb:2017avl}
	{\scshape LHCb} collaboration, \emph{{Test of lepton universality with $B^{0}
			\rightarrow K^{*0}\ell^{+}\ell^{-}$ decays}},
	\href{https://doi.org/10.1007/JHEP08(2017)055}{\emph{JHEP} {\bfseries 08}
		(2017) 055} [\href{https://arxiv.org/abs/1705.05802}{{\ttfamily
			arXiv:1705.05802}}].
	
	\bibitem{LHCb:2013ghj}
	{\scshape LHCb} collaboration, \emph{{Measurement of Form-Factor-Independent
			Observables in the Decay $B^{0} \to K^{*0} \mu^+ \mu^-$}},
	\href{https://doi.org/10.1103/PhysRevLett.111.191801}{\emph{Phys. Rev. Lett.}
		{\bfseries 111} (2013) 191801}
	[\href{https://arxiv.org/abs/1308.1707}{{\ttfamily arXiv:1308.1707}}].
	
	\bibitem{Belle:2024cis}
	{\scshape Belle, Belle-II} collaboration, \emph{{Search for Rare
			b\textrightarrow{}d\ensuremath{\ell}$^+$\ensuremath{\ell}$^-$ Transitions at
			Belle}}, \href{https://doi.org/10.1103/PhysRevLett.133.101804}{\emph{Phys.
			Rev. Lett.} {\bfseries 133} (2024) 101804}
	[\href{https://arxiv.org/abs/2404.08133}{{\ttfamily arXiv:2404.08133}}].
	
	\bibitem{LHCb:2017lpt}
	{\scshape LHCb} collaboration, \emph{{Observation of the suppressed decay
			$\Lambda^{0}_{b}\rightarrow p\pi^{-}\mu^{+}\mu^{-}$}},
	\href{https://doi.org/10.1007/JHEP04(2017)029}{\emph{JHEP} {\bfseries 04}
		(2017) 029} [\href{https://arxiv.org/abs/1701.08705}{{\ttfamily
			arXiv:1701.08705}}].
	
	\bibitem{LHCb:2015hsa}
	{\scshape LHCb} collaboration, \emph{{First measurement of the differential
			branching fraction and $C\!P$ asymmetry of the $B^\pm\to\pi^\pm\mu^+\mu^-$
			decay}}, \href{https://doi.org/10.1007/JHEP10(2015)034}{\emph{JHEP}
		{\bfseries 10} (2015) 034}
	[\href{https://arxiv.org/abs/1509.00414}{{\ttfamily arXiv:1509.00414}}].
	
	\bibitem{LHCb:2012de}
	{\scshape LHCb} collaboration, \emph{{First observation of the decay $B^{+}
			\rightarrow \pi^{+} \mu^{+} \mu^{-}$}},
	\href{https://doi.org/10.1007/JHEP12(2012)125}{\emph{JHEP} {\bfseries 12}
		(2012) 125} [\href{https://arxiv.org/abs/1210.2645}{{\ttfamily
			arXiv:1210.2645}}].
	
	\bibitem{LHCb:2018rym}
	{\scshape LHCb} collaboration, \emph{{Evidence for the decay $ {B}_S^0\to
			{\overline{K}}^{\ast 0}{\mu}^{+}{\mu}^{-} $}},
	\href{https://doi.org/10.1007/JHEP07(2018)020}{\emph{JHEP} {\bfseries 07}
		(2018) 020} [\href{https://arxiv.org/abs/1804.07167}{{\ttfamily
			arXiv:1804.07167}}].
	
	\bibitem{LHCb:2022vje}
	{\scshape LHCb} collaboration, \emph{{Measurement of lepton universality
			parameters in $B^+\to K^+\ell^+\ell^-$ and $B^0\to K^{*0}\ell^+\ell^-$
			decays}}, \href{https://doi.org/10.1103/PhysRevD.108.032002}{\emph{Phys. Rev.
			D} {\bfseries 108} (2023) 032002}
	[\href{https://arxiv.org/abs/2212.09153}{{\ttfamily arXiv:2212.09153}}].
	
	\bibitem{Isidori:2022bzw}
	G.~Isidori, D.~Lancierini, S.~Nabeebaccus and R.~Zwicky, \emph{{QED in $
			\overline{B} $\textrightarrow{}$ \overline{K}
			$\ensuremath{\ell}$^{+}$\ensuremath{\ell}$^{-}$ LFU ratios: theory versus
			experiment, a Monte Carlo study}},
	\href{https://doi.org/10.1007/JHEP10(2022)146}{\emph{JHEP} {\bfseries 10}
		(2022) 146} [\href{https://arxiv.org/abs/2205.08635}{{\ttfamily
			arXiv:2205.08635}}].
	
	\bibitem{LHCb:2022qnv}
	{\scshape LHCb} collaboration, \emph{{Test of lepton universality in $b
			\rightarrow s \ell^+ \ell^-$ decays}},
	\href{https://doi.org/10.1103/PhysRevLett.131.051803}{\emph{Phys. Rev. Lett.}
		{\bfseries 131} (2023) 051803}
	[\href{https://arxiv.org/abs/2212.09152}{{\ttfamily arXiv:2212.09152}}].
	
	\bibitem{LHCb:2014cxe}
	{\scshape LHCb} collaboration, \emph{{Differential branching fractions and
			isospin asymmetries of $B \to K^{(*)} \mu^+ \mu^-$ decays}},
	\href{https://doi.org/10.1007/JHEP06(2014)133}{\emph{JHEP} {\bfseries 06}
		(2014) 133} [\href{https://arxiv.org/abs/1403.8044}{{\ttfamily
			arXiv:1403.8044}}].
	
	\bibitem{LHCb:2021zwz}
	{\scshape LHCb} collaboration, \emph{{Branching Fraction Measurements of the
			Rare $B^0_s\rightarrow\phi\mu^+\mu^-$ and $B^0_s\rightarrow
			f_2^\prime(1525)\mu^+\mu^-$- Decays}},
	\href{https://doi.org/10.1103/PhysRevLett.127.151801}{\emph{Phys. Rev. Lett.}
		{\bfseries 127} (2021) 151801}
	[\href{https://arxiv.org/abs/2105.14007}{{\ttfamily arXiv:2105.14007}}].
	
	\bibitem{LHCb:2020lmf}
	{\scshape LHCb} collaboration, \emph{{Measurement of $CP$-Averaged Observables
			in the $B^{0}\rightarrow K^{*0}\mu^{+}\mu^{-}$ Decay}},
	\href{https://doi.org/10.1103/PhysRevLett.125.011802}{\emph{Phys. Rev. Lett.}
		{\bfseries 125} (2020) 011802}
	[\href{https://arxiv.org/abs/2003.04831}{{\ttfamily arXiv:2003.04831}}].
	
	\bibitem{Descotes-Genon:2012isb}
	S.~Descotes-Genon, J.~Matias, M.~Ramon and J.~Virto, \emph{{Implications from
			clean observables for the binned analysis of $B \to K^*\mu^+\mu^-$ at large
			recoil}}, \href{https://doi.org/10.1007/JHEP01(2013)048}{\emph{JHEP}
		{\bfseries 01} (2013) 048} [\href{https://arxiv.org/abs/1207.2753}{{\ttfamily
			arXiv:1207.2753}}].
	
	\bibitem{Artuso:2022ijh}
	M.~Artuso, G.~Isidori and S.~Stone, \emph{{New Physics in b Decays}}, World
	Scientific (5, 2022), \href{https://doi.org/10.1142/12696}{10.1142/12696}.
	
	\bibitem{LHCb:2021lvy}
	{\scshape LHCb} collaboration, \emph{{Tests of lepton universality using
			$B^0\to K^0_S \ell^+ \ell^-$ and $B^+\to K^{*+} \ell^+ \ell^-$ decays}},
	\href{https://doi.org/10.1103/PhysRevLett.128.191802}{\emph{Phys. Rev. Lett.}
		{\bfseries 128} (2022) 191802}
	[\href{https://arxiv.org/abs/2110.09501}{{\ttfamily arXiv:2110.09501}}].
	
	\bibitem{CMS:2024syx}
	{\scshape CMS} collaboration, \emph{{Test of lepton flavor universality in
			B$^{\pm}$$\to$ K$^{\pm}\mu^+\mu^-$ and B$^{\pm}$$\to$ K$^{\pm}$e$^+$e$^-$
			decays in proton-proton collisions at $\sqrt{s}$ = 13 TeV}},
	\href{https://doi.org/10.1088/1361-6633/ad4e65}{\emph{Rept. Prog. Phys.}
		{\bfseries 87} (2024) 077802}
	[\href{https://arxiv.org/abs/2401.07090}{{\ttfamily arXiv:2401.07090}}].
	
	\bibitem{CDF:2001yrm}
	{\scshape CDF} collaboration, \emph{{Search for the Decay $B_s \to \mu^+ \mu^-
			\phi$ in $p\bar{p}$ Collisions at $\sqrt{s}$ = 1.8-TeV}},
	\href{https://doi.org/10.1103/PhysRevD.65.111101}{\emph{Phys. Rev. D}
		{\bfseries 65} (2002) 111101}.
	
	\bibitem{CDF:2008zhr}
	{\scshape CDF} collaboration, \emph{{Search for the Rare Decays $B^+ \to \mu^+
			\mu^- K^+$, $B^0 \to \mu^+ \mu^- K^{*0}(892)$, and $B^0_s \to \mu^+ \mu^-
			\phi$ at CDF}}, \href{https://doi.org/10.1103/PhysRevD.79.011104}{\emph{Phys.
			Rev. D} {\bfseries 79} (2009) 011104}
	[\href{https://arxiv.org/abs/0804.3908}{{\ttfamily arXiv:0804.3908}}].
	
	\bibitem{D0:2006pmq}
	{\scshape D0} collaboration, \emph{{Search for the rare decay $B^0_{s} \to \phi
			\mu^{+} \mu^{-}$ with the D$\varnothing$ detector}},
	\href{https://doi.org/10.1103/PhysRevD.74.031107}{\emph{Phys. Rev. D}
		{\bfseries 74} (2006) 031107}
	[\href{https://arxiv.org/abs/hep-ex/0604015}{{\ttfamily hep-ex/0604015}}].
	
	\bibitem{CDF:2011grz}
	{\scshape CDF} collaboration, \emph{{Measurement of the Forward-Backward
			Asymmetry in the $B \to K^{(*)} \mu^+ \mu^-$ Decay and First Observation of
			the $B^0_s \to \phi \mu^+ \mu^-$ Decay}},
	\href{https://doi.org/10.1103/PhysRevLett.106.161801}{\emph{Phys. Rev. Lett.}
		{\bfseries 106} (2011) 161801}
	[\href{https://arxiv.org/abs/1101.1028}{{\ttfamily arXiv:1101.1028}}].
	
	\bibitem{LHCb:2013tgx}
	{\scshape LHCb} collaboration, \emph{{Differential branching fraction and
			angular analysis of the decay $B_s^0\to\phi\mu^{+}\mu^{-}$}},
	\href{https://doi.org/10.1007/JHEP07(2013)084}{\emph{JHEP} {\bfseries 07}
		(2013) 084} [\href{https://arxiv.org/abs/1305.2168}{{\ttfamily
			arXiv:1305.2168}}].
	
	\bibitem{LHCb:2015wdu}
	{\scshape LHCb} collaboration, \emph{{Angular analysis and differential
			branching fraction of the decay $B^0_s\to\phi\mu^+\mu^-$}},
	\href{https://doi.org/10.1007/JHEP09(2015)179}{\emph{JHEP} {\bfseries 09}
		(2015) 179} [\href{https://arxiv.org/abs/1506.08777}{{\ttfamily
			arXiv:1506.08777}}].
	
	\bibitem{Horgan:2015vla}
	R.R.~Horgan, Z.~Liu, S.~Meinel and M.~Wingate, \emph{{Rare $B$ decays using
			lattice QCD form factors}},
	\href{https://doi.org/10.22323/1.214.0372}{\emph{PoS} {\bfseries LATTICE2014}
		(2015) 372} [\href{https://arxiv.org/abs/1501.00367}{{\ttfamily
			arXiv:1501.00367}}].
	
	\bibitem{Bobeth:2008ij}
	C.~Bobeth, G.~Hiller and G.~Piranishvili, \emph{{CP Asymmetries in bar $B \to
			\bar{K}^* (\to \bar{K} \pi) \bar{\ell} \ell$ and Untagged $\bar{B}_s$, $B_s
			\to \phi (\to K^{+} K^-) \bar{\ell} \ell$ Decays at NLO}},
	\href{https://doi.org/10.1088/1126-6708/2008/07/106}{\emph{JHEP} {\bfseries
			07} (2008) 106} [\href{https://arxiv.org/abs/0805.2525}{{\ttfamily
			arXiv:0805.2525}}].
	
	\bibitem{Bharucha:2015bzk}
	A.~Bharucha, D.M.~Straub and R.~Zwicky, \emph{{$B\to V\ell^+\ell^-$ in the
			Standard Model from light-cone sum rules}},
	\href{https://doi.org/10.1007/JHEP08(2016)098}{\emph{JHEP} {\bfseries 08}
		(2016) 098} [\href{https://arxiv.org/abs/1503.05534}{{\ttfamily
			arXiv:1503.05534}}].
	
	\bibitem{Gao:2019lta}
	J.~Gao, C.-D.~L\"u, Y.-L.~Shen, Y.-M.~Wang and Y.-B.~Wei, \emph{{Precision
			calculations of $B \to V$ form factors from soft-collinear effective theory
			sum rules on the light-cone}},
	\href{https://doi.org/10.1103/PhysRevD.101.074035}{\emph{Phys. Rev. D}
		{\bfseries 101} (2020) 074035}
	[\href{https://arxiv.org/abs/1907.11092}{{\ttfamily arXiv:1907.11092}}].
	
	\bibitem{Li:2009tx}
	R.-H.~Li, C.-D.~Lu and W.~Wang, \emph{{Transition form factors of B decays into
			p-wave axial-vector mesons in the perturbative QCD approach}},
	\href{https://doi.org/10.1103/PhysRevD.79.034014}{\emph{Phys. Rev. D}
		{\bfseries 79} (2009) 034014}
	[\href{https://arxiv.org/abs/0901.0307}{{\ttfamily arXiv:0901.0307}}].
	
	\bibitem{Deandrea:2001qs}
	A.~Deandrea and A.D.~Polosa, \emph{{The Exclusive $B_s \to \phi$ muon+ muon-
			process in a constituent quark model}},
	\href{https://doi.org/10.1103/PhysRevD.64.074012}{\emph{Phys. Rev. D}
		{\bfseries 64} (2001) 074012}
	[\href{https://arxiv.org/abs/hep-ph/0105058}{{\ttfamily hep-ph/0105058}}].
	
	\bibitem{Dubnicka:2016nyy}
	S.~Dubni\v{c}ka, A.Z.~Dubni\v{c}kov\'a, A.~Issadykov, M.A.~Ivanov, A.~Liptaj
	and S.K.~Sakhiyev, \emph{{Decay $B_s\to \phi \ell^+ \ell^-$ in covariant
			quark model}}, \href{https://doi.org/10.1103/PhysRevD.93.094022}{\emph{Phys.
			Rev. D} {\bfseries 93} (2016) 094022}
	[\href{https://arxiv.org/abs/1602.07864}{{\ttfamily arXiv:1602.07864}}].
	
	\bibitem{Issadykov:2022imz}
	A.~Issadykov, \emph{{$B_{s}^{0} \to \bar {K}{\text{*}}{{(892)}^{0}}{{\ell }^{ +
			}}{{\ell }^{ - }}$ Decay in Covariant Confined Quark Model}},
	\href{https://doi.org/10.1134/S154747712205020X}{\emph{Phys. Part. Nucl.
			Lett.} {\bfseries 19} (2022) 460}.
	
	\bibitem{Bailey:2015dka}
	J.A.~Bailey et~al., \emph{{$B\to Kl^+l^-$ Decay Form Factors from Three-Flavor
			Lattice QCD}}, \href{https://doi.org/10.1103/PhysRevD.93.025026}{\emph{Phys.
			Rev. D} {\bfseries 93} (2016) 025026}
	[\href{https://arxiv.org/abs/1509.06235}{{\ttfamily arXiv:1509.06235}}].
	
	\bibitem{Ball:2004rg}
	P.~Ball and R.~Zwicky, \emph{{$B_{d,s} \to \rho, \omega, K^*, \phi$ decay
			form-factors from light-cone sum rules revisited}},
	\href{https://doi.org/10.1103/PhysRevD.71.014029}{\emph{Phys. Rev. D}
		{\bfseries 71} (2005) 014029}
	[\href{https://arxiv.org/abs/hep-ph/0412079}{{\ttfamily hep-ph/0412079}}].
	
	\bibitem{Wu:2006rd}
	Y.-L.~Wu, M.~Zhong and Y.-B.~Zuo, \emph{{$B_{(s)}, D_{(s)} \to \pi, K,
			\eta, \rho, K^*, \omega, \phi$ Transition Form Factors and Decay Rates with
			Extraction of the CKM parameters $|V_{ub}|$, $|V_{cs}|$, $|V_{cd}|$}},
	\href{https://doi.org/10.1142/S0217751X06033209}{\emph{Int. J. Mod. Phys. A}
		{\bfseries 21} (2006) 6125}
	[\href{https://arxiv.org/abs/hep-ph/0604007}{{\ttfamily hep-ph/0604007}}].
	
	\bibitem{Cheng:2017bzz}
	W.~Cheng, X.-G.~Wu and H.-B.~Fu, \emph{{Reconsideration of the $B \to K^*$
			transition form factors within the QCD light-cone sum rules}},
	\href{https://doi.org/10.1103/PhysRevD.95.094023}{\emph{Phys. Rev. D}
		{\bfseries 95} (2017) 094023}
	[\href{https://arxiv.org/abs/1703.08677}{{\ttfamily arXiv:1703.08677}}].
	
	\bibitem{Wang:2017jow}
	Y.-M.~Wang, Y.-B.~Wei, Y.-L.~Shen and C.-D.~L\"u, \emph{{Perturbative
			corrections to B \textrightarrow{} D form factors in QCD}},
	\href{https://doi.org/10.1007/JHEP06(2017)062}{\emph{JHEP} {\bfseries 06}
		(2017) 062} [\href{https://arxiv.org/abs/1701.06810}{{\ttfamily
			arXiv:1701.06810}}].
	
	\bibitem{Lu:2018cfc}
	C.-D.~L\"u, Y.-L.~Shen, Y.-M.~Wang and Y.-B.~Wei, \emph{{QCD calculations of $B
			\to \pi, K$ form factors with higher-twist corrections}},
	\href{https://doi.org/10.1007/JHEP01(2019)024}{\emph{JHEP} {\bfseries 01}
		(2019) 024} [\href{https://arxiv.org/abs/1810.00819}{{\ttfamily
			arXiv:1810.00819}}].
	
	\bibitem{Gao:2021sav}
	J.~Gao, T.~Huber, Y.~Ji, C.~Wang, Y.-M.~Wang and Y.-B.~Wei, \emph{{B
			\textrightarrow{} D\ensuremath{\ell}\ensuremath{\nu}$_{\ell}$ form factors
			beyond leading power and extraction of |V$_{cb}$| and R(D)}},
	\href{https://doi.org/10.1007/JHEP05(2022)024}{\emph{JHEP} {\bfseries 05}
		(2022) 024} [\href{https://arxiv.org/abs/2112.12674}{{\ttfamily
			arXiv:2112.12674}}].
	
	\bibitem{Cui:2022zwm}
	B.-Y.~Cui, Y.-K.~Huang, Y.-L.~Shen, C.~Wang and Y.-M.~Wang, \emph{{Precision
			calculations of B$_{d,s}$ \textrightarrow{} \ensuremath{\pi}, K decay form
			factors in soft-collinear effective theory}},
	\href{https://doi.org/10.1007/JHEP03(2023)140}{\emph{JHEP} {\bfseries 03}
		(2023) 140} [\href{https://arxiv.org/abs/2212.11624}{{\ttfamily
			arXiv:2212.11624}}].
	
	\bibitem{Wang:2007an}
	W.~Wang, R.-H.~Li and C.-D.~Lu, \emph{{Radiative charmless $B_{(s)}
			\to V \gamma$ and $B_{(s)}\to A \gamma$ decays in pQCD
			approach}},  \href{https://arxiv.org/abs/0711.0432}{{\ttfamily
			arXiv:0711.0432}}.
	
	\bibitem{Xiao:2013lia}
	Z.-J.~Xiao and X.~Liu, \emph{{The two-body hadronic decays of $B_c$ meson in
			the perturbative QCD approach: A short review}},
	\href{https://doi.org/10.1007/s11434-014-0418-z}{\emph{Chin. Sci. Bull.}
		{\bfseries 59} (2014) 3748}
	[\href{https://arxiv.org/abs/1401.0151}{{\ttfamily arXiv:1401.0151}}].
	
	\bibitem{Jin:2020jtu}
	S.-P.~Jin, X.-Q.~Hu and Z.-J.~Xiao, \emph{{Study of $B_s\to K^{(*)}\ell^+
			\ell^-$ decays in the PQCD factorization approach with lattice QCD input}},
	\href{https://doi.org/10.1103/PhysRevD.102.013001}{\emph{Phys. Rev. D}
		{\bfseries 102} (2020) 013001}
	[\href{https://arxiv.org/abs/2003.12226}{{\ttfamily arXiv:2003.12226}}].
	
	\bibitem{Jin:2020qfp}
	S.-P.~Jin and Z.-J.~Xiao, \emph{{Study of $B_s\to{\phi}{\ell}^+{\ell}^{-}$
			Decays in the PQCD Factorization Approach with Lattice QCD Input}},
	\href{https://doi.org/10.1155/2021/3840623}{\emph{Adv. High Energy Phys.}
		{\bfseries 2021} (2021) 3840623}
	[\href{https://arxiv.org/abs/2011.11409}{{\ttfamily arXiv:2011.11409}}].
	
	\bibitem{Soni:2020bvu}
	N.R.~Soni, A.~Issadykov, A.N.~Gadaria, J.J.~Patel and J.N.~Pandya, \emph{{Rare
			$b \rightarrow d$ decays in covariant confined quark model}},
	\href{https://doi.org/10.1140/epja/s10050-022-00685-y}{\emph{Eur. Phys. J. A}
		{\bfseries 58} (2022) 39} [\href{https://arxiv.org/abs/2008.07202}{{\ttfamily
			arXiv:2008.07202}}].
	
	\bibitem{Lu:2011jm}
	C.-D.~Lu and W.~Wang, \emph{{Analysis of $B\to K^*_J (\to K \pi) \mu^+\mu^-$ in
			the higher kaon resonance region}},
	\href{https://doi.org/10.1103/PhysRevD.85.034014}{\emph{Phys. Rev. D}
		{\bfseries 85} (2012) 034014}
	[\href{https://arxiv.org/abs/1111.1513}{{\ttfamily arXiv:1111.1513}}].
	
	\bibitem{Ahmady:2019hag}
	M.~Ahmady, S.~Keller, M.~Thibodeau and R.~Sandapen, \emph{{Reexamination of the
			rare decay $B_s \to \phi \mu^+ \mu^-$ using holographic light-front QCD}},
	\href{https://doi.org/10.1103/PhysRevD.100.113005}{\emph{Phys. Rev. D}
		{\bfseries 100} (2019) 113005}
	[\href{https://arxiv.org/abs/1910.06829}{{\ttfamily arXiv:1910.06829}}].
	
	\bibitem{Li:2018rax}
	S.-P.~Li, X.-Q.~Li, Y.-D.~Yang and X.~Zhang, \emph{{$
			{R}_{D^{\left(*\right)}},{R}_{K^{\left(*\right)}} $ and neutrino mass in the
			2HDM-III with right-handed neutrinos}},
	\href{https://doi.org/10.1007/JHEP09(2018)149}{\emph{JHEP} {\bfseries 09}
		(2018) 149} [\href{https://arxiv.org/abs/1807.08530}{{\ttfamily
			arXiv:1807.08530}}].
	
	\bibitem{Barman:2018jhz}
	B.~Barman, D.~Borah, L.~Mukherjee and S.~Nandi, \emph{{Correlating the
			anomalous results in $b \to s$ decays with inert Higgs doublet dark matter
			and muon $(g-2)$}},
	\href{https://doi.org/10.1103/PhysRevD.100.115010}{\emph{Phys. Rev. D}
		{\bfseries 100} (2019) 115010}
	[\href{https://arxiv.org/abs/1808.06639}{{\ttfamily arXiv:1808.06639}}].
	
	\bibitem{DelleRose:2019ukt}
	L.~Delle~Rose, S.~Khalil, S.J.D.~King and S.~Moretti, \emph{{$R_K$ and
			$R_{K^*}$ in an Aligned 2HDM with Right-Handed Neutrinos}},
	\href{https://doi.org/10.1103/PhysRevD.101.115009}{\emph{Phys. Rev. D}
		{\bfseries 101} (2020) 115009}
	[\href{https://arxiv.org/abs/1903.11146}{{\ttfamily arXiv:1903.11146}}].
	
	\bibitem{Ordell:2019zws}
	A.~Ordell, R.~Pasechnik, H.~Ser\^odio and F.~Nottensteiner,
	\emph{{Classification of anomaly-free 2HDMs with a gauged U(1)' symmetry}},
	\href{https://doi.org/10.1103/PhysRevD.100.115038}{\emph{Phys. Rev. D}
		{\bfseries 100} (2019) 115038}
	[\href{https://arxiv.org/abs/1909.05548}{{\ttfamily arXiv:1909.05548}}].
	
	\bibitem{Marzo:2019ldg}
	C.~Marzo, L.~Marzola and M.~Raidal, \emph{{Common explanation to the
			$R_{K^{(*)}}$, $R_{D^{(*)}}$ and $\epsilon^\prime/\epsilon$ anomalies in a
			3HDM+$\nu_R$ and connections to neutrino physics}},
	\href{https://doi.org/10.1103/PhysRevD.100.055031}{\emph{Phys. Rev. D}
		{\bfseries 100} (2019) 055031}
	[\href{https://arxiv.org/abs/1901.08290}{{\ttfamily arXiv:1901.08290}}].
	
	\bibitem{Iguro:2018qzf}
	S.~Iguro and Y.~Omura, \emph{{Status of the semileptonic $B$ decays and muon
			g-2 in general 2HDMs with right-handed neutrinos}},
	\href{https://doi.org/10.1007/JHEP05(2018)173}{\emph{JHEP} {\bfseries 05}
		(2018) 173} [\href{https://arxiv.org/abs/1802.01732}{{\ttfamily
			arXiv:1802.01732}}].
	
	\bibitem{Iguro:2023jju}
	S.~Iguro, \emph{{Conclusive probe of the charged Higgs solution of P5' and
			RD(*) discrepancies}},
	\href{https://doi.org/10.1103/PhysRevD.107.095004}{\emph{Phys. Rev. D}
		{\bfseries 107} (2023) 095004}
	[\href{https://arxiv.org/abs/2302.08935}{{\ttfamily arXiv:2302.08935}}].
	
	\bibitem{Aslam:2009cv}
	M.J.~Aslam, C.-D.~Lu and Y.-M.~Wang, \emph{{$B \to K^*_0(1430) \ell^+ \ell^-$
			decays in supersymmetric theories}},
	\href{https://doi.org/10.1103/PhysRevD.79.074007}{\emph{Phys. Rev. D}
		{\bfseries 79} (2009) 074007}
	[\href{https://arxiv.org/abs/0902.0432}{{\ttfamily arXiv:0902.0432}}].
	
	\bibitem{Trifinopoulos:2019lyo}
	S.~Trifinopoulos, \emph{{B -physics anomalies: The bridge between R -parity
			violating supersymmetry and flavored dark matter}},
	\href{https://doi.org/10.1103/PhysRevD.100.115022}{\emph{Phys. Rev. D}
		{\bfseries 100} (2019) 115022}
	[\href{https://arxiv.org/abs/1904.12940}{{\ttfamily arXiv:1904.12940}}].
	
	\bibitem{Shaw:2019fin}
	A.~Shaw, \emph{{Looking for $B\rightarrow X_s \ell^+\ell^-$ in a nonminimal
			universal extra dimensional model}},
	\href{https://doi.org/10.1103/PhysRevD.99.115030}{\emph{Phys. Rev. D}
		{\bfseries 99} (2019) 115030}
	[\href{https://arxiv.org/abs/1903.10302}{{\ttfamily arXiv:1903.10302}}].
	
	\bibitem{Altmannshofer:2014cfa}
	W.~Altmannshofer, S.~Gori, M.~Pospelov and I.~Yavin, \emph{{Quark flavor
			transitions in $L_\mu-L_\tau$ models}},
	\href{https://doi.org/10.1103/PhysRevD.89.095033}{\emph{Phys. Rev. D}
		{\bfseries 89} (2014) 095033}
	[\href{https://arxiv.org/abs/1403.1269}{{\ttfamily arXiv:1403.1269}}].
	
	\bibitem{Bhattacharya:2014wla}
	B.~Bhattacharya, A.~Datta, D.~London and S.~Shivashankara, \emph{{Simultaneous
			Explanation of the $R_K$ and $R(D^{(*)})$ Puzzles}},
	\href{https://doi.org/10.1016/j.physletb.2015.02.011}{\emph{Phys. Lett. B}
		{\bfseries 742} (2015) 370}
	[\href{https://arxiv.org/abs/1412.7164}{{\ttfamily arXiv:1412.7164}}].
	
	\bibitem{Crivellin:2015lwa}
	A.~Crivellin, G.~D'Ambrosio and J.~Heeck, \emph{{Addressing the LHC flavor
			anomalies with horizontal gauge symmetries}},
	\href{https://doi.org/10.1103/PhysRevD.91.075006}{\emph{Phys. Rev. D}
		{\bfseries 91} (2015) 075006}
	[\href{https://arxiv.org/abs/1503.03477}{{\ttfamily arXiv:1503.03477}}].
	
	\bibitem{Falkowski:2015zwa}
	A.~Falkowski, M.~Nardecchia and R.~Ziegler, \emph{{Lepton Flavor
			Non-Universality in B-meson Decays from a U(2) Flavor Model}},
	\href{https://doi.org/10.1007/JHEP11(2015)173}{\emph{JHEP} {\bfseries 11}
		(2015) 173} [\href{https://arxiv.org/abs/1509.01249}{{\ttfamily
			arXiv:1509.01249}}].
	
	\bibitem{Bhattacharya:2016mcc}
	B.~Bhattacharya, A.~Datta, J.-P.~Gu\'evin, D.~London and R.~Watanabe,
	\emph{{Simultaneous Explanation of the $R_K$ and $R_{D^{(*)}}$ Puzzles: a
			Model Analysis}}, \href{https://doi.org/10.1007/JHEP01(2017)015}{\emph{JHEP}
		{\bfseries 01} (2017) 015}
	[\href{https://arxiv.org/abs/1609.09078}{{\ttfamily arXiv:1609.09078}}].
	
	\bibitem{Falkowski:2018dsl}
	A.~Falkowski, S.F.~King, E.~Perdomo and M.~Pierre, \emph{{Flavourful $Z'$
			portal for vector-like neutrino Dark Matter and $R_{K^{(*)}}$}},
	\href{https://doi.org/10.1007/JHEP08(2018)061}{\emph{JHEP} {\bfseries 08}
		(2018) 061} [\href{https://arxiv.org/abs/1803.04430}{{\ttfamily
			arXiv:1803.04430}}].
	
	\bibitem{Dwivedi:2019uqd}
	S.~Dwivedi, D.~Kumar~Ghosh, A.~Falkowski and N.~Ghosh, \emph{{Associated
			$Z^\prime$ production in the flavorful $U(1)$ scenario for $R_{K^{(*)}}$}},
	\href{https://doi.org/10.1140/epjc/s10052-020-7810-4}{\emph{Eur. Phys. J. C}
		{\bfseries 80} (2020) 263}
	[\href{https://arxiv.org/abs/1908.03031}{{\ttfamily arXiv:1908.03031}}].
	
	\bibitem{Capdevila:2020rrl}
	B.~Capdevila, A.~Crivellin, C.A.~Manzari and M.~Montull, \emph{{Explaining
			$b\to s\ell^+\ell^-$ and the Cabibbo angle anomaly with a vector triplet}},
	\href{https://doi.org/10.1103/PhysRevD.103.015032}{\emph{Phys. Rev. D}
		{\bfseries 103} (2021) 015032}
	[\href{https://arxiv.org/abs/2005.13542}{{\ttfamily arXiv:2005.13542}}].
	
	\bibitem{Hiller:2014yaa}
	G.~Hiller and M.~Schmaltz, \emph{{$R_K$ and future $b \to s \ell \ell$ physics
			beyond the standard model opportunities}},
	\href{https://doi.org/10.1103/PhysRevD.90.054014}{\emph{Phys. Rev. D}
		{\bfseries 90} (2014) 054014}
	[\href{https://arxiv.org/abs/1408.1627}{{\ttfamily arXiv:1408.1627}}].
	
	\bibitem{Gripaios:2014tna}
	B.~Gripaios, M.~Nardecchia and S.A.~Renner, \emph{{Composite leptoquarks and
			anomalies in $B$-meson decays}},
	\href{https://doi.org/10.1007/JHEP05(2015)006}{\emph{JHEP} {\bfseries 05}
		(2015) 006} [\href{https://arxiv.org/abs/1412.1791}{{\ttfamily
			arXiv:1412.1791}}].
	
	\bibitem{Becirevic:2017jtw}
	D.~Be\v{c}irevi\'c and O.~Sumensari, \emph{{A leptoquark model to accommodate
			$R_K^\mathrm{exp} < R_K^\mathrm{SM}$ and $R_{K^\ast}^\mathrm{exp} <
			R_{K^\ast}^\mathrm{SM}$}},
	\href{https://doi.org/10.1007/JHEP08(2017)104}{\emph{JHEP} {\bfseries 08}
		(2017) 104} [\href{https://arxiv.org/abs/1704.05835}{{\ttfamily
			arXiv:1704.05835}}].
	
	\bibitem{Cornella:2019hct}
	C.~Cornella, J.~Fuentes-Martin and G.~Isidori, \emph{{Revisiting the vector
			leptoquark explanation of the B-physics anomalies}},
	\href{https://doi.org/10.1007/JHEP07(2019)168}{\emph{JHEP} {\bfseries 07}
		(2019) 168} [\href{https://arxiv.org/abs/1903.11517}{{\ttfamily
			arXiv:1903.11517}}].
	
	\bibitem{DaRold:2019fiw}
	L.~Da~Rold and F.~Lamagna, \emph{{A vector leptoquark for the B-physics
			anomalies from a composite GUT}},
	\href{https://doi.org/10.1007/JHEP12(2019)112}{\emph{JHEP} {\bfseries 12}
		(2019) 112} [\href{https://arxiv.org/abs/1906.11666}{{\ttfamily
			arXiv:1906.11666}}].
	
	\bibitem{Popov:2019tyc}
	O.~Popov, M.A.~Schmidt and G.~White, \emph{{$R_2$ as a single leptoquark
			solution to $R_{D^{(*)}}$ and $R_{K^{(*)}}$}},
	\href{https://doi.org/10.1103/PhysRevD.100.035028}{\emph{Phys. Rev. D}
		{\bfseries 100} (2019) 035028}
	[\href{https://arxiv.org/abs/1905.06339}{{\ttfamily arXiv:1905.06339}}].
	
	\bibitem{Datta:2019bzu}
	A.~Datta, J.L.~Feng, S.~Kamali and J.~Kumar, \emph{{Resolving the $(g-2)_{\mu}$
			and $B$ Anomalies with Leptoquarks and a Dark Higgs Boson}},
	\href{https://doi.org/10.1103/PhysRevD.101.035010}{\emph{Phys. Rev. D}
		{\bfseries 101} (2020) 035010}
	[\href{https://arxiv.org/abs/1908.08625}{{\ttfamily arXiv:1908.08625}}].
	
	\bibitem{Crivellin:2019dwb}
	A.~Crivellin, D.~M\"uller and F.~Saturnino, \emph{{Flavor Phenomenology of the
			Leptoquark Singlet-Triplet Model}},
	\href{https://doi.org/10.1007/JHEP06(2020)020}{\emph{JHEP} {\bfseries 06}
		(2020) 020} [\href{https://arxiv.org/abs/1912.04224}{{\ttfamily
			arXiv:1912.04224}}].
	
	\bibitem{Iguro:2021kdw}
	S.~Iguro, J.~Kawamura, S.~Okawa and Y.~Omura, \emph{{TeV-scale vector
			leptoquark from Pati-Salam unification with vectorlike families}},
	\href{https://doi.org/10.1103/PhysRevD.104.075008}{\emph{Phys. Rev. D}
		{\bfseries 104} (2021) 075008}
	[\href{https://arxiv.org/abs/2103.11889}{{\ttfamily arXiv:2103.11889}}].
	
	\bibitem{Huang:2018rys}
	Z.-R.~Huang, M.A.~Paracha, I.~Ahmed and C.-D.~L\"u, \emph{{Testing Leptoquark
			and $Z^{\prime}$ Models via $B\to K_{1}(1270,1400)\mu^{+}\mu^{-}$ Decays}},
	\href{https://doi.org/10.1103/PhysRevD.100.055038}{\emph{Phys. Rev. D}
		{\bfseries 100} (2019) 055038}
	[\href{https://arxiv.org/abs/1812.03491}{{\ttfamily arXiv:1812.03491}}].
    
\bibitem{MunirBhutta:2020ber}
F.~Munir~Bhutta, Z.-R.~Huang, C.-D.~L\"u, M.A.~Paracha and W.~Wang, \emph{{New
  physics in b\textrightarrow{}s\ensuremath{\ell}\ensuremath{\ell}
  anomalies and its implications for the complementary neutral current
  decays}}, \href{https://doi.org/10.1016/j.nuclphysb.2022.115763}{\emph{Nucl.
  Phys. B} {\bfseries 979} (2022) 115763}
  [\href{https://arxiv.org/abs/2009.03588}{{\ttfamily arXiv:2009.03588}}].

\bibitem{Bhutta:2024zwj}
F.M.~Bhutta, A.~Rehman, M.J.~Aslam, I.~Ahmed and S.~Ishaq, \emph{{Angular
  observables of the four-fold $B \to K_{1}(1270,1400)(\to V P)
  \ell^{+}\ell^{-}$ decays in and beyond the Standard Model}},
  [\href{https://arxiv.org/abs/2410.20633}{{\ttfamily arXiv:2410.20633}}].

\bibitem{Ishaq:2013toa}
S.~Ishaq, F.~Munir and I.~Ahmed, \emph{{Lepton polarization asymmetries in $B
  /to K_{1}l^{+}l^{-}$ decay as a searching tool for new physics}},
  \href{https://doi.org/10.1007/JHEP07(2013)006}{\emph{JHEP} {\bfseries 07}
  (2013) 006}.
    
	\bibitem{Das:2018orb}
	D.~Das, B.~Kindra, G.~Kumar and N.~Mahajan, \emph{{$B\to
			K^\ast_2(1430)\ell^+\ell^-$ distributions at large recoil in the Standard
			Model and beyond}},
	\href{https://doi.org/10.1103/PhysRevD.99.093012}{\emph{Phys. Rev. D}
		{\bfseries 99} (2019) 093012}
	[\href{https://arxiv.org/abs/1812.11803}{{\ttfamily arXiv:1812.11803}}].
	
	\bibitem{Mohapatra:2021izl}
	M.K.~Mohapatra and A.~Giri, \emph{{Implications of light $Z'$ on semileptonic
			$B(B_s)\to T\{K_2^*(1430)(f_2'(1525))\}{\ell}^+{\ell}^-$
			decays at large recoil}},
	\href{https://doi.org/10.1103/PhysRevD.104.095012}{\emph{Phys. Rev. D}
		{\bfseries 104} (2021) 095012}
	[\href{https://arxiv.org/abs/2109.12382}{{\ttfamily arXiv:2109.12382}}].

	\bibitem{Rajeev:2020aut}
	N.~Rajeev, N.~Sahoo and R.~Dutta, \emph{{Angular analysis of $B_s\, \to\,
			f_{2}'\,(1525)\,(\to K^+\,K^-)\,\mu^+ \,\mu^-$ decays as a probe to lepton
			flavor universality violation}},
	\href{https://doi.org/10.1103/PhysRevD.103.095007}{\emph{Phys. Rev. D}
		{\bfseries 103} (2021) 095007}
	[\href{https://arxiv.org/abs/2009.06213}{{\ttfamily arXiv:2009.06213}}].

    \bibitem{Isgur:1990yhj}
    N.~Isgur and M.~B.~Wise, \emph{Weak transition form factors between heavy mesons}, 
    \href{https://doi.org/10.1016/0370-2693(90)91219-2}{\emph{Phys. Lett. B} 
    \textbf{237} (1990) 527}.

    \bibitem{Geng:2001vy}
	C.Q.~Geng, C.-W.~Hwang and C.C.~Liu, \emph{{Study of rare $B^+_{c} \to D_{d,s}^{(*)}
			\ell^+ \ell^-$ decays}},
	\href{https://doi.org/10.1103/PhysRevD.65.094037}{\emph{Phys. Rev. D}
		{\bfseries 65} (2002) 094037}
	[\href{https://arxiv.org/abs/hep-ph/0110376}{{\ttfamily hep-ph/0110376}}].
    
	\bibitem{LHCb:2019tea}
	{\scshape LHCb} collaboration, \emph{{Measurement of the $B_c^-$ meson
			production fraction and asymmetry in 7 and 13 TeV $pp$ collisions}},
	\href{https://doi.org/10.1103/PhysRevD.100.112006}{\emph{Phys. Rev. D}
		{\bfseries 100} (2019) 112006}
	[\href{https://arxiv.org/abs/1910.13404}{{\ttfamily arXiv:1910.13404}}].
	
	\bibitem{LHCb:2023lyb}
	{\scshape LHCb} collaboration, \emph{{A search for rare B \textrightarrow{}
			D\ensuremath{\mu}$^{+}$\ensuremath{\mu}$^{-}$ decays}},
	\href{https://doi.org/10.1007/JHEP02(2024)032}{\emph{JHEP} {\bfseries 02}
		(2024) 032} [\href{https://arxiv.org/abs/2308.06162}{{\ttfamily
			arXiv:2308.06162}}].
	
	\bibitem{Wang:2014yia}
	W.-F.~Wang, X.~Yu, C.-D.~L\"u and Z.-J.~Xiao, \emph{{Semileptonic decays
			$B_c^+$ \textrightarrow{} $D_{(s)}^{(*)}(l^+\nu_l,l^+l^-,\nu\bar{\nu}$) in
			the perturbative QCD approach}},
	\href{https://doi.org/10.1103/PhysRevD.90.094018}{\emph{Phys. Rev. D}
		{\bfseries 90} (2014) 094018}
	[\href{https://arxiv.org/abs/1401.0391}{{\ttfamily arXiv:1401.0391}}].

    \bibitem{LHCb:2013kwl}
	R.~Aaij \emph{et al.} [LHCb Collaboration], \emph{{Observation of $B^+_c \rightarrow J/\psi D_s^+$ and $B^+_c \rightarrow J/\psi D_s^{*+}$ decays}},
	\href{https://doi.org/10.1103/PhysRevD.87.112012}{\emph{Phys. Rev. D}
		{\bfseries 87} (2013) 112012}
	[\href{https://arxiv.org/abs/1304.4530}{{\ttfamily arXiv:1304.4530}}].
    
	\bibitem{Ebert:2010dv}
	D.~Ebert, R.N.~Faustov and V.O.~Galkin, \emph{{Rare Semileptonic Decays of $B$
			and $B_c$ Mesons in the Relativistic Quark Model}},
	\href{https://doi.org/10.1103/PhysRevD.82.034032}{\emph{Phys. Rev. D}
		{\bfseries 82} (2010) 034032}
	[\href{https://arxiv.org/abs/1006.4231}{{\ttfamily arXiv:1006.4231}}].
	
	\bibitem{Azizi:2008vv}
	K.~Azizi, F.~Falahati, V.~Bashiry and S.M.~Zebarjad, \emph{{Analysis of the
			Rare $B_c \to D^*_{s,d} \ell^+ \ell^-$ Decays in QCD}},
	\href{https://doi.org/10.1103/PhysRevD.77.114024}{\emph{Phys. Rev. D}
		{\bfseries 77} (2008) 114024}
	[\href{https://arxiv.org/abs/0806.0583}{{\ttfamily arXiv:0806.0583}}].
	
	\bibitem{Ivanov:2024iat}
	M.~A.~Ivanov, J.~N.~Pandya, P.~Santorelli and N.~R.~Soni, \emph{{Decay $B_c^+ \to D_{(s)}^{(*)+} \ell^+\ell^-$ within covariant confined quark model}},
	\href{https://doi.org/10.1103/PhysRevD.110.096003}{\emph{Phys. Rev. D}
		{\bfseries 110} (2024) 096003}
	[\href{https://arxiv.org/abs/2404.15085}{{\ttfamily arXiv:2404.15085}}].
	
	\bibitem{Yilmaz:2012ah}
	U.O.~Yilmaz, \emph{{Study of $B_c \to D_s^* \ell^+ \ell^-$ in Single Universal
			Extra Dimension}},
	\href{https://doi.org/10.1103/PhysRevD.85.115026}{\emph{Phys. Rev. D}
		{\bfseries 85} (2012) 115026}
	[\href{https://arxiv.org/abs/1204.1261}{{\ttfamily arXiv:1204.1261}}].
	
	\bibitem{Maji:2020zlq}
	P.~Maji, S.~Mahata, P.~Nayek, S.~Biswas and S.~Sahoo, \emph{{Investigation of
			rare semileptonic $ B_c \to (D_{s,d}^{(*)} ) \mu^+ \mu^-$ decays with
			non-universal $ Z'$ effect}},
	\href{https://doi.org/10.1088/1674-1137/44/7/073106}{\emph{Chin. Phys. C}
		{\bfseries 44} (2020) 073106}
	[\href{https://arxiv.org/abs/2003.12272}{{\ttfamily arXiv:2003.12272}}].
	
	\bibitem{Maji:2020wer}
	P.~Maji, S.~Biswas, P.~Nayek and S.~Sahoo, \emph{{Charged Higgs contribution on
			$B_c \rightarrow (D_s, D_s^*) l^+l^-$}},
	\href{https://doi.org/10.1093/ptep/ptaa048}{\emph{PTEP} {\bfseries 2020}
		(2020) 053B07} [\href{https://arxiv.org/abs/2003.07041}{{\ttfamily
			arXiv:2003.07041}}].
	
	\bibitem{Mohapatra:2021ynn}
	M.K.~Mohapatra, N.~Rajeev and R.~Dutta, \emph{{Combined analysis of
			$B_c\to D_s^{(*)}{\mu}^+{\mu}^-$ and
			$B_c\to D_s^{(*)}{\nu}{\bar\nu}$
			decays within $Z'$ and leptoquark new physics models}},
	\href{https://doi.org/10.1103/PhysRevD.105.115022}{\emph{Phys. Rev. D}
		{\bfseries 105} (2022) 115022}
	[\href{https://arxiv.org/abs/2108.10106}{{\ttfamily arXiv:2108.10106}}].
	
	\bibitem{Dutta:2019wxo}
	R.~Dutta, \emph{{Model independent analysis of new physics effects on $B_c \to
			(D_s,\,D^{\ast}_s)\,\mu^+\mu^-$ decay observables}},
	\href{https://doi.org/10.1103/PhysRevD.100.075025}{\emph{Phys. Rev. D}
		{\bfseries 100} (2019) 075025}
	[\href{https://arxiv.org/abs/1906.02412}{{\ttfamily arXiv:1906.02412}}].
	
	\bibitem{Li:2023mrj}
	Y.-S.~Li and X.~Liu, \emph{{Angular distribution of the FCNC process
			$B_c\to D_s^*(\to D_s{\pi}){\ell}^+{\ell}^-$}},
	\href{https://doi.org/10.1103/PhysRevD.108.093005}{\emph{Phys. Rev. D}
		{\bfseries 108} (2023) 093005}
	[\href{https://arxiv.org/abs/2309.08191}{{\ttfamily arXiv:2309.08191}}].

    \bibitem{LHCb:2024rto}
     R.~Aaij \textit{et al.} [LHCb Collaboration],
    \emph{Test of lepton flavour universality with $B_s^0 \rightarrow \phi \ell^+\ell^-$ decays},
    \href{https://arxiv.org/abs/2410.13748}{{\ttfamily arXiv:2410.13748}}
    [hep-ex].

	\bibitem{Mohapatra:2024lmp}
	M.K.~Mohapatra, A.K.~Yadav and S.~Sahoo, \emph{{Signature of (axial)vector
			operators in $B_c\to D_s^{(*)} \mu^+ \mu^-$ decays}},
	[\href{https://arxiv.org/abs/2409.01269}{{\ttfamily arXiv:2409.01269}}].

\bibitem{Zaki:2023mcw}
M.~Zaki, M.A.~Paracha and F.M.~Bhutta, \emph{{Footprints of New Physics in the
  angular distribution of
  $B_c\to D_s^*(\to D_s{\gamma},(D_s{\pi})){\ell}^+{\ell}^{-}$
  decays}}, \href{https://doi.org/10.1016/j.nuclphysb.2023.116236}{\emph{Nucl.
  Phys. B} {\bfseries 992} (2023) 116236}
  [\href{https://arxiv.org/abs/2303.01145}{{\ttfamily arXiv:2303.01145}}].

	\bibitem{SinghChundawat:2022ldm}
	N.R.~Singh~Chundawat, \emph{{New physics in
			$B\to K^*{\tau}^+{\tau}^-$: A model independent
			analysis}}, \href{https://doi.org/10.1103/PhysRevD.107.055004}{\emph{Phys.
			Rev. D} {\bfseries 107} (2023) 055004}
	[\href{https://arxiv.org/abs/2212.01229}{{\ttfamily arXiv:2212.01229}}].

    \bibitem{Ciuchini:2022wbq}
	M.~Ciuchini, M.~Fedele, E.~Franco, A.~Paul, L.~Silvestrini and M.~Valli, \emph{{Constraints on lepton universality violation from rare B decays}},
	\href{https://doi.org/10.1103/PhysRevD.107.055036}{\emph{Phys. Rev. D}
		{\bfseries 107} (2023) 055036}
	[\href{https://arxiv.org/abs/2212.10516}{{\ttfamily arXiv:2212.10516}}].

    \bibitem{Wen:2023pfq}
	Q.~Wen and F.~Xu, \emph{{Global fits of new physics in $b\to s$ after the $R_K^{(*)}$ 2022 release}},
	\href{https://doi.org/10.1103/PhysRevD.108.095038}{\emph{Phys. Rev. D}
		{\bfseries 108} (2023) 095038}
	[\href{https://arxiv.org/abs/2305.19038}{{\ttfamily arXiv:2305.19038}}].

    \bibitem{Buchalla:1995vs}
	G.~Buchalla, A.~J.~Buras and M.~E.~Lautenbacher, \emph{{Weak decays beyond leading logarithms}},
	\href{https://doi.org/10.1103/RevModPhys.68.1125}{\emph{Rev. Mod. Phys.} {\bfseries 68} (1996) 1125}
	[\href{https://arxiv.org/abs/hep-ph/9512380}{{\ttfamily hep-ph/9512380}}].

    \bibitem{Buras:1998raa}
	A.~J.~Buras, \emph{{Weak Hamiltonian, CP violation and rare decays}},
	in *Les Houches Summer School in Theoretical Physics, Session 68: Probing the Standard Model of Particle Interactions*,
	[\href{https://arxiv.org/abs/hep-ph/9806471}{{\ttfamily arXiv:hep-ph/9806471}}].

	\bibitem{Faessler:2002ut}
	A.~Faessler, T.~Gutsche, M.A.~Ivanov, J.G.~Korner and V.E.~Lyubovitskij,
	\emph{{The Exclusive rare decays $B \to$ K(K*) $\bar{\ell} \ell$ and $B_c
			\to$ D(D*) $\bar{\ell} \ell$ in a relativistic quark model}},
	\href{https://doi.org/10.1007/s1010502c0018}{\emph{Eur. Phys. J. direct}
		{\bfseries 4} (2002) 18}
	[\href{https://arxiv.org/abs/hep-ph/0205287}{{\ttfamily hep-ph/0205287}}].
	
	\bibitem{Bouchard:2013eph}
	{\scshape HPQCD} collaboration, \emph{{Rare decay $B \to K \ell^+ \ell^-$ form
			factors from lattice QCD}},
	\href{https://doi.org/10.1103/PhysRevD.88.054509}{\emph{Phys. Rev. D}
		{\bfseries 88} (2013) 054509}
	[\href{https://arxiv.org/abs/1306.2384}{{\ttfamily arXiv:1306.2384}}].
	
	\bibitem{ParticleDataGroup:2022pth}
	{\scshape Particle Data Group} collaboration, \emph{{Review of Particle
			Physics}}, \href{https://doi.org/10.1093/ptep/ptac097}{\emph{PTEP} {\bfseries
			2022} (2022) 083C01}.
	
	\bibitem{Altmannshofer:2008dz}
	W.~Altmannshofer, P.~Ball, A.~Bharucha, A.J.~Buras, D.M.~Straub and M.~Wick,
	\emph{{Symmetries and Asymmetries of $B \to K^{*} \mu^{+} \mu^{-}$ Decays in
			the Standard Model and Beyond}},
	\href{https://doi.org/10.1088/1126-6708/2009/01/019}{\emph{JHEP} {\bfseries
			01} (2009) 019} [\href{https://arxiv.org/abs/0811.1214}{{\ttfamily
			arXiv:0811.1214}}].
	
	\bibitem{UTfit:2006vpt}
	{\scshape UTfit} collaboration, \emph{{The Unitarity Triangle Fit in the
			Standard Model and Hadronic Parameters from Lattice QCD: A Reappraisal after
			the Measurements of $\Delta m_s$ and BR($B \to \tau \nu_\tau$)}},
	\href{https://doi.org/10.1088/1126-6708/2006/10/081}{\emph{JHEP} {\bfseries
			10} (2006) 081} [\href{https://arxiv.org/abs/hep-ph/0606167}{{\ttfamily
			hep-ph/0606167}}].
	
	\bibitem{Blake:2016olu}
	T.~Blake, G.~Lanfranchi and D.M.~Straub, \emph{{Rare $B$ Decays as Tests of the
			Standard Model}},
	\href{https://doi.org/10.1016/j.ppnp.2016.10.001}{\emph{Prog. Part. Nucl.
			Phys.} {\bfseries 92} (2017) 50}
	[\href{https://arxiv.org/abs/1606.00916}{{\ttfamily arXiv:1606.00916}}].
	
	\bibitem{Alok:2023yzg}
	A.K.~Alok, N.R.~Singh~Chundawat and A.~Mandal, \emph{{Investigating the
			potential of $R_{K^{(*)}}^{\tau\mu}$ to
			probe lepton flavor universality violation}},
	\href{https://doi.org/10.1016/j.physletb.2023.138289}{\emph{Phys. Lett. B}
		{\bfseries 847} (2023) 138289}
	[\href{https://arxiv.org/abs/2303.16606}{{\ttfamily arXiv:2303.16606}}].
	
	\bibitem{Alguero:2021anc}
	M.~Alguer\'o, B.~Capdevila, S.~Descotes-Genon, J.~Matias and M.~Novoa-Brunet,
	\emph{{$b\rightarrow s\ell ^+\ell ^-$ global fits after $R_{K_S}$ and
			$R_{K^{*+}}$}},
	\href{https://doi.org/10.1140/epjc/s10052-022-10231-1}{\emph{Eur. Phys. J. C}
		{\bfseries 82} (2022) 326}
	[\href{https://arxiv.org/abs/2104.08921}{{\ttfamily arXiv:2104.08921}}].
	
	\bibitem{Crivellin:2019dun}
	A.~Crivellin, D.~M\"uller and C.~Wiegand, \emph{{$b\to s\ell^+\ell^-$
			transitions in two-Higgs-doublet models}},
	\href{https://doi.org/10.1007/JHEP06(2019)119}{\emph{JHEP} {\bfseries 06}
		(2019) 119} [\href{https://arxiv.org/abs/1903.10440}{{\ttfamily
			arXiv:1903.10440}}].
	
	\bibitem{Bobeth:2016llm}
	C.~Bobeth, A.J.~Buras, A.~Celis and M.~Jung, \emph{{Patterns of Flavour
			Violation in Models with Vector-Like Quarks}},
	\href{https://doi.org/10.1007/JHEP04(2017)079}{\emph{JHEP} {\bfseries 04}
		(2017) 079} [\href{https://arxiv.org/abs/1609.04783}{{\ttfamily
			arXiv:1609.04783}}].
	
\end{thebibliography}
\providecommand{\url}[1]{#1}\providecommand{\href}[2]{#2}\begingroup\raggedright\endgroup



\end{document}